\documentclass{aa}  

\usepackage{graphicx}
\usepackage{txfonts}
\usepackage{amsmath}	
\usepackage{amssymb}
\usepackage{placeins}
\usepackage{caption}
\usepackage{lscape}
\usepackage{hyperref}
\usepackage{threeparttablex}
\hypersetup{
  colorlinks   = true, %Colours links instead of ugly boxes
  urlcolor     = blue, %Colour for external hyperlinks
  linkcolor    = blue, %Colour of internal links
  citecolor   = blue %Colour of citations
}
\usepackage{natbib}
\bibpunct{(}{)}{;}{a}{}{,}

\begin{document}

   \title{Search for giant planets in M67 V: a warm Jupiter orbiting the turn-off star S1429}

   \author{Luis Thomas\thanks{email: lthomas@usm.lmu.de}
          \inst{1,2}
          \and
          Roberto Saglia\inst{2,1}
          \and
          Luca Pasquini\inst{3}
          \and
          Anna Brucalassi\inst{4}
          \and
          P. Bonifacio\inst{5}
          \and
          Jos\'e Renan de Medeiros\inst{6}
          \and
          Izan de Castro Le\~ao\inst{6}
          \and
          Bruno Leonardo Canto Martins\inst{6,4}
          \and
          Henrik Lukas Ruh\inst{7}
          \and
          L.R. Bedin\inst{8}
           \and
           M. Libralato\inst{8}
           \and
           K. Biazzo\inst{9}
   }

   \institute{Universit\"ats-Sternwarte M\"unchen, Fakult\"at für Physik,  Ludwig-Maximilians-Universit\"at M\"unchen, Scheinerstr. 1, D-81679 M\"unchen, Germany\\
         \and
         Max-Planck-Institut f\"ur extraterrestrische Physik, Giessenbachstrasse 1, D-85748 Garching, Germany\\
         \and
         ESO-European Southern Observatory, Karl-Schwarzschild-Strasse 2, 85748 Garching bei M\"unchen, Germany\\
         \and
         Istituto Nazionale di Astrofisica, Osservatorio Astrofisico di Arcetri, 50125 Firenze, Italy\\
         \and
         GEPI, Observatoire de Paris, Universit\'{e} PSL, CNRS,  5 Place Jules Janssen, 92190 Meudon, France\\
         \and
         Departamento de F\'{i}sica Te\'{o}rica e Experimental, Universidade Federal do Rio Grande do Norte, Campus Universit\'{a}rio, Natal, RN, 59072-970, Brazil.\\
         \and
         Georg-August-Universit\"at, Institut f\"ur Astrophysik und Geophysik,
Friedrich-Hund-Platz 1, 37077 G\"ottingen\\
        \and
Instituto Nazionale di Astrofisica, Osservatorio Astronomico di Padova,
35122 Padova, Italy\\
\and
Instituto Nazionale di Astrofisica, Osservatorio Astronomico di Roma, via Frascati 33, Monte Porzio Catone, Italy\\
             }

   \date{Received January 15, 2024; accepted February 28, 2024}

  \abstract

   {Planets orbiting members of open or globular clusters offer a great opportunity to study exoplanet populations systematically as stars within clusters provide a mostly homogeneous sample at least in chemical composition and stellar age.
However, even though there have been coordinated efforts to search for exoplanets in stellar clusters, only a small number of planets has been detected.
One successful example is the seven-year radial velocity (RV) survey "Search for giant planets in M67" of 88 stars in the open cluster M67 which led to the discovery of five giant planets, including three close-in ($P < 10$ days) hot-Jupiters.}
   {In this work we continue and extend the observation of stars in M67 with the aim to search for additional planets.}
   {We conducted spectroscopic observations with the HPF, HARPS, HARPS-North, and SOPHIE spectrographs of 11 stars in M67. Six of our targets showed a variation or long-term trends in their RV during the original survey, while the other five were not observed in the original sample bringing the total number of stars to 93.}
   {An analysis of the radial velocities revealed one additional planet around the turn-off point star S1429 and gave solutions for the orbits of stellar companions around S2207 and YBP2018. S1429 b is a warm Jupiter on a likely circular orbit with a period of $77.48_{-0.19}^{+0.18}$ days and a minimum mass $\text{M} \sin i = 1.80 \pm 0.2$ M$_\text{J}$. We update the hot-Jupiter occurrence rate in M67 to include the five new stars, deriving  $4.2_{-2.3}^{+4.1} \%$ when considering all stars, and $5.4_{-3.0}^{+5.1}  \%$ if binary star systems are removed.}
   {}
   \keywords{Open clusters and associations: individual: M67 -- Techniques:
radial velocities -- Planets and satellites: detection
               }

   \maketitle
%
%-------------------------------------------------------------------

\section{Introduction}
With over 5000 exoplanets discovered to date, we are now able to see distinct features in the distribution of the exoplanet population with interesting implications for planet formation and evolution theories. Already the discovery of the first exoplanet, a gas giant orbiting a sun-like star in a very close orbit \citep{mayor1995jupiter} immediately challenged solar-system centric formation theories. With estimated occurrence rates of $ \sim 0.5 - 1 \%$ \citep{wright2012frequency,petigura2018california,zhou2019two}, we now know that these so-called hot-Jupiters (HJs) are not as common as their plethora of early detection suggested. However, the positions very close to their host stars still warrant an explanation. While in-situ formation for these planets can not be ruled out completely, the prevailing theory is that giant planets form beyond the ice line where feeding zones are larger and solid materials are more abundant and then migrate inwards after formation.
\\
Migration mechanisms are divided into two main groups: disk migration \citep{goldreich1980disk,lin1986tidal,ida2008toward} and high-eccentricity tidal migration \citep{rasio1996dynamical,wu2011secular,petrovich2015hot,bitsch2020eccentricity}. These two types of migrations shape the orbital architecture of their planetary systems in different ways and would therefore produce different populations of hot-Jupiter systems. Among others, the ellipticity distribution of warm-Jupiters as potential progenitors of hot-Jupiters, the obliquity distribution of hot-Jupiters, as well as the prevalence of close-in and farther-out companions compared to the population of colder giants can all be used as tracers of the two migration mechanisms \citep{dawson2018origins,fortney2021hot}. For example, the existence of nearby companions in hot-Jupiter systems would point to a more quiescent mechanism such as disc migration. In contrast, the absence of such planets indicates a more dynamic migration history such as high-eccentricity migration.
\\
Considering the properties of the host star adds another layer of complexity as the distribution of exoplanets can change around different types of stars. There is a well-established positive correlation between giant planet occurrence and the host star metallicity \citep{gonzalez1997stellar,santos2004spectroscopic,fischer2005planet}, which might even be stronger for hot-Jupiters than farther out giants 
\citep{jenkins2017new,petigura2018california}. The giant planet occurrence rate increases with stellar mass up until $\sim 2 \, M_\odot$ after which it drops rapidly \citep{johnson2010giant,reffert2015precise,ghezzi2018retired}. Both trends also seem to hold true for giant planets around evolved stars although the position of the occurrence peak is at a slightly lower mass $\sim 1.68 M_\odot$ \citep{wolthoff2022precise}. Contrary to early observations and predictions of a paucity of close-in giant planets around evolved stars \citep{sato2005radial,sato2008planetary,kennedy2009stellar,currie2009semimajor,villaver2009orbital}, recent analysis of the TESS data for these stars have determined the hot-Jupiter occurrence rate to be comparable to that of planets hosted by Main Sequence stars. This suggests that the effect of stellar evolution on the presence of giant planets, through planet engulfment, does not play a role until the star evolves enough along the giant branch so that its radius fills a significant part of the planetary orbit \citep{grunblatt2019giant,temmink2023occurrence}. 
\\
One way to look into the correlation between host star properties and planet occurrence more systematically would be to examine a large sample of stars in the same cluster and compare between planet hosts and stars without a planet. As stars in clusters should have similar ages and chemical compositions \citep{pasquini2004detailed,randich2005flames,de2007chemical}, they can function as a homogeneous sample to test the effects of stellar mass and evolution history on their planets, helping to distinguish between theories of planet formation, different evolution mechanisms like migration and investigate the significance of stellar companions and encounters in the dense cluster environment \citep[see e.g.][]{shara2016,hamers2017}. Additionally, it is easier to narrow down the properties of stars within clusters in particular the stellar age. 
\\
Even though the number of confirmed planets inside stellar clusters is still limited some key findings have come from observations of exoplanets in young open clusters. The existence of two super-Neptunes in the $\sim 10$ Myr old Upper Scorpius OB association indicates that some close-in planets must either form in-situ or migrate within the first 10 Myr which would have to be caused by interaction with the disc \citep{mann2016zodiacalb,david2016neptune}. The planetary systems of K2-100 and V1298Tau offer a rare opportunity to observe planets in an early stage of their evolution where the high UV radiation from their young host stars is thought to significantly shape the evolution of the young planets going forward \citep{barragan2019radial,david2019four,david2019warm}.
\\
Although in recent years the Kepler \citep{borucki2010kepler} and K2 \citep{howell2014k2} missions have been very successful in detecting planets around cluster member stars \citep{meibom2013same,obermeier2016k2,mann2016zodiacal,mann2017zodiacal,livingston2018three,rizzuto2018zodiacal,vanderburg2018zodiacal,livingston2019k2}, the first detections of planetary companions in star clusters came from radial velocity surveys: a long-period giant planet in the Hyades cluster \citep{sato2007planetary} and two substellar mass objects in NGC 2423 \citep{lovis2007planets}. Further RV searches revealed more planets, including two hot-Jupiters in the Praesepe open cluster \citep{quinn2012two,malavolta2016gaps} and a hot-Jupiter in the Hyades cluster \citep{quinn2014hd}. Together with non-detections \citep{paulson2004searching,takarada2020radial} these early surveys revealed giant planet occurrence rates comparable to that of field stars when corrected for the metallicity dependence. 
\\
A notable exception is the open cluster M67 that was the focus of the program "Search for giant planets in M67" \citep{pasquini2012search,brucalassi2014three,brucalassi2016search,brucalassi2017search} which this work is an extension of. This was a survey of 88 stars in the open cluster M67 observed over the span of 7 years between 2008 and 2015 with four different spectrographs (HARPS, HARPS-North, SOPHIE, and HRS). Additionally, observations of 14 giant stars in M67 with the CORALIE spectrograph between 2003 and 2005 were included. The scientific motivation for these observations was to study the impact of stellar properties on giant planet formation in a homogeneous sample. M67 is a good target as it is well studied to have age and metallicity close to the solar values \citep{randich2006element,yadav2008ground,onehag2011m67,onehag2014abundances} and has a large number of stars for an open cluster with a well populated main-sequence, turn-off point and giant branches. The main results are the detection of five exoplanets in M67 including three hot-Jupiters and two longer-period giant planets. In addition, 14 new candidates for binary or sub-stellar companions were reported. Considering only single stars an unusually high hot-Jupiter occurrence rate of $5.7^{+5.5}_{-3.0} \%$ was calculated. Recently, several works have looked into the formation of hot-Jupiters specifically in the environment of open clusters but none were able to produce a higher hot-Jupiter occurrence rate for M67 \citep{wang2020hot,li2023making,li2023makingb}.
\\
In this paper, we are reporting observations from an extension of the "Search for Giant Planets in M67" survey resulting in the detection of one additional planet and two new binary candidates in the open cluster. With our observations, we add five new targets to the original sample of 88 stars \citep{pasquini2012search}. All five of our new targets are at the turn-off point of M67 on their way to evolving off the main sequence. We chose to focus on turn-off point stars in this extension as they were the least represented group of stars in the original sample with 58 main-sequence ($\sim 10 \%$), 23 giant ($\sim 50 \%$), and only 7 turn-off point stars ($\sim 8 \%$). In brackets are the fractions of observed stars in the sample to the total number of stars of the given type in M67 from \cite{pasquini2012search}. The giant stars are overrepresented due to their larger brightness which makes them the easiest targets to observe.
\\
The goal of these observations was to contribute to the picture of planet occurrence across all types of stars in M67 to investigate the impact of stellar evolution on their planets. Furthermore, we also conducted observation for six targets that were identified in \citep{brucalassi2017search} to have long-term trends in their radial velocities indicative of a stellar companion to try to get orbital solutions for these systems.
\\
The paper is structured as follows. Section \ref{two} lists the conducted observations and describes the reduction and RV extraction procedure. In Section \ref{three} we describe the derivation of the properties for the observed stars, the analysis steps, and the results of our search for periodic signals in the RV data. Section \ref{four} includes a discussion of the validity of our detected planetary signal around S1429 as well as an assessment of the distribution of properties for the six exoplanets in M67. Lastly, we summarize the results in Section \ref{five}.
\section{Observations} \label{two}
\subsection{HPF observations and RV Calculation} \label{obsrv}
Between December 2019 and March 2022, we continued the observations of 11 stars in M67 with the Habitable Planet Finder spectrograph \citep[HPF,][]{mahadevan2012habitable,mahadevan2014habitable,mahadevan2018habitable}. We included six stars that showed potential long-term RV variations from \cite{brucalassi2017search} and five stars at the turn-off point of M67 which had no previously published data. 
\\
HPF is a high-resolution fiber-fed spectrograph operating in the near-infrared (808-1278 nm) installed at the 10m Hobby-Eberle-Telescope (HET) in the McDonald Observatory in Texas \citep{het1998,het2021}. It has three separate fibers for science, sky, and simultaneous calibration observations \citep{Kanodia_2018}. For our observations, we did not use the option to inject light of a laser frequency comb in the calibration fiber during the exposures for simultaneous calibration to avoid contamination of the science spectrum as most of our targets are relatively faint. However, we collected a spectrum with the sky fiber at the same time as our observations which was used to correct the spectra for sky emission. The exposure time was set to 900 seconds for 9 of the 11 targets. The two brighter stars S448 and S1557 have slightly shorter exposure times at 150s and 600s respectively.
\\
For the reduction we use GOLDILOCKS\footnote{\url{https://github.com/grzeimann/Goldilocks_Documentation}} which is an automated pipeline for HPF aiming to provide high-quality data reduction in a short timeframe. GOLDILOCKS corrects for bias, and non-linearity and calculates the slope and error images following the procedures from \cite{ninan2018habitable}. Next, it extracts the 1-D spectra by measuring the traces of the three fibers from a master Alpha Bright (flat field image), building a fiber profile along the trace, and using the simple formula of \cite{horne1986optimal} to calculate the spectra as a function of the column.
\\
Table \ref{tab:obs} gives an overview of the number of spectra that were acquired for each target.
\begin{table}
    \centering
    \caption{Number of observations of each star in our sample with the different spectrographs (HPF: Habitable Planet Finder spectrograph, H: HARPS, S: SOPHIE, HN: HARPS-North). Only observations that were not reported in \protect\cite{brucalassi2017search} are listed. Stars that were not in the original sample are marked in bold. The number in brackets shows the total number of observations over the entire survey for the given target.}
    \begin{tabular}{c|c|c|c|c|c}
    \hline
    \hline
      Star   & HPF& H & S & HN & Total\\
      \hline
       S488 & 6 & & & & 6 (43)\\
       S815 & 5 & & & & 5 (33)\\
       \textbf{S995} & \textbf{11} & \textbf{2} & \textbf{2} & & \textbf{15}\\
        \textbf{S1083} & \textbf{32} & \textbf{3} & \textbf{2} & \textbf{4} & \textbf{41}\\
       \textbf{S1268} & \textbf{38} & \textbf{2} & \textbf{2} & \textbf{4} & \textbf{46}\\
       \textbf{S1429} & \textbf{39} & \textbf{4} & \textbf{2} & \textbf{9} & \textbf{54}\\
       S1557 & 5 & & & & 5 (42) \\
       \textbf{S2207} & \textbf{14} & \textbf{4} & \textbf{2} & \textbf{6} & \textbf{26} \\
       YBP778 & 3 & 1 & & & 4 (27)\\
       YBP1062 & 3 & 1 & & & 4 (26)\\
       YBP1137 & 4 & & & & 4 (17)\\
       YBP2018 & 4 & 1 & &  & 5 (39)\\
       \hline
    \end{tabular}
    \label{tab:obs}
\end{table}
\\
The radial velocities are derived using a least-squares fitting algorithm implemented in the python program SpEctrum Radial Velocity AnaLyser \citep[SERVAL, ][]{zechmeister2018spectrum}. SERVAL was originally developed for the CARMENES spectrograph but has since been adapted to work with many other spectrographs such as HARPS, HARPS-N, ESPRESSO, SOPHIE, and HPF. It first constructs a high signal-to-noise template by coadding all available spectra for each star and then uses the template matching method \citep{anglada2012harps} to derive the radial velocity shifts by minimizing the $\chi ^2$ statistic. 
\\
For HPF, both science and sky fibers are divided by the blaze function of the instrument\footnote{\url{https://github.com/grzeimann/Goldilocks_Documentation/blob/master/hpf_blaze_spectra.fits}}. We use a cubic spline to interpolate the sky emission at the wavelength solution of the science channel and subtract the scaled sky emission from the science flux. Large residuals may remain in strongly contaminated regions. Therefore, we additionally mask regions of strong sky emission. The sky emission mask is generated in a similar manner as described in \citep{metcalf2019}. We first generate a master sky-background spectrum by co-adding sky spectra and fitting the continuum sky background. We then identify sky emission lines as regions that deviate more than five sigma from the continuum. As large parts of the HPF spectra are severely contaminated by tellurics we only use the spectral orders 4,5,6,15,16,17 and 26 from the 28 available orders to calculate the radial velocities. 
Spectra are also corrected for the barycentric motion using the python package barrycorrpy \citep{Kanodia_2018}. We then create templates for each spectral order by coadding all observations for each order. The final RV is calculated as a weighted average of the radial velocity shifts of the individual orders.
\\
SERVAL also offers the option of using synthetic stellar spectra (e.g. PHOENIX models) as a template for the RV calculation. While the method of using a synthetic template typically has a lower precision than the template construction from the observations given a large enough number of good-quality spectra, in certain cases it is preferable to have an accurately calibrated value for the radial velocity. For a subset of our sample, we are interested in studying long-term trends in their radial velocities as a sign of a potential binary system. In this case, a m/s precision is not needed and it is advantageous to have an accurate radial velocity to compare the data from different instruments taken over the span of many years. For these targets, we follow a similar approach as the previous works in this project outlined in \cite{pasquini2012search}. 
\\
We calculate the offset between our HPF and HARPS spectra which can then be used to correct the HPF radial velocities to the zero-point of HARPS. For this, we only used data from stars that showed no long-term RV variations to calculate the offset as others either did not have HARPS spectra at all or had a significant time gap between the HARPS and HPF spectra which made the offset calculation dependent on the chosen model for the RV trend. To calculate the offset we derived RVs for S1083, S1268, S1429, HD32923, S1557, YBP778 and YBP1062 from the HPF spectra using a PHOENIX stellar template with parameters that most closely matched the turn-off point stars ($T_{eff} = 6000 \ K$, $\log g = 4.0$, [Fe/H] $=0$) to compare to the DRS reduced HARPS RVs. 
\\
For S1429 we subtracted the best fit Keplerian orbit from Section \ref{sec1429} before calculating the offsets. S1083 and S1268 showed no significant periodicity in the periodogram analysis and thus did not have any corrections applied before calculating the offsets of their HPF measurements. As these three objects were the focus of our planet search in this sample of stars they had the most observations and therefore also had the biggest impact in determining the offset.
\\
Additionally, we used HPF spectra for four more stars: HD32923, S1557, YBP778, and YPB1062. HD32923 is a known RV standard star with little variation and trends in its radial velocities. It is therefore well suited to compare the offsets between different instruments. The other three stars either had HARPS data taken at the same epochs as we observed with HPF (YBP778 and YBP1062) or in the case of S1557 did not have HARPS data from the same epoch, but showed no long-term RV trends and therefore archival HARPS data could be used for the offset calculation. In total we used four HPF spectra of HD32923 and compared them to archival HARPS data of the star, four HPF spectra of S1557, with 13 HARPS spectra from the previous observing campaign that show no significant RV variation, three HPF spectra of YBP778, and two of YBP1062 both of which have one HARPS spectrum taken during the time of the HPF observations.
\\
We calculate the offset between each HPF datapoint and the weighted average of the HARPS DRS RV for all planets individually. The final offset given from a weighted average of the offsets of all individual data points was calculated as $466 \pm 13$ m/s. The error was derived from the scatter of the data points. The offset was subtracted from all HPF RVs of the targets that showed long-term trends to have accurately calibrated values for the analysis. 
\\
For the three stars in our sample which were examined for potential planets, we used the standard method of constructing the template from our observations to get the highest possible precision in our radial velocities. During the planet search, individual offsets are fitted for each spectrograph (see Section \ref{analysissec}). These are then used to correct the relative HPF radial velocities to the zero-point of HARPS.
\\
The HPF RVs for all stars, corrected to the zero-point of HARPS with one of the two methods described above are reported in Appendix \ref{tables}.
\subsection{Additional spectroscopic observations}
We also included unpublished spectroscopic observations from the HARPS, HARPS-North, and SOPHIE spectrographs in our analysis. The number of observations from each spectrograph is shown in Table \ref{tab:obs}. HARPS is the high-resolution spectrograph at the ESO 3.6m telescope \citep{mayor2003} covering the optical wavelength range between 378-691 nm. We observed in the high-efficiency mode with a resolving power of 90,000 (compared to 115,000 in the high-resolution mode), to get a better signal-to-noise on our relatively faint targets. 
\\
We also performed observations with HARPS-North (HARPS-N) which is a copy of the HARPS spectrograph installed at the 3.6m Telescopio Nazionale Galileo (TNG) on La Palma Canary Island. Lastly, we included 10 observations from the SOPHIE spectrograph \citep{perruchot2008sophie,bouchy2013} at the Observatoire Haute Provence 1.93 m telescope collected using the high-efficiency mode (R 40,000). 
\\
For all three spectrographs, we used similar automated pipelines to reduce their spectra and extract radial velocities by cross-correlating them with a G2-type mask obtained from the Sun's spectrum \citep{pepe2002}. The resulting radial velocities are thus including the systemic velocity. For two of the five targets that have either SOPHIE and/or HARPS-North observations, we found large variations indicative of stellar companions. Unlike, the HPF RVs derived from the PHOENIX templates we don't have a lot of spectra from HARPS-North and SOPHIE and we do not expect large differences between the zero-points of SOPHIE and HARPS-North compared to HARPS as they are all calculated with the same technique and mask. We therefore do not apply any corrections for the analysis of these two datasets.
\\
For the other three stars, we looked for planetary signals so we fit the datasets from each spectrograph with individual offsets and excess noise terms as the precision required is higher than for the two stars with large long-term RV variations (see Section \ref{analysissec}).
\\
All radial velocities from the three spectrographs are reported in Appendix \ref{tables}.
\section{Analysis} \label{three}
\subsection{Stellar Parameters} \label{ari}
\begin{figure}
    \centering
    \includegraphics[width=1\columnwidth]{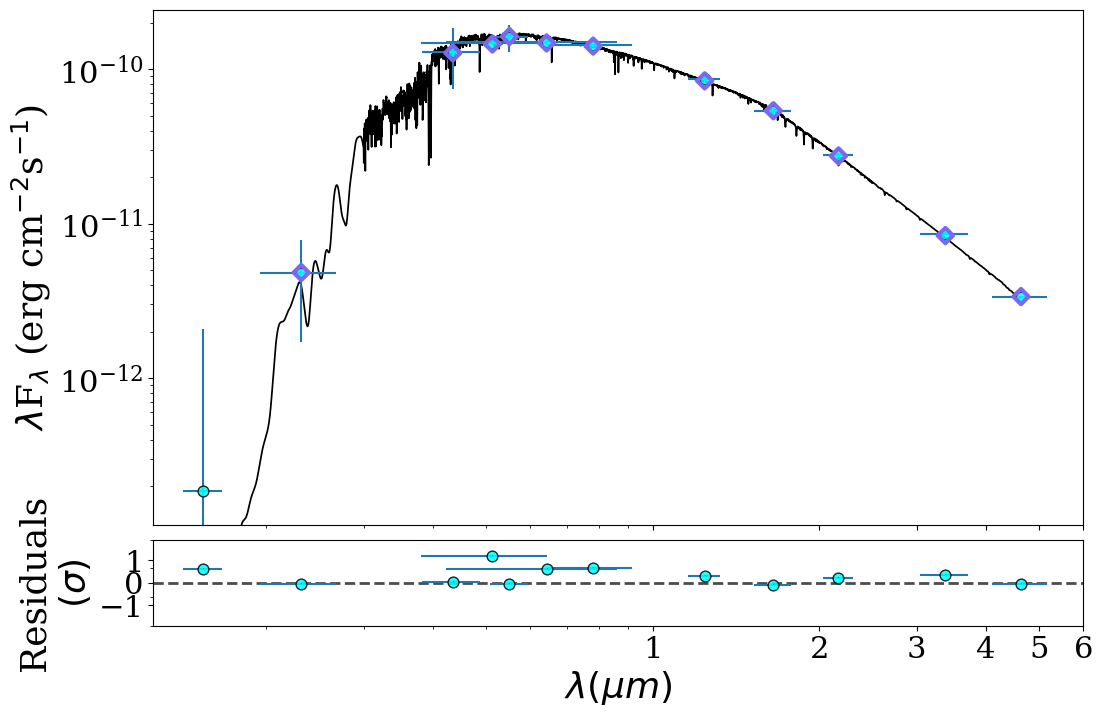} 
    \includegraphics[width=1\columnwidth]{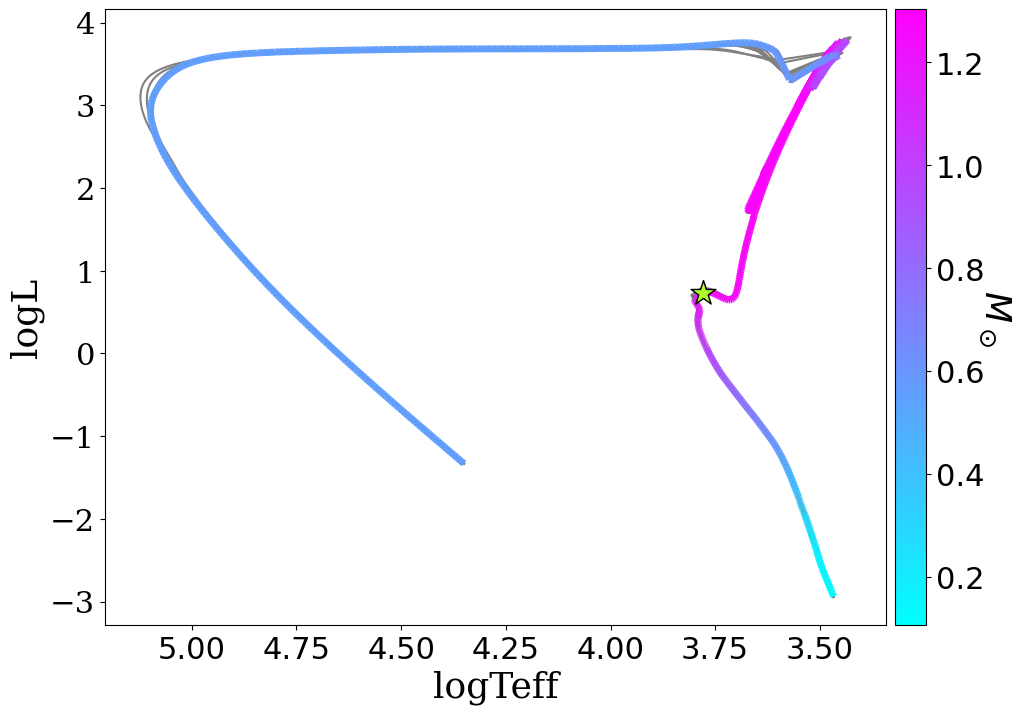} 
    \caption{Upper panel: SED fit for S1429. Datapoints in blue are flux values from available broadband photometry with the error in the x-axis direction corresponding to the passband of the filter and the y-error corresponding to the 1-$\sigma$ uncertainties of the flux. The best-fitting Phoenix model is plotted in black. Purple diamonds represent the flux value of the synthetic photometry from the Phoenix model in each passband. Below the residuals normalized to the errors of the photometry are shown. Lower panel: isochrone fit for S1429.}
    \label{fig:sed}
\end{figure}

To derive the stellar parameters we model the stellar energy distribution (SED) by fitting PHOENIX \citep{husser2013new}, BT-Settl \citep{Allard2012}, Kurucz \citep{1993KurCD..13.....K} and Castelli and Kurucz \citep{Castelli2004} model grids convolved with the filter response functions of available broadband photometry using the python code ARIADNE \citep{vines2022}. For our SED fitting, we use magnitudes from 2MASS JHKs, Johnson UBV, GAIA DR2 G, RP and BP, ALL-WISE W1 and W2, and where available GALEX NUV and FUV.
During the fitting process, the model grids are interpolated in T$_{\text{eff}}$ - log g - [Fe/H] space. To obtain the observed fluxes, the fluxes in each grid are multiplied with a normalization factor $(R_*/D)^2$, dependent on the stellar radius $R_*$ and the distance to the star $D$. Additional free parameters in the fitting are the interstellar extinction $A_v$ and a term for the excess uncertainty of each photometric measurement \citep{vines2022}.  
\\
For the spectroscopic parameters T$_{\text{eff}}$, log g and [Fe/H], we use normal priors centered around the values from the APOGEE data of the SDSS Data Release 16 \citep{jonsson2020apogee} with three times the uncertainty as the variance. For the other values, we use the default priors of ARIADNE which takes uniform priors between 0.5 and 20 $R_\odot$ and between 0 and the maximum line-of-sight extinction from the SFD galactic dust map for $R_*$ and $A_v$, a normally distributed prior for the distance from the Bailer-Jones estimates from GAIA EDR3 and Gaussian priors with three times the squared uncertainty as the variance for the excess noise terms. The nested sampling algorithm dynesty is then used to sample the posterior distribution of the parameters and estimate the Bayesian evidence. The final values for the parameters are then calculated using Bayesian model averaging where the weighted average of each parameter is computed based on the relative probabilities of the models. This is done to account for potential biases and offsets that are produced by uncertainties in the evolutionary tracks of these models. An example of the SED for S1429 with the best fitting model is shown in Figure \ref{fig:sed}. 
\\
The mass and age of the stars are estimated using the isochrones \citep{morton2015isochrones} software package to compare interpolated MESA Isochrones and Stellar Tracks \citep[MIST,][]{dotter2016mesa,choi2016mesa} evolutionary models to the parameters derived from the SED fit. We limit the age range to the estimated age of M67 ($3.5 - 5.0$ Gyr). All derived parameters for the stars in our sample are listed in Table \ref{tab:star}.
\begin{table*}
    \centering
    \caption{Stellar parameters derived from the SED analysis and MIST Isochrones fit for the eleven stars in our sample.}
    \resizebox{\textwidth}{!}{
    \begin{tabular}{c|c|c|c|c|c|c|c|c|c|c|c|c}
    \hline
    \hline
      Parameters   & S488 & S815 & S995 & S1083 & S1268 & S1429 & S1557 & S2207 & YBP778 & YBP1062 & YBP1137 & YBP2018 \\
      \hline
       $\alpha$ (J2000)  &  	08:50:12.30 & 	08:50:54.39 & 08:51:20.12& 08:51:27.42 & 08:51:49.95 & 08:52:03.56 & 08:52:16.56 & 08:51:32.45 & 08 51 20.80 & 08:51:36.08 &  	08:51:12.02 & 08:51:39.41\\
        $\delta$ (J2000) & 11:51:24.50 & 11:56:29.08 & 11:46:41.75 &11:53:26.58 & 11:49:31.20 & 11:41:23.91 & 11:19:38.23 & 11:47:52.37 & 11 45 02.51 & 11:47:47.04 & 11:48:26.73 & 12:00:57.18\\
        Spec.type & M0III  & F9V & F9V &F9.5V & F9.5V & F9.5V & K5.5V & G1V & G0V & G3V & G7V & G4V\\
       m$_V$ [mag]&  8.9 & 12.9 & 12.8 & 12.8 & 12.7 & 12.8 & 10.1 & 12.8 & 13.1 & 14.5 & 14.9 &14.6\\ 
       M$_*$ [M$_\odot$] & $1.26 \pm 0.13$  & $1.2 \pm 0.1$ & $1.24 \pm 0.04$ &$1.26 \pm 0.04$ & $1.24 \pm 0.04$ &$1.26 \pm 0.04$ & $1.24 \pm 0.04$ & $1.26 \pm 0.05$ & $1.27 \pm 0.05$ & $1.01 \pm 0.04$ & $0.94 \pm 0.04$ & $1.03 \pm 0.04$\\ 
       R$_*$ [R$_\odot$] &  $51.6 \pm 1.5$ & $2.08 \pm 0.4$  & $2.03 \pm 0.05$ &$2.13 \pm 0.05$ &$2.16 \pm 0.05$ &$2.13 \pm 0.04$ & $17.6 \pm 0.4$ & $2.26 \pm 0.08$ & $1.8 \pm 0.04$ & $1.03 \pm 0.04$ & $0.91 \pm 0.04$ & $1.03 \pm 0.03$\\ 
       $\log g$ [cgs]& $1.35 \pm 0.11$  & $4.00 \pm 0.18$ & $4.03 \pm 0.19$ & $4.01 \pm 0.18$ & $3.93 \pm 0.19$ &$4.01 \pm 0.16$ & $2.16 \pm 0.13$ & $3.8 \pm 0.2$ &$4.16 \pm 0.21$ & $4.6 \pm 0.6$ & $4.6 \pm 0.5$ & $4 \pm 1$\\ 
       $T_{eff}$ [K]& $3870 \pm 30$ & $6050 \pm 50$ &$6050 \pm 60$&$6000 \pm 60$ & $5960 \pm 50$ &$6000 \pm 40$ & $4350 \pm 30$ & $5800 \pm 100$ & $5900 \pm 60$ & $5740 \pm 70$ & $5540 \pm 80$ & $5680 \pm 80$\\ 
       $\left[\textrm{Fe/H}\right]$ [dex] & $-0.05 \pm 0.02$  & $-0.06 \pm 0.2$ & $-0.03 \pm 0.05$ &$-0.008 \pm 0.04$ & $-0.03 \pm 0.03$ & $-0.013 \pm 0.023$ & $0.002 \pm 0.015$ & $-0.02 \pm 0.04$ & $0.00 \pm 0.04$ & $-0.08 \pm 0.14$ & $-0.09 \pm 0.17$ & $-0.02 \pm 0.4$\\
       Age [Gyr] & $4.3 \pm 1.1$ & $4.3 \pm 0.6$ & $4.5 \pm 0.4$ &$4.4 \pm 0.4$ & $4.4 \pm 0.4$ & $4.3 \pm 0.4$ & $4.9 \pm 0.4$ & $4.3 \pm 0.5$ & $4.0 \pm 0.6$ & $4.6 \pm 1.4$ & $3.8 \pm 1.2$ & $4 \pm 1$\\ 
       D [pc] & $841 \pm 11$ & $884 \pm 12$ & $861 \pm 10$ & $887 \pm 13$ & $817 \pm 6$ & $870 \pm 8$ & $836 \pm 10$ & $840 \pm 22$ & $840 \pm 11$ & $843 \pm 24$ & $834 \pm 15$ & $874 \pm 16$\\
       A$_V$ [mag]& $0.04 \pm 0.02$ & $0.05 \pm 0.02$ & $0.04 \pm 0.02$ &$0.04 \pm 0.03$ & $0.05 \pm 0.03$ & $0.034 \pm 0.019$ & $0.048 \pm 0.024$ & $0.05 \pm 0.04$ & $0.03 \pm 0.03$ & $0.05 \pm 0.03$ & $0.03 \pm 0.04$ & $0.07 \pm 0.06$\\ 
       \hline
    \end{tabular}
    \label{tab:star}
    }
\end{table*}

\subsection{Periodogram Analysis and Orbital Solutions} \label{analysissec}
\subsubsection{RVSearch}
To search for periodic signals in our data, we use the Python package RVSearch \citep{rosenthal2021california} which is specifically designed for uninformed "blind" searches. RVSearch uses a $\Delta$BIC goodness-of-fit periodogram to search for periodicities in the data by comparing the fit to the radial velocity data of a single-planet Keplerian model from RadVel \citep{fulton2018radvel} and a model without a planet over a period grid.
After the periodogram is constructed the program performs a linear fit to a log-scale histogram of the periodogram's power values to compute an empirical false-alarm probability (FAP) \citep{rosenthal2021california}. If a signal exceeds the given threshold for the false-alarm probability (in this work FAP $ \leq 0.1 \% $) RVSearch performs a maximum a posteriori fit to derive the orbital parameters for the current planet configuration. During this fit all parameters including eccentricity are left free, however, two hardbound priors constraining $K > 0 $ and $0 \leq e<1$ are given.
\\
In the next step, another planet is added to the model, and the periodogram search is repeated this time with the known Keplerian orbit from the previous fit as the comparative model for the $\Delta$BIC calculation. This sequence is performed iteratively until no further signals in the periodogram exceeding the specified false-alarm probabilities are found. After no more significant signals are found, RVSearch samples the posteriors of the current best-fit models via an affine-invariant sampling that is implemented in RadVel using the emcee package \citep{foreman2013emcee} to derive the parameter uncertainties. 
\\
One of the advantages of the iterative approach of RVSearch over the Lomb-Scargle-periodogram analysis is that it fits for instrument-specific parameters such as the RV-offset and the instrument jitter throughout the search. This lets us combine the data from multiple instruments without the need to correct all of them to a common zero-point enabling us to use the more precise coadded templates from Serval to derive the RVs of our planet host candidates. It also lets us consider the difference in the precision of the various instruments through an individual jitter term for each spectrograph.

\subsubsection{Stars included in the original Sample}
\begin{figure*}
    \centering
    \includegraphics[width=1.75\columnwidth]{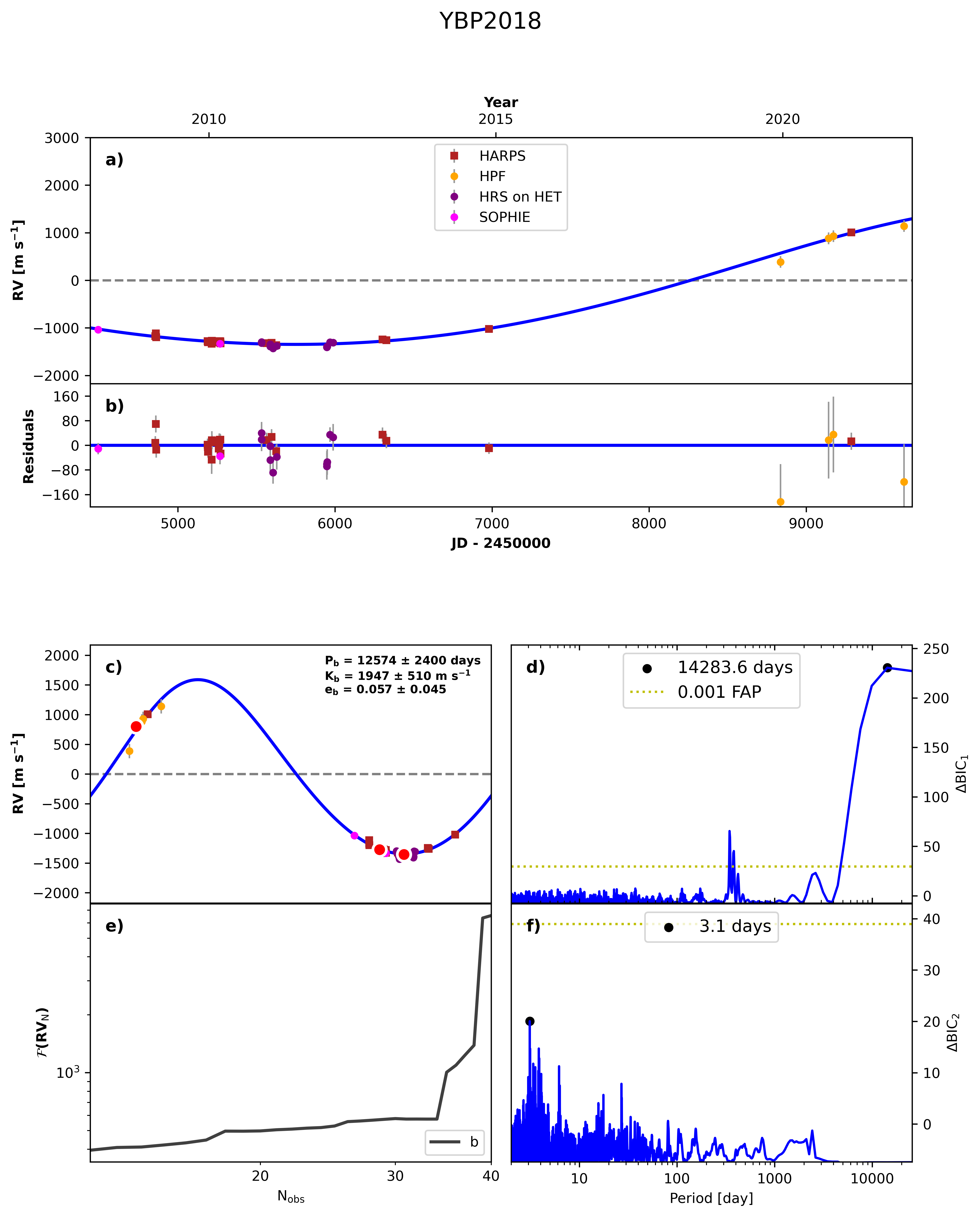} 
    \caption{RVSearch summary plot for YBP 2018 showing the radial velocity time-series with the best-fit model (Panel a) and residuals (Panel b). Datapoints labeled HRS are from the old high-resolution spectrograph that was installed at the HET. All of the radial velocity observations are corrected to the zero-point of HARPS and therefore treated as coming from one instrument. Panel c shows the phase-folded RV curve with the values derived from the MCMC. Panel d and f show the periodogram of the one and two planet models and Panel e plots the significance of the radial velocity signal as a function of the number of observations.}
    \label{fig:2018}
\end{figure*}

We included seven stars from the original 88 in our sample to be observed with HPF. All of them showed either long-term trends or potential planetary signals in their radial velocities which we were looking to track with HPF and if possible determine a minimum mass and period for the companion. 
\\
To observe the long-term trend it was preferable to sacrifice precision in order to get relatively accurate calibrated radial velocities as the fitting of the offset with RVSearch is not very constrained when the observation window spans less than one period, especially for the relatively low number of observations we have for this work.
\\
We therefore derived our HPF RVs using the same Phoenix template from Section \ref{obsrv} and corrected them to the zero-point of HARPS using the calculated offset. In the following periodogram analysis, all observations are treated as coming from the same instrument. Out of the seven datasets, three did not show a significant peak in the periodogram from RVSearch. For S488, S1557, and YBP1137 the quality of the HPF spectra was not high enough as the error on the RV points was as large as the amplitude of the trend we wanted to observe.
\\
For the other four targets, RVSearch did detect a significant periodicity in the data. However, all of the peaks in the periodograms were at periods larger than the observational baseline. This caused the MCMC to fail in returning a well-sampled posterior in three cases (S815, YBP778, and YBP1062) which resulted in unrealistic errors for the periods that were larger than the periods themselves. 
A similar behavior has previously been noted in \cite{laliotis2023doppler} where the treatment of signals with periods larger than the observational baseline showed to be problematic. While the periodogram can only fully resolve periods shorter than the observational baseline the MCMC is not subject to this limitation and thus fails to return a well-sampled posterior. In Appendix \ref{appb} we show the plotted radial velocities and periodograms including the data from \cite{brucalassi2017search} for the targets where no significant periodic signals were found.
\\
For YBP2018 the MCMC did converge and produced a distinct solution albeit with relatively large errors. Figure \ref{fig:2018} summarizes the results from RVSearch. In \cite{brucalassi2017search} YBP2018 was presented as a potential planet candidate with peak-to-peak RV variation of 400 m/s and a tentative Keplerian solution with a period of $\sim 2487$ days. With our new data, we can rule out a planetary companion as it proves that the star has a much larger orbital period of $12574 \pm 2400$ days and an RV amplitude of $1947 \pm 510$ m/s. As seen in Panel e of Figure \ref{fig:2018} the significance of the RV signal strongly increases after the inclusion of our new data in the fit. Given the mass of $1.03$ M$_\odot$ for YBP2018 the companion has a minimum mass of $0.22 \pm 0.06$ M$_\odot$.
\subsubsection{Potential stellar companions for S995 and S2207}
\begin{figure*}
    \centering
    \includegraphics[width=1.75\columnwidth]{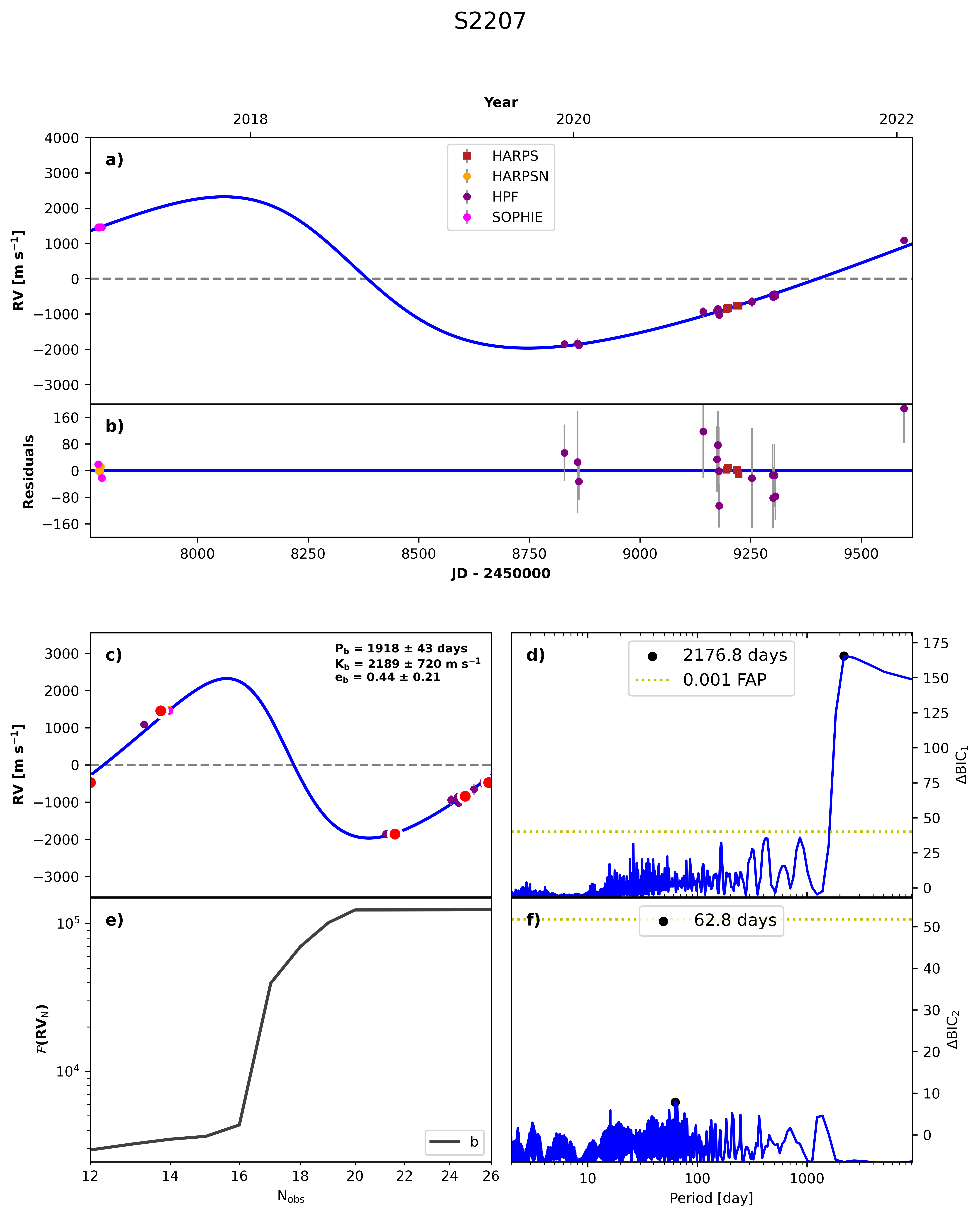}
    \caption{Same as Figure \ref{fig:2018} but for S2207.}
    \label{fig:2207}
\end{figure*}

S995 and S2207 are the first two stars in this paper which were not in the original sample of \cite{pasquini2012search}. We started observations for both stars in December 2019. After the first year of observations, a large trend in the radial velocities of both stars indicative of a stellar companion was identified. Thus, they were only sparsely observed going forward as the focus was set on the other three stars as potential planet hosts in the sample. In total, we collected 11 spectra for S995 and 14 spectra for S2207 with HPF. Additionally, we have 2 SOPHIE and 2 HARPS observations for S995 as well as 6 HARPS-North, 2 SOPHIE, and 4 HARPS spectra for S2207. 
\\
We used the same approach that was described in the previous section for the seven stars of the original sample to search for periodicity and fit the potential orbits. The HPF RVs for both stars were calculated with a PHOENIX template and corrected to the zero-point of HARPS. The periodogram for S995 has the highest peak at 365 days with similar peaks at integer fractions and multiples of the signal ($\sim 90$ days, $\sim 730$ days). No peak exceeds the $0.1 \%$ false-positive threshold and looking at the RV data the potential orbit looks to be highly eccentric (see Figure \ref{fig:995}). Additional data is needed to find the correct period and orbital configuration for this potential companion.
\\
For S2207 the orbital fitting worked revealing a stellar companion with a period of $1882 \pm 24$ days, an eccentricity of $0.5 \pm 0.13$, and an amplitude of $2847 \pm 860$ m/s. Figure \ref{fig:2207} shows the summary of the results for this star. Given the derived mass for S2207 the companion would have a minimum mass of $0.17 \pm 0.05$ M$_\odot$ The large error on the mass of the companion is due to the incomplete sampling of the curve and can be improved with future observations.
\subsubsection{RV analysis of S1083 and S1268} \label{s1083}
\begin{figure*}
    \centering
    \includegraphics[width=1\columnwidth]{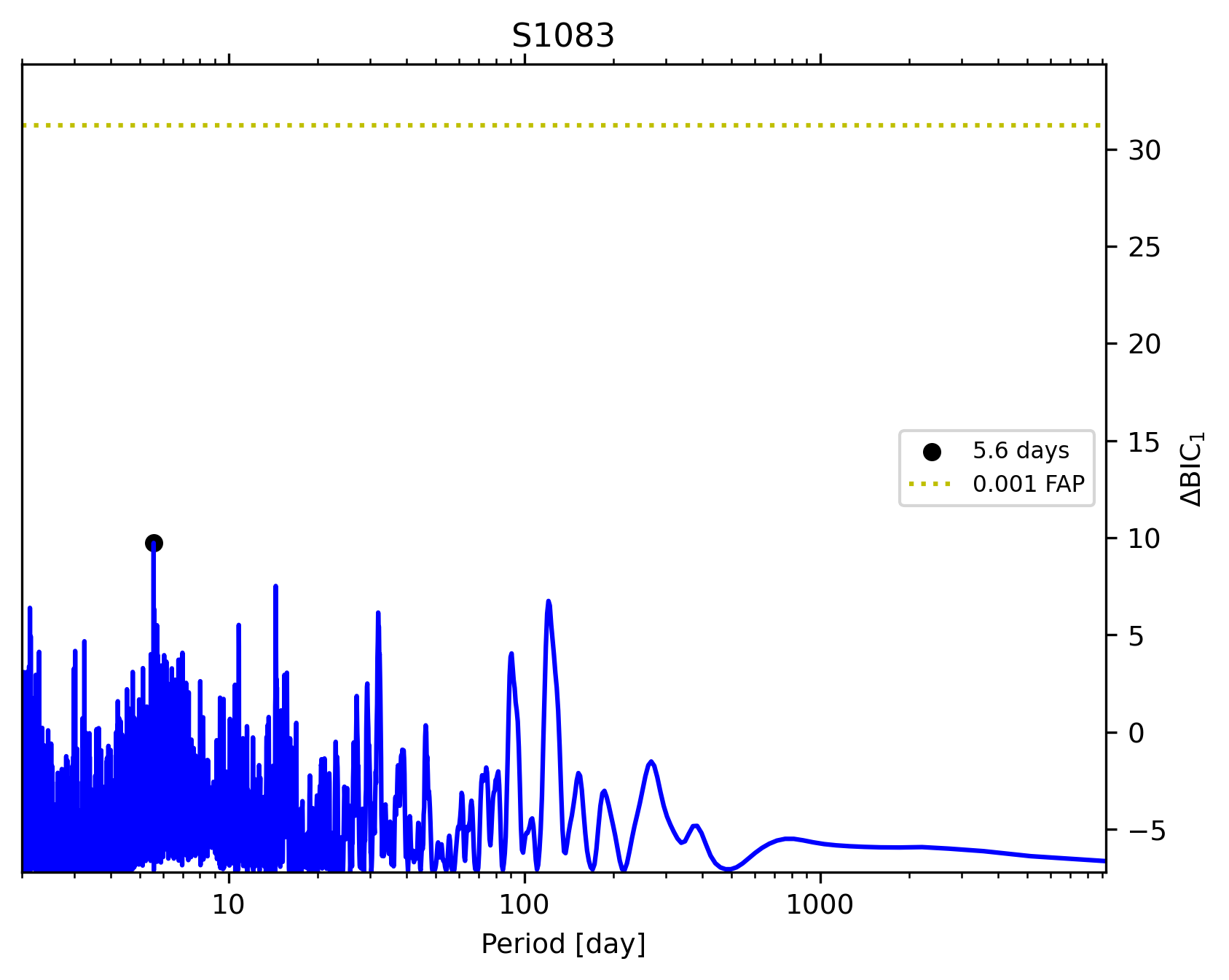} 
    \includegraphics[width=1\columnwidth]{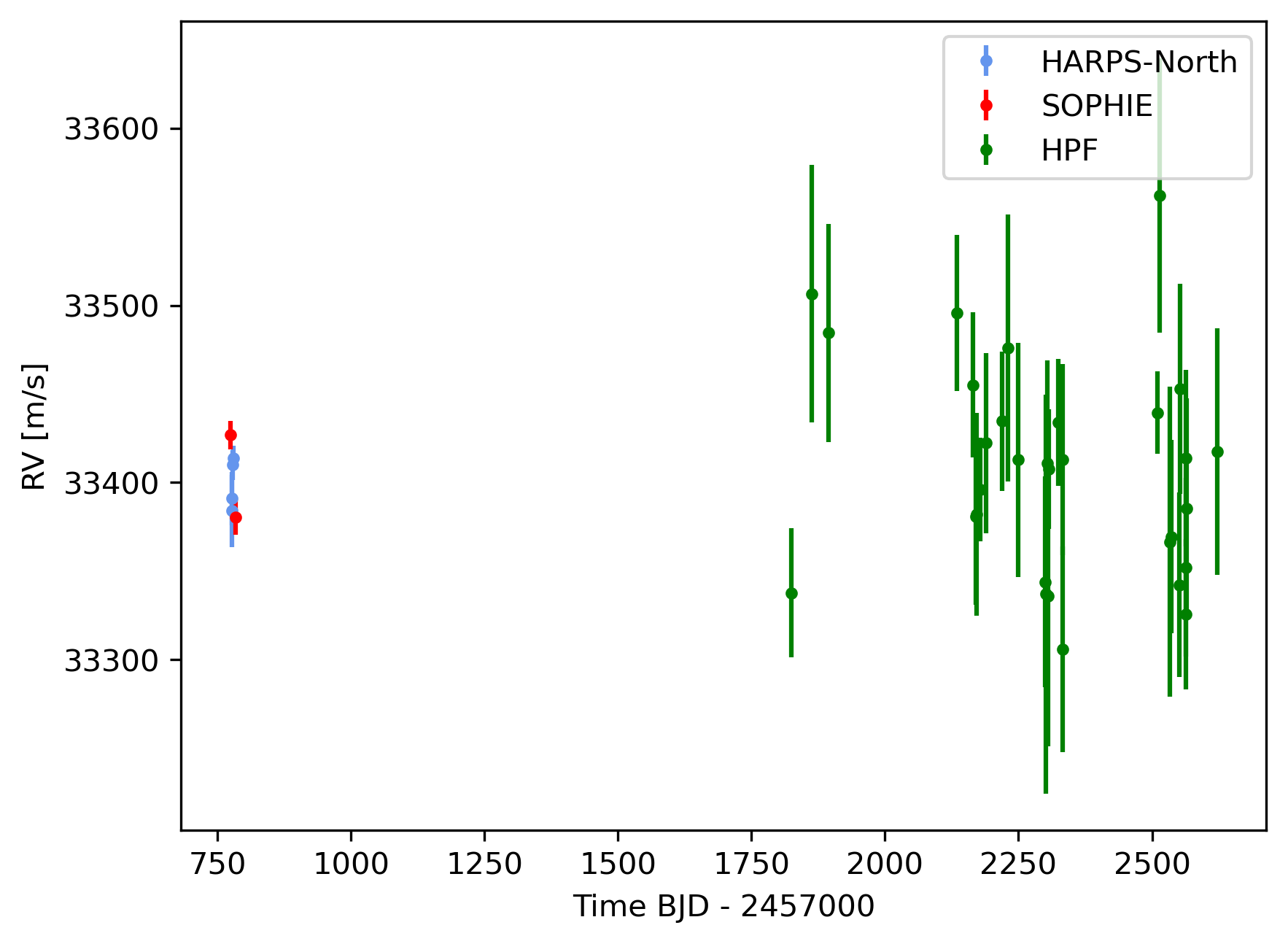} 
    \caption{Periodograms (left) and radial velocity time-series (right) of S1083. The period value for the highest peak and the $\Delta$BIC value for the $0.1 \%$ false-alarm probability are highlighted in black and yellow in the periodogram plot.}
    \label{fig:1083}
\end{figure*}

\begin{figure*}
    \centering
    \includegraphics[width=1\columnwidth]{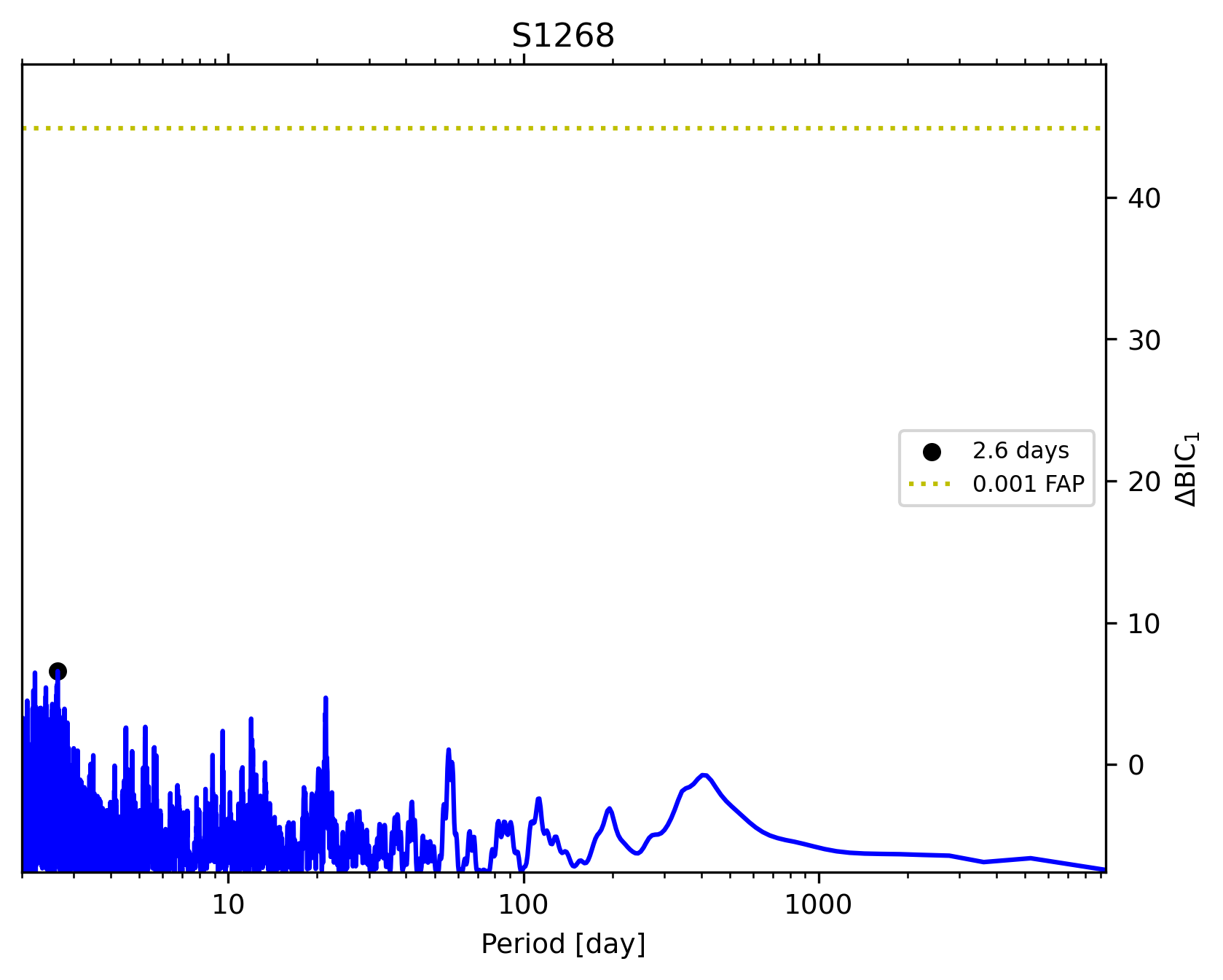} 
    \includegraphics[width=1\columnwidth]{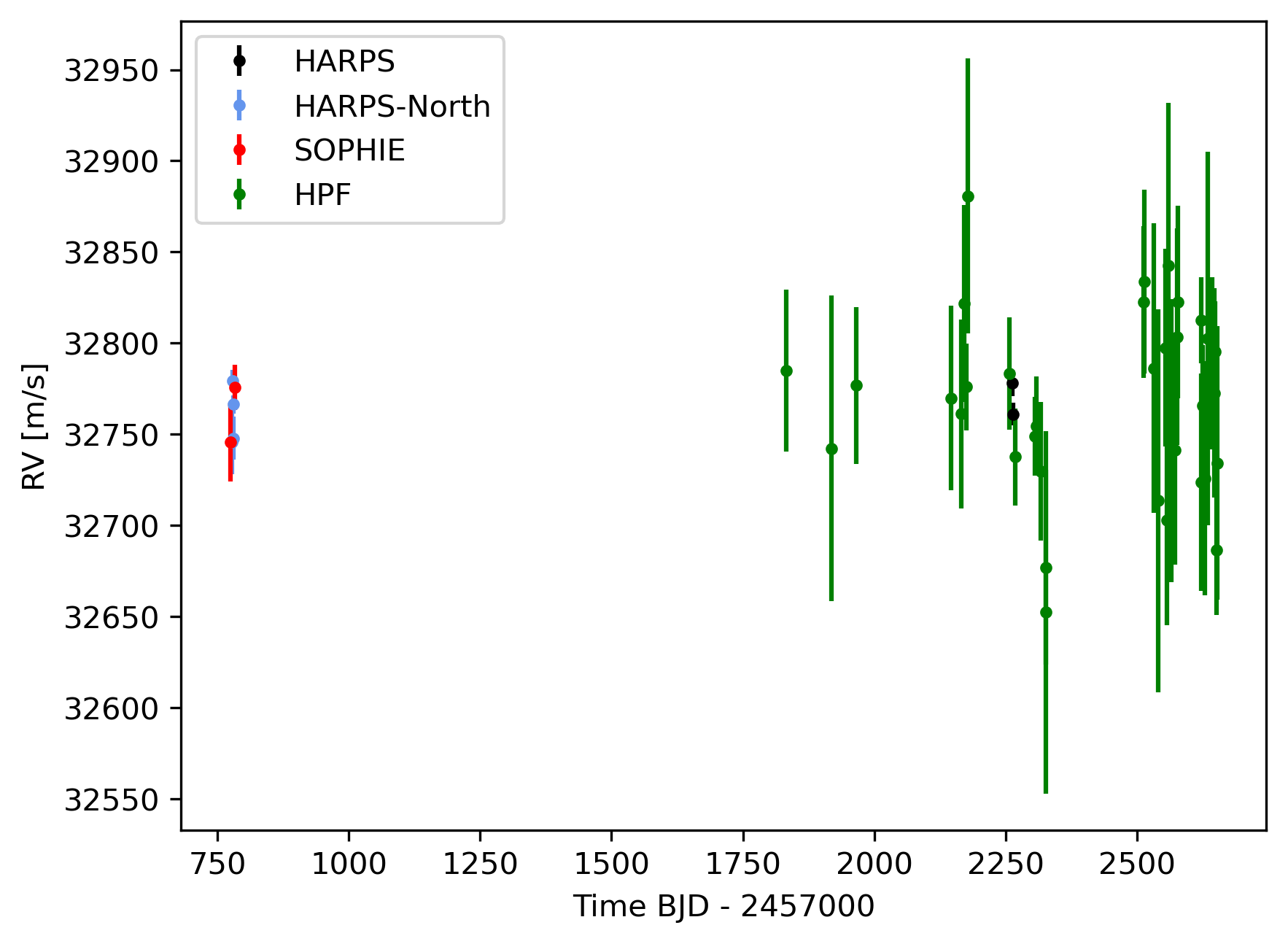} 
    \caption{Same as Figure \ref{fig:1083} but for S1268.}
    \label{fig:1268}
\end{figure*}

The two stars S1083 and S1268 showed no large RV variations or trends in their early RV data and were thus selected as candidates to host an exoplanet and observed more thoroughly. We collected 31 and 37 spectra for S1083 and S1268 respectively. Additionally, 4 HARPS-North, 2 SOPHIE, and 3 HARPS observations are available for S1083, and 4 HARPS-North, 2 SOPHIE, and 2 HARPS observations for S1268. The HARPS spectra for S1083 show some anomalies which suggest that something went wrong either in the observation or the reduction. Two of the HARPS spectra are shifted by 550 m/s against the third one. Both of those observations also have unusually small error bars at 1 m/s. We therefore chose to ignore the HARPS observations for S1083 in the analysis.
\\
As explained in Section 2.1 we used a template created from all HPF observations for the calculation of the HPF radial velocity values with SERVAL to achieve the highest possible precision. During the periodogram analysis with RVSearch we treat each of the 4 datasets separately with individual offsets and jitter terms. The results of the periodogram analysis together with plots of the RVs are shown in Figure \ref{fig:1083} and \ref{fig:1268}. Neither star has a significant periodicity in their RV data. The rms of the radial velocities is 64 m/s for S1083 and 49 m/s for S1268. 
\\
To investigate the completeness of our observations we use the injection-recovery test implemented in RVSearch. We inject 3000 synthetic planet signals drawn from a log-uniform period and M $\sin i$ distribution into our data and test whether they are recovered by the planet search algorithm. We chose an upper bound of 50,000 days for the period and a lower bound of 0.1 m/s for the RV amplitude of our synthetic planets. As shown in Figure \ref{fig:recov} our data was barely to sensitive planets larger than Jupiter in very close orbits and the sensitivity dropped quickly with increasing periods. We are therefore not able to draw definitive conclusions on the presence of planetary companions around S1083 and S1268.
\subsubsection{Potential planetary signal around S1429} \label{sec1429}
\begin{figure*}
    \centering
    \includegraphics[width=1.6\columnwidth]{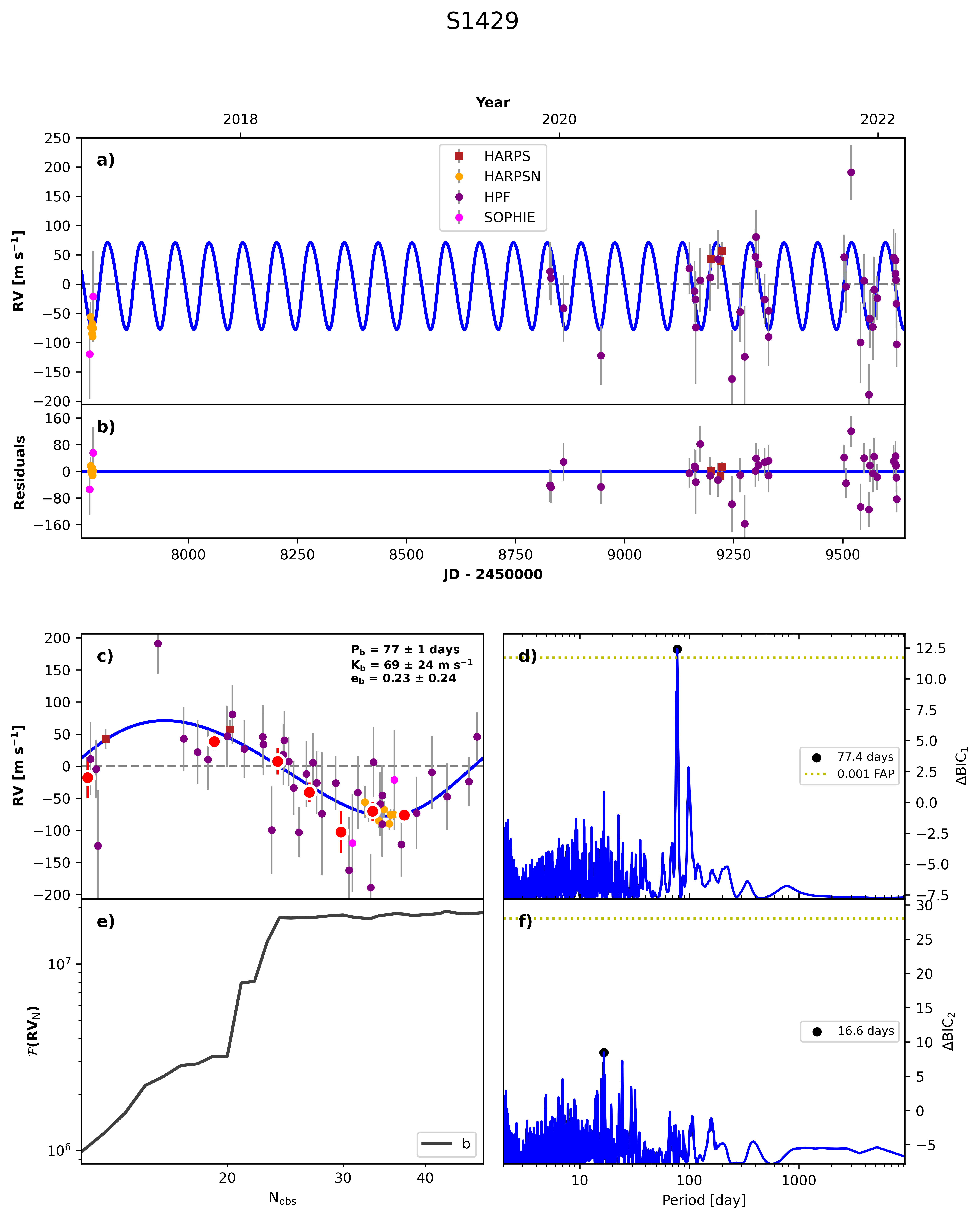} 
    \caption{Same as Figure \ref{fig:2018} but for S1429. For S1429 observations coming from the different instruments are fitted with separate offsets and jitter terms to take into account differences between the spectrographs.}
    \label{fig:14292}
\end{figure*}

\begin{figure*}
    \centering
    \includegraphics[width=0.8\columnwidth]{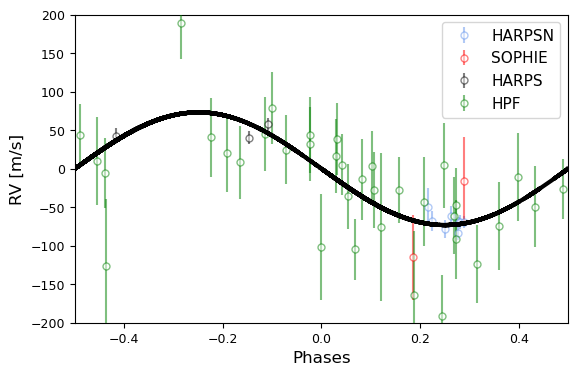} 
    \includegraphics[width=0.8\columnwidth]{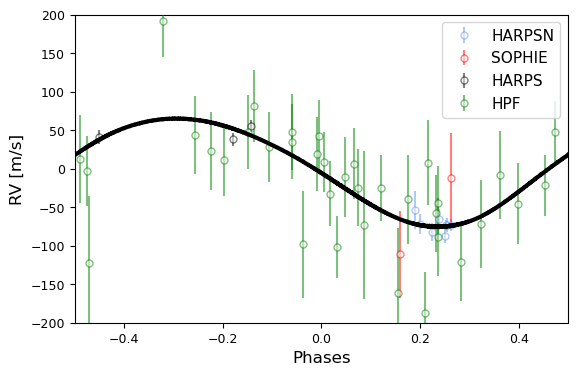} 
    \caption{Phase-folded radial velocity orbits for S1429 derived from the fit with Juliet. The left shows the orbit for a circular model while on the right eccentricity is left to vary freely between 0 and 1.}
    \label{fig:14293}
\end{figure*}

S1429 is the most observed star in our sample. Our dataset includes 9 spectra from HARPS-North observed between January 22nd and January 28th, 2017, 2 spectra from SOPHIE observed on the 19th and 27th of January 2017, 4 datapoints from HARPS between the 13th of December 2020 and the 8th of January 2021 and lastly 37 HPF spectra taken between the 11th December of 2019 and the 13th of February 2022. We excluded one data point for each of the HARPS, HARPS-North, and HPF spectra which have a much lower SNR than the average for the respective dataset. The initial radial velocities showed no large variations indicative of stellar companions which made S1429 one of the three main targets for a planetary companion. As with S1083 and S1268, we constructed a high signal-to-noise template using all 36 HPF spectra and fit individual offsets and jitter terms for each instrument during the periodogram analysis.
\\
 The periodogram shows a clear peak exceeding the $0.1\%$ false alarm probability at 77 days (see Figure \ref{fig:14292}). This peak is also the strongest periodic signal when the HPF data is analysed without the other datasets although the false-positive probability in this case is larger than $0.1\%$. Already when combining only the HPF and HARPS-N spectra which are the two largest datasets for this object the threshold of $0.1\%$ FAP is exceeded. Since the iterative periodogram search did not return a second significant periodic signal RVSearch fits a single-planet Keplerian model to the data and samples the posterior distribution with an MCMC. The resulting fit is shown in Figure \ref{fig:14292}. 
\\
The fit gives a non-zero eccentricity of $0.23$. However, it is also consistent with zero at the 1-$\sigma$ level. To investigate whether this eccentricity is significant or the product of a bias towards non-zero eccentricities in the fitting of Keplerian orbits, we reanalyze our RV data with the python package Juliet \citep{espinoza2019juliet}. Juliet is built to compare models fit to photometric or radial velocity data for exoplanet detection by sampling Bayesian posteriors and evidence using nested sampling algorithms. The Bayesian evidence ($\ln Z$) can be used to compare different models with a $\Delta \ln Z > 2$ being the threshold for a moderate preference and a $\Delta \ln Z > 5$ for a strong preference of one model over the other \citep{trotta2008bayes}. One of the drawbacks of nested sampling is that because it focuses on evidence calculations the posterior distribution is a by-product and might not be explored as efficiently. To compensate \cite{higson2019dynamic} proposed a modification to the standard nested sampling algorithm where instead of keeping the number of live points constant they are dynamically changed to adjust the focus during the run. This dynamic nested sampling algorithm is implemented in Juliet through the python code dynesty \citep{speagle2020dynesty}.
\\
We perform two separate fits, one with eccentricity and argument of periastron $\omega$ fixed to $0$ and $90^{\circ}$ respectively and one where both parameters are varied freely over a range of uniform priors. We impose a Gaussian prior on the relative offsets $\gamma$ of the radial velocities between the datasets centered around the values derived from the RVSearch periodogram analysis with the variance taken as the standard deviation of the RVs for the respective spectrographs. For the radial velocity amplitude $K$, the period and the time of superior conjunction $T_c$ we use uniform priors. The results of the two fits are shown in Table \ref{tab:juliet}. The Bayesian evidence slightly favors the circular model with a $\Delta \ln Z$ of $1.8$ however with this small of a difference in evidence the preference for the circular model is barely statistically significant. This is likely due to the incomplete sampling at the maximum of the phase-folded radial velocity curve (see Figure \ref{fig:14293}) which limits the precision with which we can determine the eccentricity. The resulting corner plots for the two fits are shown in Appendix \ref{corn}. For all parameters, the mean values are very close to the maximum of the distribution. However, the eccentricity could not be constrained as well as the other parameters and has a slightly wider distribution.
\begin{table*}
    \centering
    \begin{threeparttable}[b]
    \centering
    \caption{Summary of the Juliet fit to the RV data of S1429.}
    \begin{tabular}{l|l|c|c}
    \hline
    \hline
    Parameters & Prior & Circular fit & Eccentric fit \\
    \hline
        $P$ [d]        &      $\mathcal{U}[60, 90]$       & $77.48_{-0.19}^{+0.18}$   &  $77.51_{-0.20}^{+0.20}$\\
        $T_0$ (BJD$_{\rm TDB}$)   & $\mathcal{U}[2459220.00, 2459260.00]$  & $2459231.1_{-1.2}^{+1.3}$   &  $2459233.7_{-3.0}^{+3.4}$\\
        $e$               & $\mathcal{U}[0,0.99]$          & 0 (fixed)  &  $0.15_{-0.10}^{+0.11}$\\
        $\omega_{*}$ [deg]   & $\mathcal{U}[0,360]$       & 90 (fixed)   & $210_{-60}^{+70}$ \\
        $K$ [m~s$^{-1}$]      & $\mathcal{U}[0,150]$       & $74_{-8}^{+8}$   & $72_{-8}^{+9}$\\
        $\gamma_{\rm HPF}$ (m s$^{-1}$)   & $\mathcal{N}[-19.967,72]$   & $ -18_{-9}^{+9}$ \tnote{a)}& $-21_{-9}^{+9}$ \tnote{a)}\\
        $\gamma_{\rm HARPS}$ (m s$^{-1}$)  & $\mathcal{N}[34022.9,8]$    & $34022_{-6}^{+6}$& $34024_{-6}^{+6}$\\
        $\gamma_{\rm SOPHIE}$ (m s$^{-1}$)  & $\mathcal{N}[34083.4,50]$    & $34078_{-30}^{+31}$ & $34073_{-30}^{+30}$ \\
        $\gamma_{\rm HARPS-N}$ (m s$^{-1}$) & $\mathcal{N}[34148.2,10]$     & $34142_{-8}^{+8}$ & $34146_{-8}^{+8}$\\
        $\sigma_{\rm HPF}$ (m s$^{-1}$)  & $\mathcal{L}[1e-3,100]$   & $ 34_{-9}^{+9}$ &  $34_{-8}^{+9}$\\
        $\sigma_{\rm HARPS}$ (m s$^{-1}$)   & $\mathcal{L}[1e-3,100]$   & $ 1.5_{-1.6}^{+15}  $ & $0.17_{-0.17}^{+11}$ \\
        $\sigma_{\rm SOPHIE}$ (m s$^{-1}$)  & $\mathcal{L}[1e-3,100]$    & $ 54_{-21}^{+27}  $ & $55_{-21}^{+26} $\\
        $\sigma_{\rm HARPS-N}$ (m s$^{-1}$)  & $\mathcal{L}[1e-3,100]$    & $ 0.06 _{-0.06}^{+1.3}  $ & $0.11_{-0.11}^{+1.9}  $ \\
        $\ln Z$  &    & $-262.8$ & $-264.6$\\
    \hline 
    \end{tabular}
    \begin{tablenotes}
       \item [a)] The offsets $\gamma_{\rm HPF}$ do not include the systemic velocity of the star; this is applied in Table \ref{tab:rv1429}.
     \end{tablenotes}
    \end{threeparttable}
    \label{tab:juliet}
\end{table*}
\section{Discussion} \label{four}
\subsection{Validation of the S1429 Signal}
\begin{figure}
    \centering
    \includegraphics[width=0.6\columnwidth]{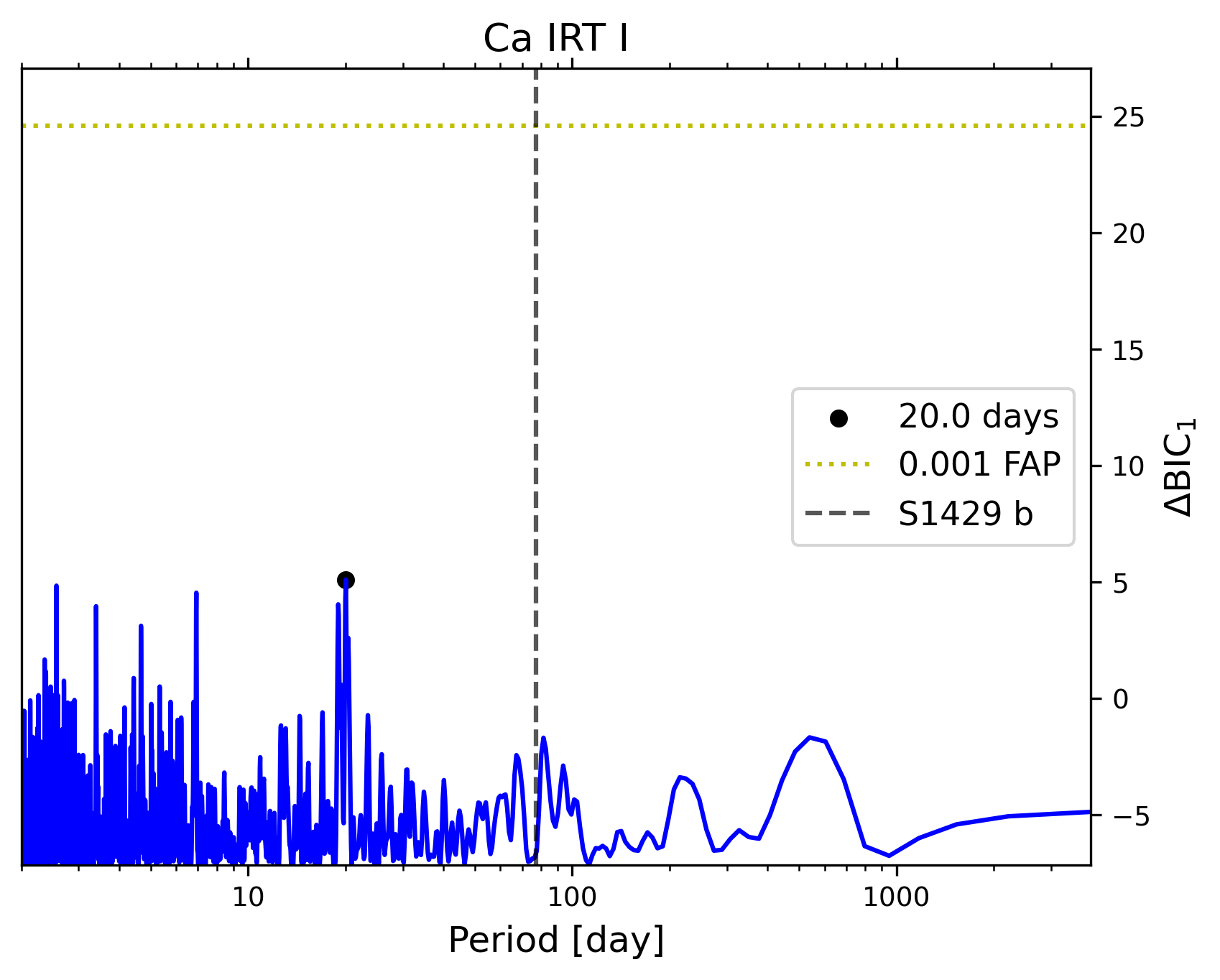} 
    \includegraphics[width=0.6\columnwidth]{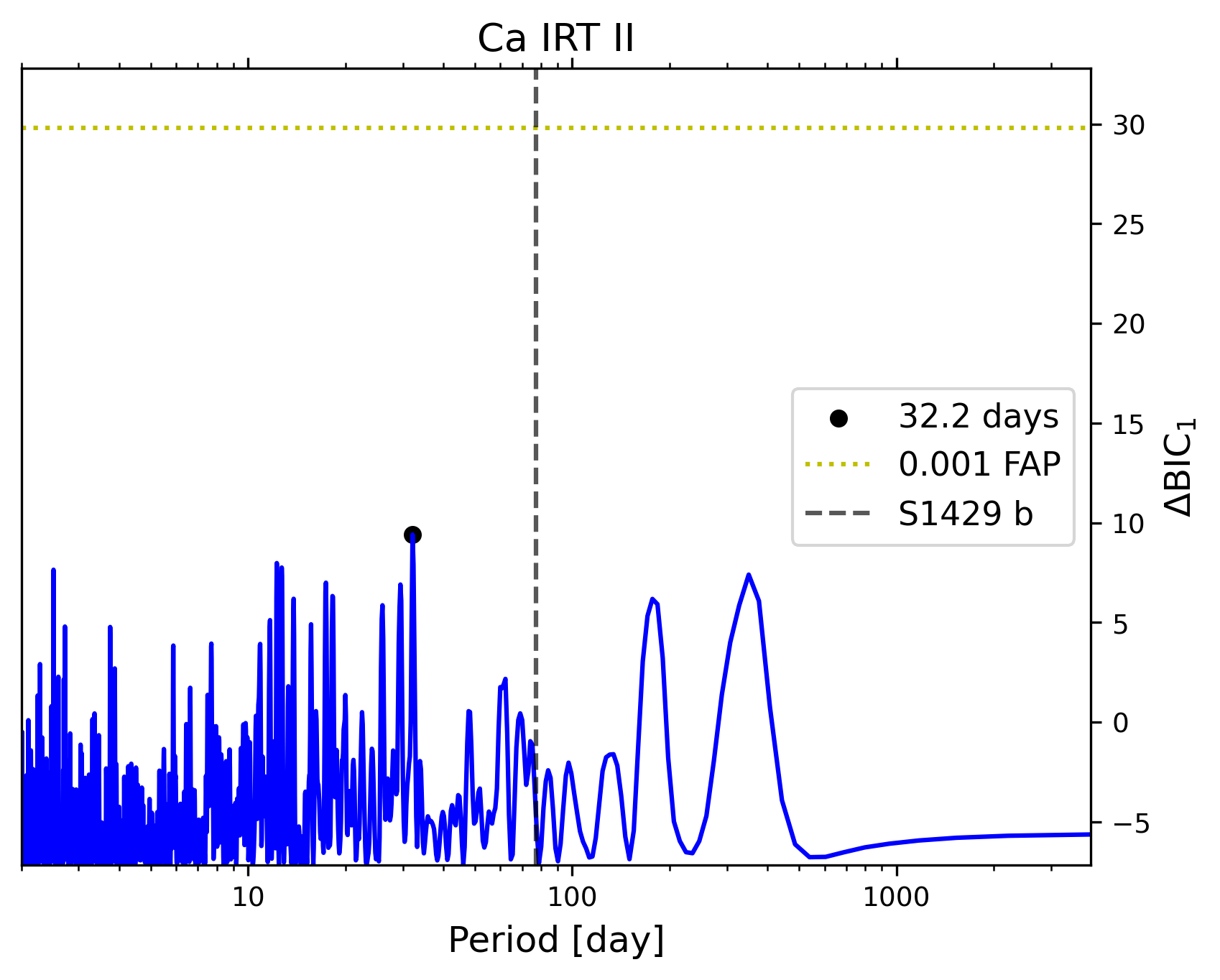} 
    \includegraphics[width=0.6\columnwidth]{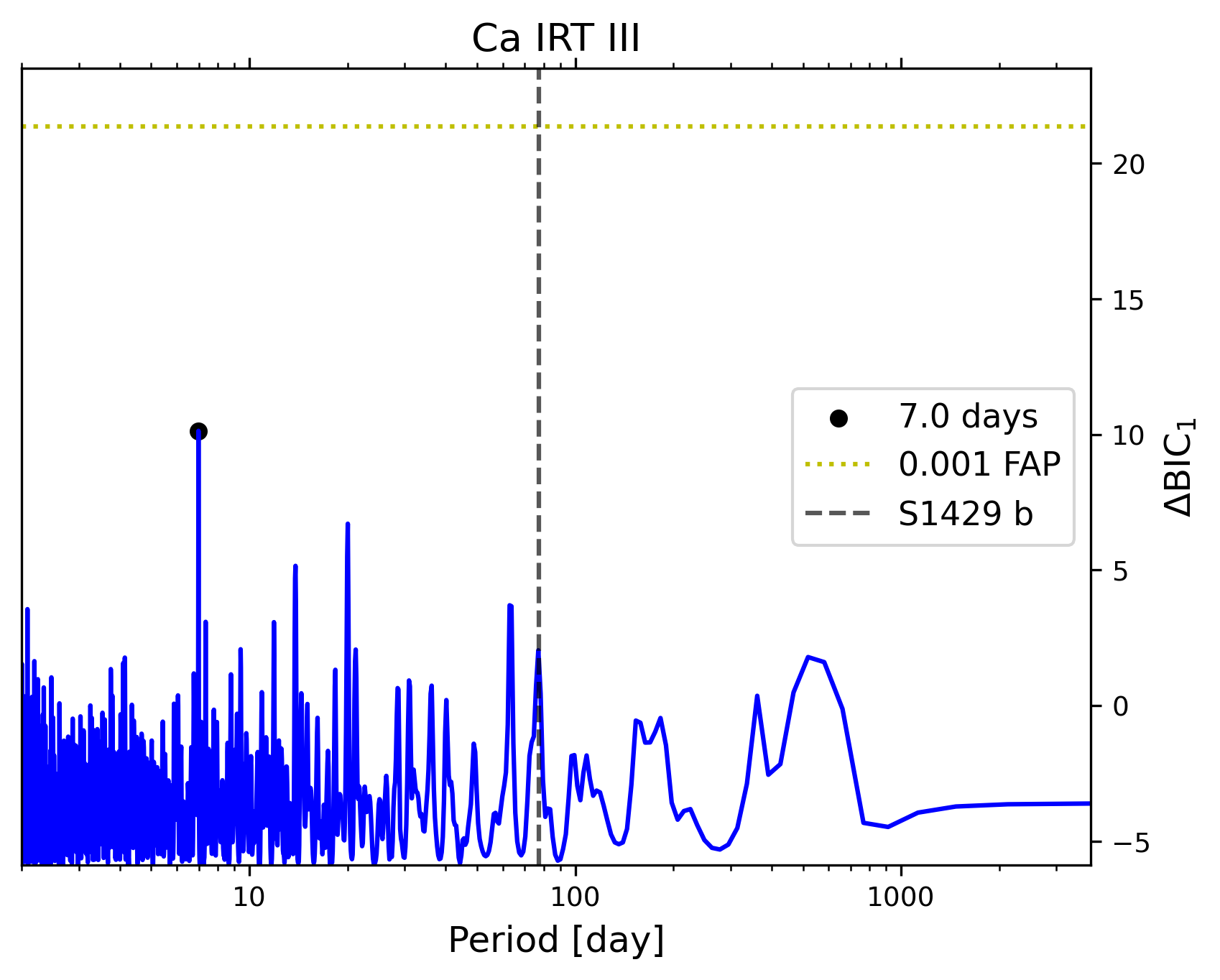} 
    \includegraphics[width=0.6\columnwidth]{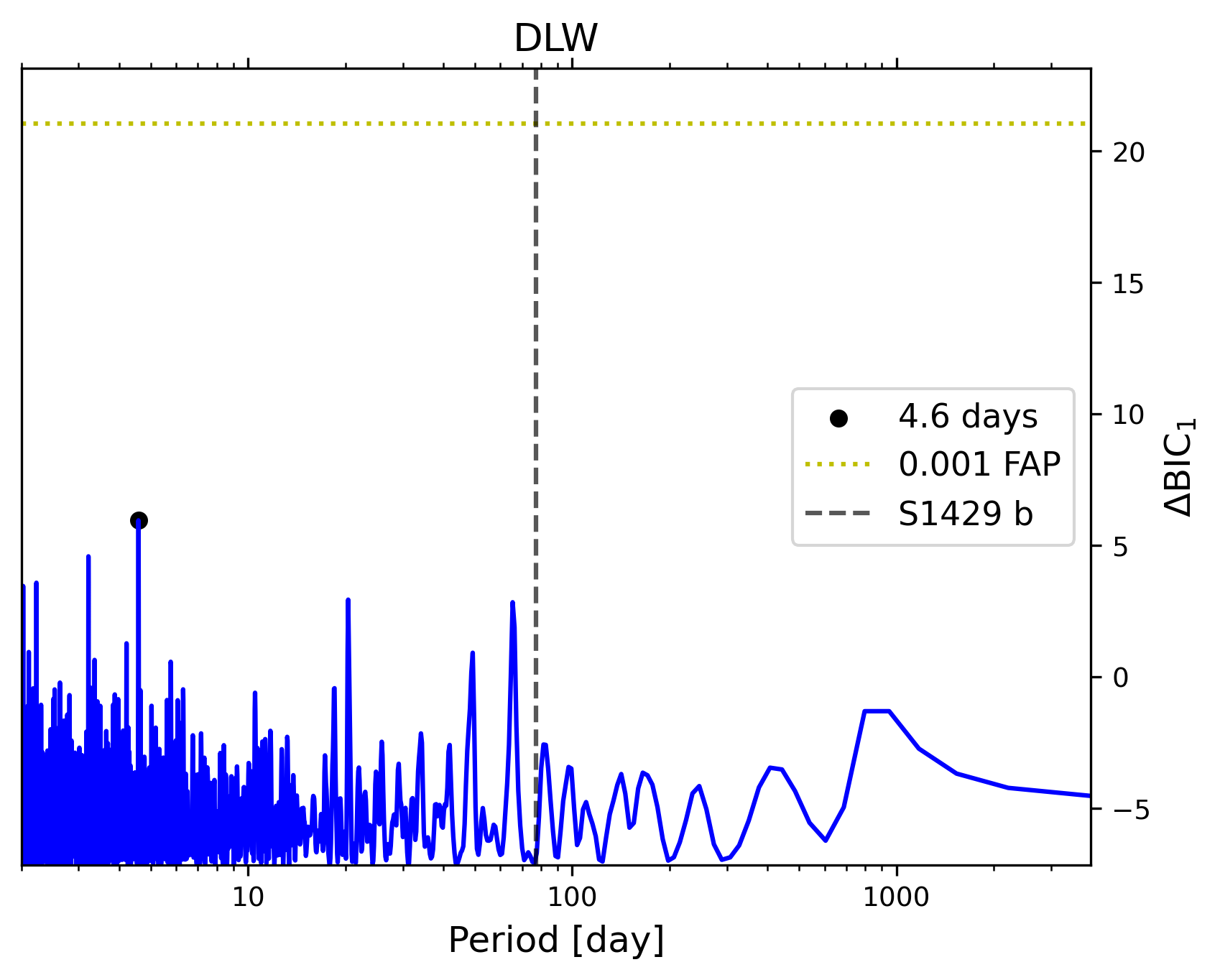} 
    \includegraphics[width=0.6\columnwidth]{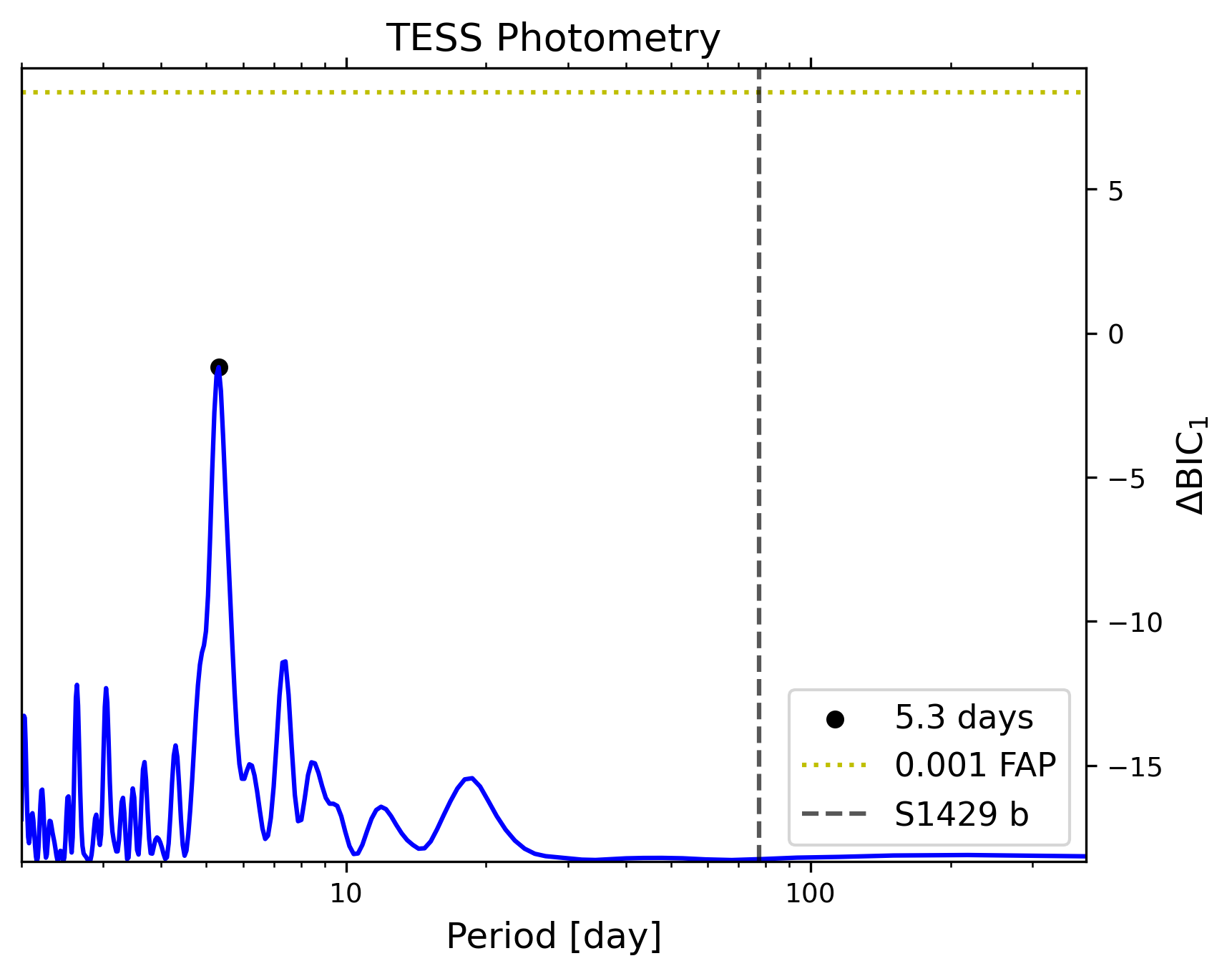} 
    \caption{Periodograms of the activity indicators derived from the HPF spectra and the photometric data from TESS. The first three plots show the periodicity in the signals of the calcium triplet lines. The fourth and fifth panels show the periodograms for the differential line width from SERVAL and the TESS photometry. The horizontal yellow line indicates the 0.1 \% false-alarm probability threshold and the vertical black line shows the location of the potential planet signal in the RVs of S1429.}
    \label{fig:act}
\end{figure}

In order to verify that the periodic signal in the RV of S1429 is caused by a planetary companion and not due to the effects of stellar activity, we examine several activity indicators. For each one, we check if we can find a periodic variation either exactly at the period found in the radial velocity data or an integer alias of that period. The first two activity indicators that we use are the Ca II infrared triplet (IRT) and the differential line width (dLW). Both are computed from the HPF spectra with SERVAL. The Ca II IRT has shown to be a good indicator for tracking magnetic activity on the stellar surface which changes the depth of the line cores at a period corresponding to the rotational period of the star \citep{fuhrmeister2019carmenes}. The differential line width is calculated by SERVAL as a proxy for the mean stellar line profile, where variations of the line widths and shapes can be caused either by processes intrinsic to the star (pulsation, magnetic activity) or systematic effects of the instrument \citep[for an in-depth explanation see][]{zechmeister2018spectrum}. In either case, a correlation of the dLW with the radial velocity variations would point to a source that is not planetary in nature. To ensure that we treat all of our spectroscopic signals the same, we use RVSearch to compute periodograms with the same uninformed approach that was used for the radial velocities to find periodicities in the Ca II IRT and dLW data.
\\
For some of our data, the signal in the third line in the Ca triplet was too low to fit the line. Therefore, only 19 of the 36 possible observations could be used for the periodogram. The results are shown in Figure \ref{fig:act}. Both activity indicators show no periodicity at the orbital period of our planet candidate or at any integer aliases. The first two Ca II triplet lines have their peaks at 20 days and 32 days respectively with the 20-day period being a 365-day alias of the 32-day period for the second Ca line. The third Ca II IRT line and the differential line width also have peaks at the 20-day period but for both, those are slightly lower than their highest peaks at 7 and 4.6 days. However, none of the signals are strong enough to be statistically significant. This is somewhat expected as stars in M67 are shown to have low amounts of chromospheric activity due to the relatively old age of the open cluster \citep{pace2004age}.  
\\
\\
In addition to these two spectroscopic activity indices, we also examine the TESS light curve of S1429 to check for periodic variations in the brightness which could be caused by star spots rotating with the star. Light curves of S1429 were produced as TESS-SPOC High-Level Science Product \citep{caldwell2020tess} by the TESS Science Processing Operation Center pipeline \citep[SPOC,][]{jenkins2016tess} from the 10-minute cadence full frame images. The SPOC lightcurves give two flux values, the Simple Aperture Photometry (SAP) flux and the PDCSAP flux \citep[Presearch Data Conditioning Simple Aperture Photometry;][]{smith2012kepler,stumpe2012kepler,stumpe2014multiscale}, which corrects the SAP data for instrumental systematics. As the PDCSAP lightcurves are optimised to detect transit signals from exoplanets it is possible that the applied corrections also remove longer periodic stellar trends and not only instrumental systematics \citep{mathys2023long}. In cases where the expected rotation period is larger than 20 days, it can be better to use the SAP lightcurves if the stellar signal is not masked by instrumental systematics. However, on inspection of the lightcurve, the SAP flux shows strong trends from instrumental systematics which makes the detection of potential stellar rotation signals difficult. We therefore decide to use the PDCSAP lightcurves for our periodogram analysis which we download with the python package lightkurve \citep{2018ascl.soft12013L}. We again use RVSearch to look for periodicities in the photometric data (see Figure \ref{fig:act}). The highest peak is at 5.3 days but a lower peak around the 20-day signal that was seen for the spectroscopic activity indicators is visible. 
\\
Additionally, we search the light curve for transit events. Given the derived period and $T_0$ with their respective uncertainties, we expect the transit to occur at the beginning of the second half of Sector 45 between 2459539.06 BJD and 2459543.04 BJD. For the first 24 hours of this window, there is a gap in the light curve but in the rest, no transits can be found. Given the properties of the S1429 system, this would indicate an orbital inclination $<88.5^\circ$
\\
The rotational velocity of S1429 in the literature is given as $v \sin i \sim 5$ km/s \citep{jonsson2020apogee}. Assuming an inclination of $ \sim 90^{\circ}$ combined with the derived stellar radius of 2.13 $R_\odot$ would result in a rotational period of $\sim 21$ days. The 20-day signal is therefore a realistic candidate for the rotational period of the star although it should be noted that while the uncertainty for $v \sin i$ is not given by \cite{jonsson2020apogee} it is typically of the order of a couple km/s so we expect this estimated period to be of limited accuracy.
\\
As all of our activity indicators do not overlap with the period derived from the radial velocity analysis we conclude that our signal is most likely caused by the presence of a planet hereinafter referred to as S1429 b. 

\subsection{Planet properties in M67}
\begin{figure}
    \centering
    \includegraphics[width=0.9\columnwidth]{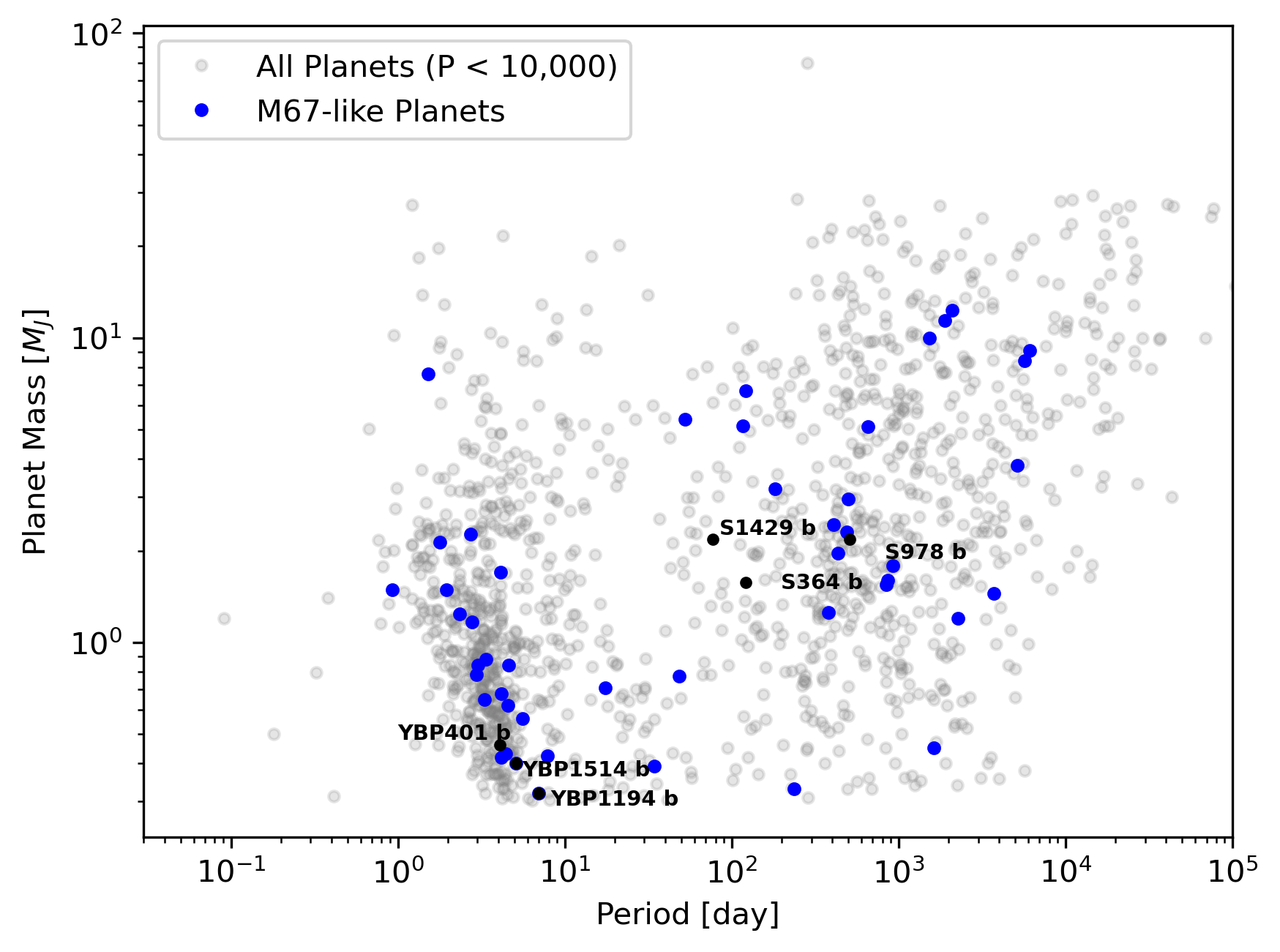} 
    \includegraphics[width=0.9\columnwidth]{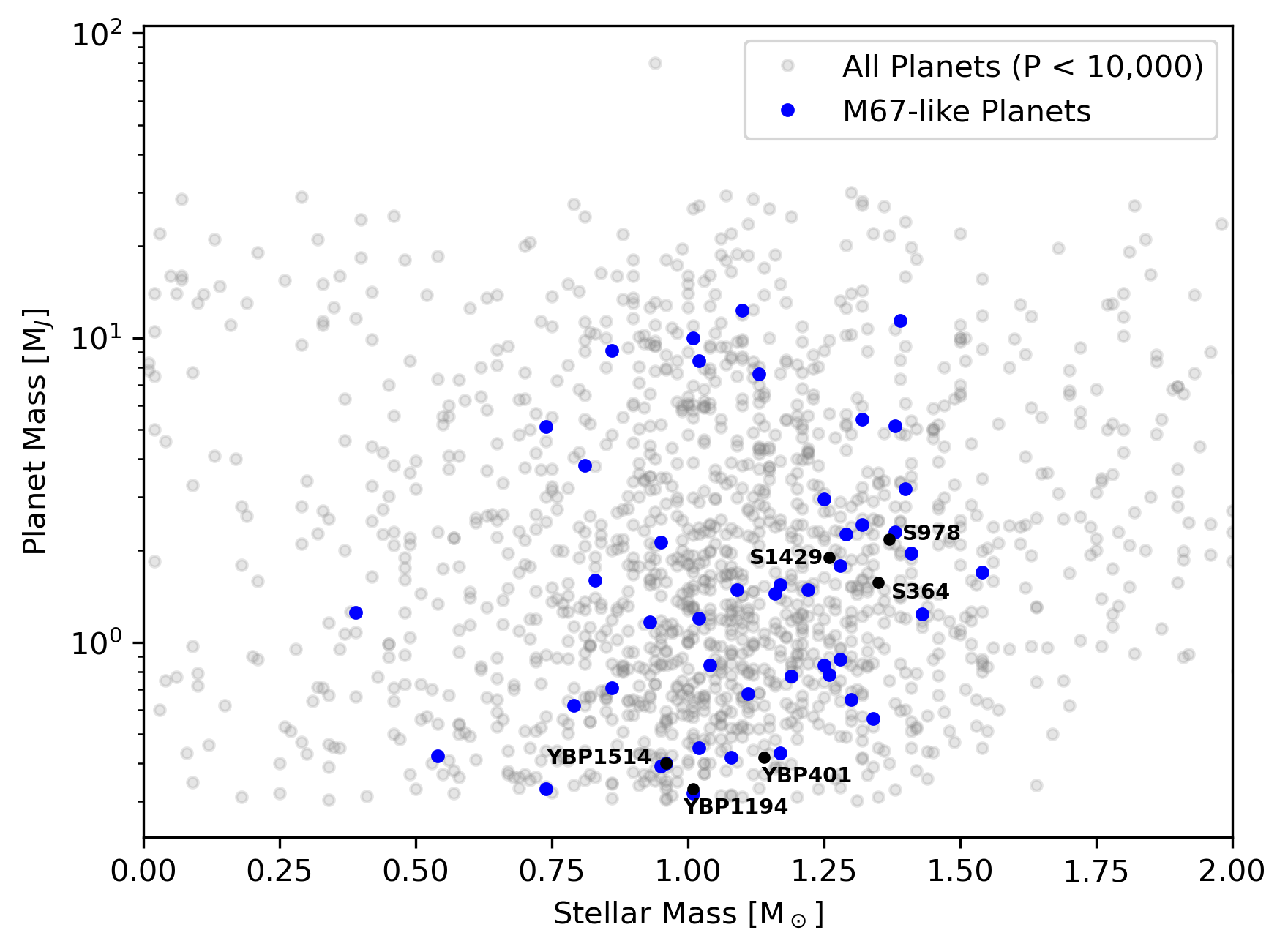} 
    \includegraphics[width=0.9\columnwidth]{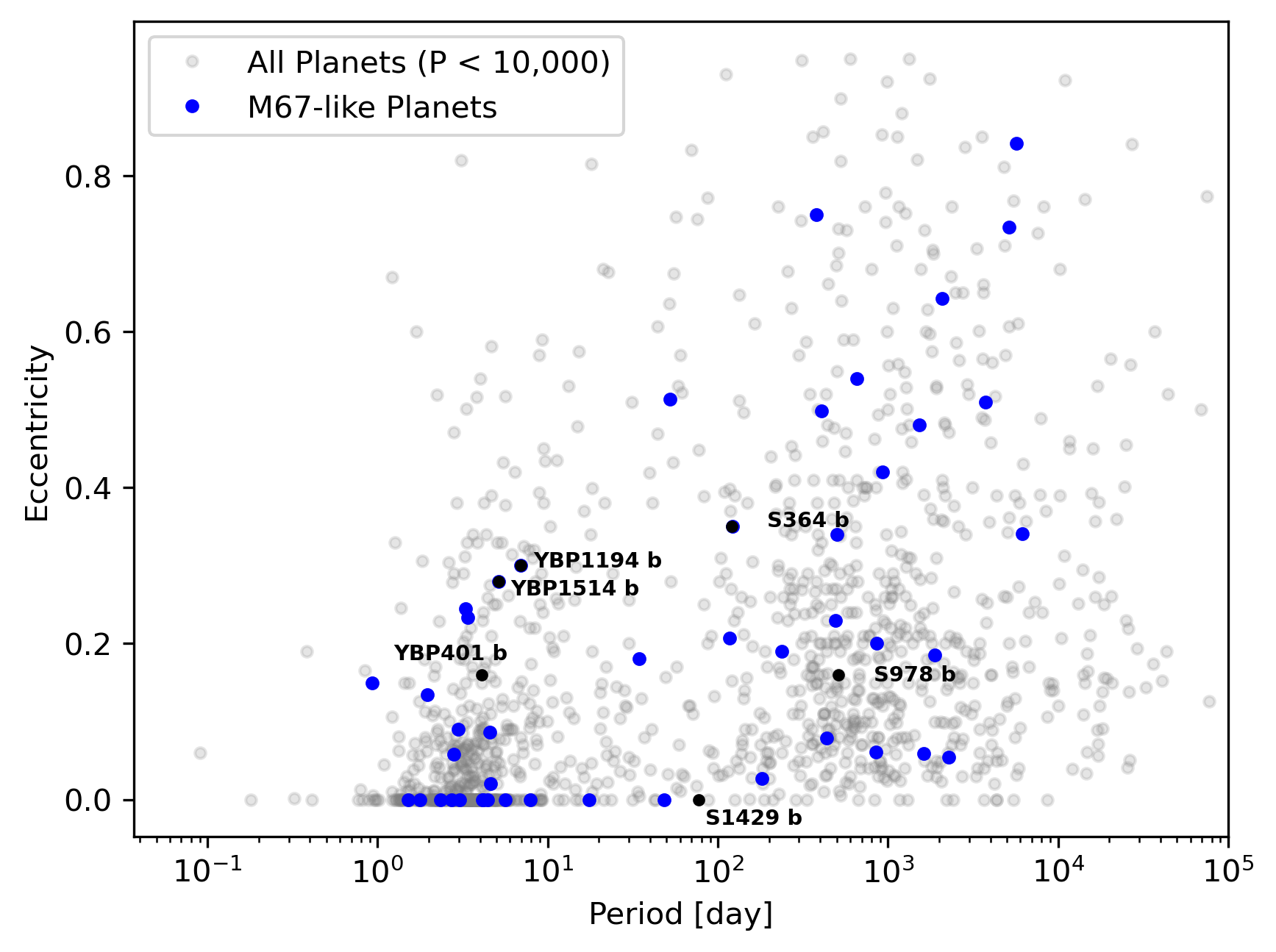} 
    \caption{The population of known giant exoplanets ($\text{M} \ \sin i > 0.3 \ \text{M}_J$) with either a true mass or M $sin i$ measurement in planet mass-period (upper left) and planet mass-stellar mass (upper right) and eccentricity-period (lower panel) space using data from the NASA Exoplanet Archive. Light grey is the full giant planet population while blue are planets around stars with similar chemical composition and age as the stars in M67 ($-0.1 <$ Fe/H $ < 0.1$ and $3.8 <$ age $<4.5$). The six planets discovered in M67 are overplotted in black.}
    \label{fig:pop}
\end{figure}

S1429 b is the sixth planet to be discovered in M67 with the other five being presented in \cite{brucalassi2017search}. Given the mass of the host star from Section \ref{ari}, we calculate a minimum mass of $1.80 \pm 0.2$ M$_J$ for S1429 b assuming a circular orbit. Using the orbital period and stellar properties, we derive a semi-major axis of $0.384 \pm 0.004$ au \footnote{Given the distance of $\sim 1$ kpc this makes the reflex astrometric wobbling of the parent star undetectable even with the 5th Gaia data release} and an equilibrium temperature of $683 \pm 9$ K. With these properties, S1429 b falls into the class of warm-Jupiters (P $\sim 10 - 200$ days). Similarly to hot-Jupiters, these objects are thought to form at larger separations and then migrate inwards making them a secondary probe to test different migration theories. There are however, some key differences between the two populations such as the prevalence of close-by and long-period companions or the distribution of eccentricities which indicates that multiple different mechanisms are responsible for the variety of close-in giant planets \citep[for a summary see eg.][]{dawson2018origins}. In total, we have now found three hot-Jupiters and three planets that belong to the warm/cold Jupiter class in M67. 
\\
The main advantage of studying planets inside star clusters is the homogeneity in chemical composition and age of the host stars, making it easier to discover the underlying reason for certain trends in the exoplanet population. 
Figure \ref{fig:pop} shows the location of all the discovered M67 planets in the mass-period distribution of known giant exoplanets (M $\sin i>0.3\text{M}_J$) with host stars similar to the M67 stars. For this plot, we downloaded data on all known exoplanets from the NASA Exoplanet Archive and filtered them according to the age ($3.8-4.5$ Gyr) and metallicity ($-0.1 -0.1$) of M67. A notable difference in planetary mass between the three hot-Jupiters and the three farther out planets in M67 can be seen, with the planets at larger separation having a $4 -5 $ times larger minimum mass. The same trend is visible in the sample of planets around M67-like stars and even the whole exoplanet population where the short-period planets clump around lower mass values, while planets at longer periods have a more homogenous mass distribution with a similar number of low and high-mass planets.
\\
This trend of an increase in planetary mass with larger orbital periods was reported in some of the early statistical works \citep{udry2003statistical}. However, recent studies using the California Legacy Survey (CLS) catalog which is based on a blind RV search of 719 FGKM stars spanning over three decades \citep{rosenthal2021california} have suggested that both hot-Jupiters and farther out giant planets follow a similar mass-distribution with the early differences in the distribution coming from observational biases preferably detecting more massive planets at larger separation \citep{zink2023hot}. 
\\
We also find a correlation between the mass of the planet and the mass of its host star in M67 (see Figure \ref{fig:pop}), with more massive host stars hosting more massive planets. While a similar trend might be present for the M67-like stars, in the full exoplanet population there seems to be no dependence of the planetary mass on the host star mass. If this correlation is real and not caused by a small/biased sample, this would suggest that it only holds true for a certain subset of stars e.g. solar metallicity stars with stars outside of this subset showing a different or no correlation between the masses.  
\\
However, as the detection of less massive planets gets more difficult for both larger separations and larger host star masses, the lack of lower-mass giant planets at larger orbital periods and around more massive stars in our sample may stem from a detection bias. We use the same injection recovery tests presented in Section \ref{s1083} to investigate the completeness of the datasets for the planet hosts in M67. The completeness contour plots for each of the six planet hosts in M67 are shown in Figure \ref{fig:recov}.
The results of our injection recovery test show that the sensitivity of our datasets for planets with M $\sin i$ below one Jupiter mass drops rapidly for increasing periods and most smaller giants could not be recovered at periods larger than $10 - 20$ days. This suggests that the larger masses we found for the farther out planets could stem from a detection bias towards more massive planets. Similarly, the sensitivity of our RV data around the three more massive stars (S364, S978, and S1429) is significantly lower for less massive planets especially in the cases of S364 and S1429 compared to the lower-mass stars. Therefore, the correlation between planetary and stellar mass is also expected given our data.
\\
Nevertheless, it is still noteworthy that all three hot-Jupiters in M67 have minimum masses significantly below one Jupiter mass as these types of planets are detectable in all our datasets. 
\\
An anomaly for the planets in M67 is the rather large eccentricity for the three hot-Jupiters. Figure \ref{fig:pop} shows the eccentricity distribution of giant planets (M $\sin i>0.3\text{M}_J$) as a function of the orbital period. While moderate eccentricities for the warm/cold Jupiters in M67 are expected as this population of exoplanets is known to have a broad eccentricity distribution \citep{dong2021toi}, hot-Jupiters tend to favor circular orbits due to tidal circularization effects caused by the proximity to the host star. Considering all exoplanets with periods between 1 and 10 days, the average eccentricity is $\sim 0.07$. This value does not change significantly when only considering the sample of "M67-like" stars. The three hot-Jupiters in M67 all have larger than average eccentricities with YBP1194 b ($e = 0.31 \pm 0.08$) and YBP1514 b ($e = 0.27 \pm 0.09$) having the most eccentric orbits out of the planets in the sample of M67-like stars. However, looking at the eccentricities of close-in planets the distribution seems to get broader with increasing periods. Indeed when we split the hot-Jupiter population into two period bins ($1 <$ P $<5$ and $5 <$ P $<10$) the average eccentricity is considerably higher for the longer periods at $0.12$ compared to $0.05$. The distribution is also slightly broader with a higher standard deviation ($0.15$ vs. $0.10$). Considering that both YBP1194 b and YBP1514 b have periods larger than 5 days the higher eccentricity values do not necessarily point towards a drastically different formation and evolution history of the M67 giants compared to planets around field stars. Studies have shown that the majority of hot-Jupiters have a long-period giant companion \citep{knutson2014friends,bryan2016statistics,zink2023hot} which could disturb the orbit of the hot-Jupiter, especially at larger separations where the tidal forces from the star weaken. In the case of M67, we did not detect any hints of long-term RV variations around the three hot-Jupiter hosts although as seen in Figure \ref{fig:recov} the present data was only sensitive to planetary mass objects out to periods $\sim 2000 - 3000$ days and even then only for masses significantly above one Jupiter mass. Since we can not rule out the presence of outer perturbing giant planets, a dedicated follow-up to search for outer companions could give insights into whether the same migration mechanisms as in field stars are responsible for the shaping of hot-Jupiter systems in M67 or if other dynamical interactions such as stellar flybys \citep{malmberg2011,shara2016,wang2022hot} in the dense cluster are at play.
\subsection{Planet population at different evolutionary stages}
\begin{figure*}
    \centering
    \includegraphics[width=0.9\columnwidth]{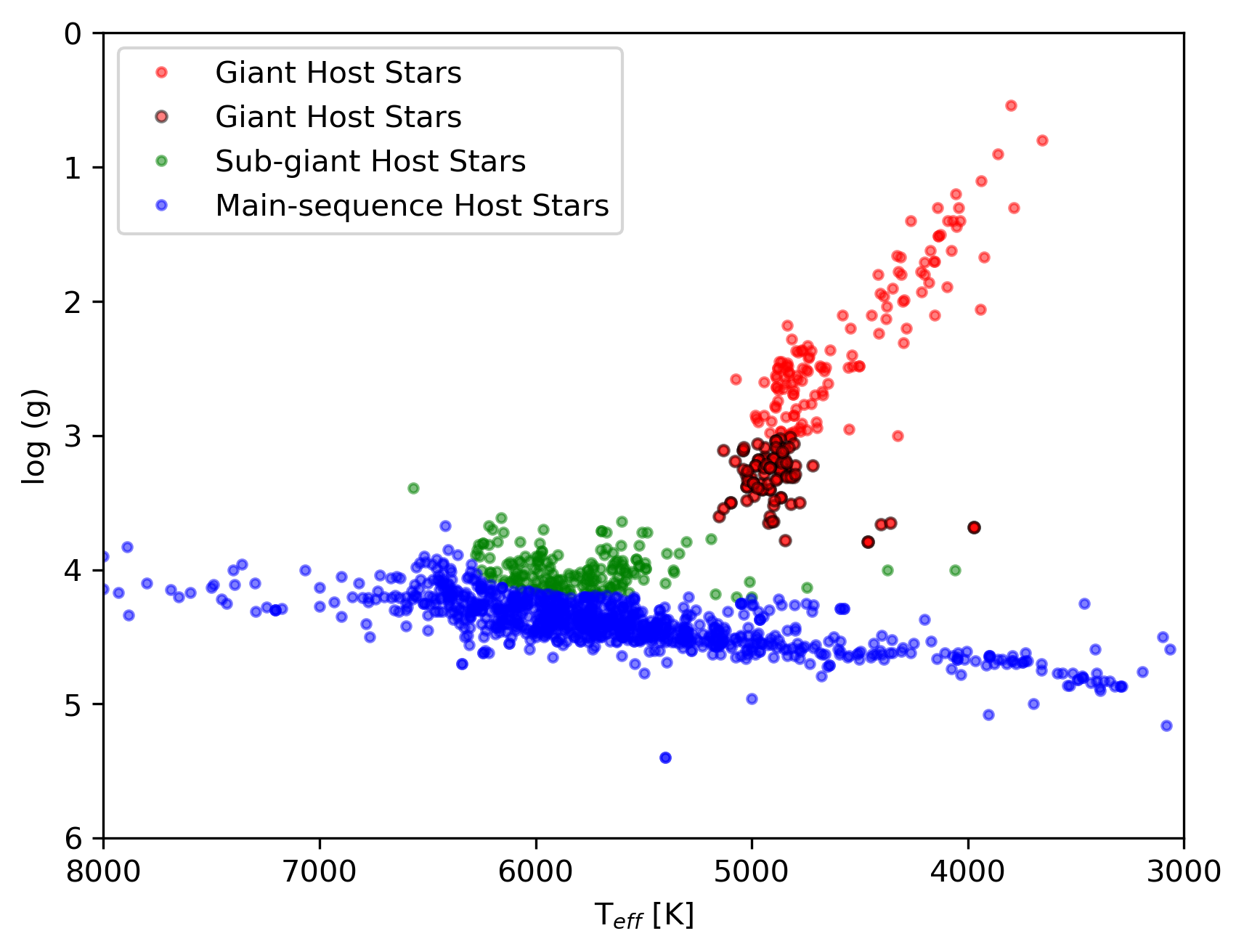} 
    \includegraphics[width=0.9\columnwidth]{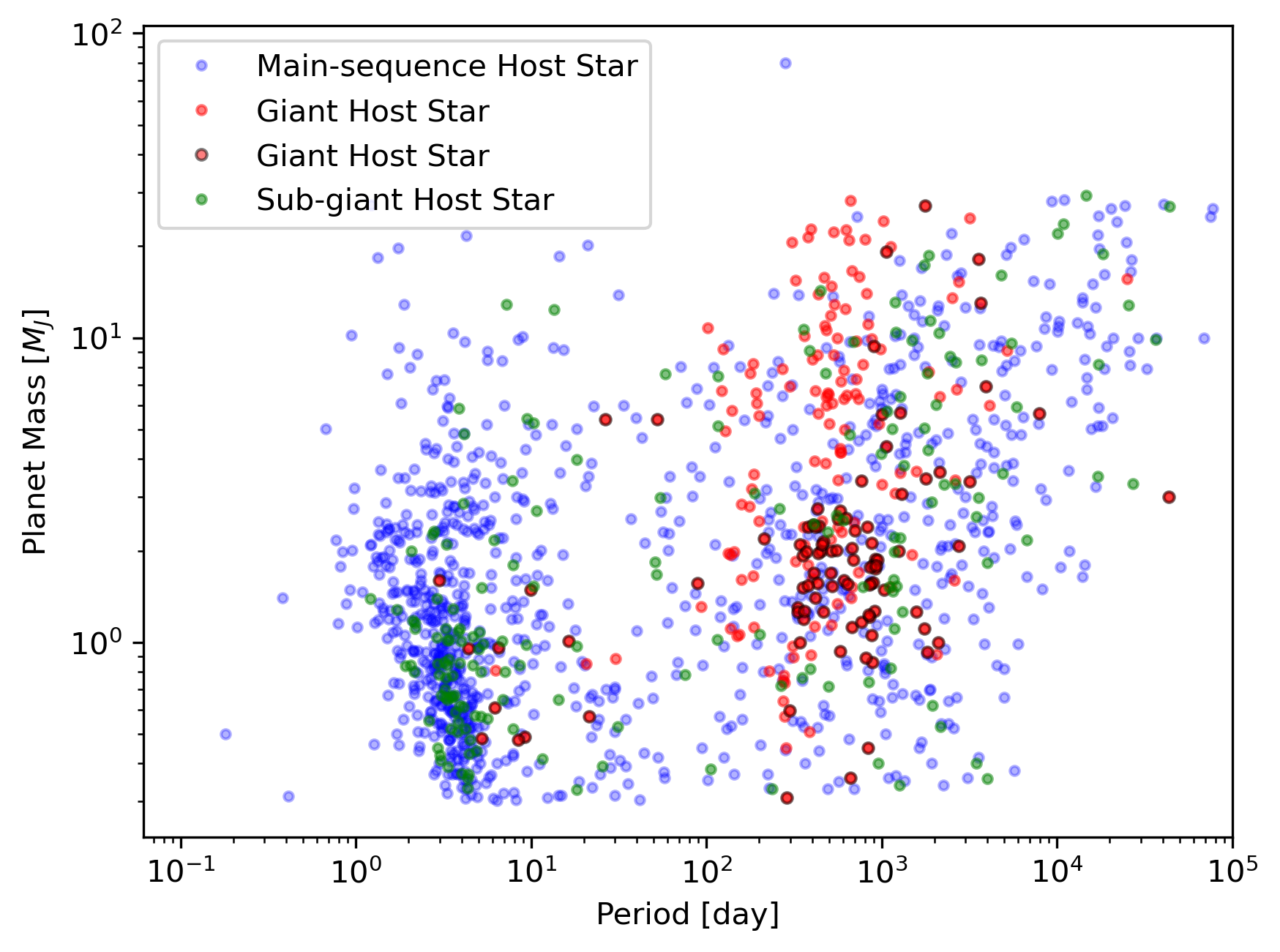} 
    \includegraphics[width=0.9\columnwidth]{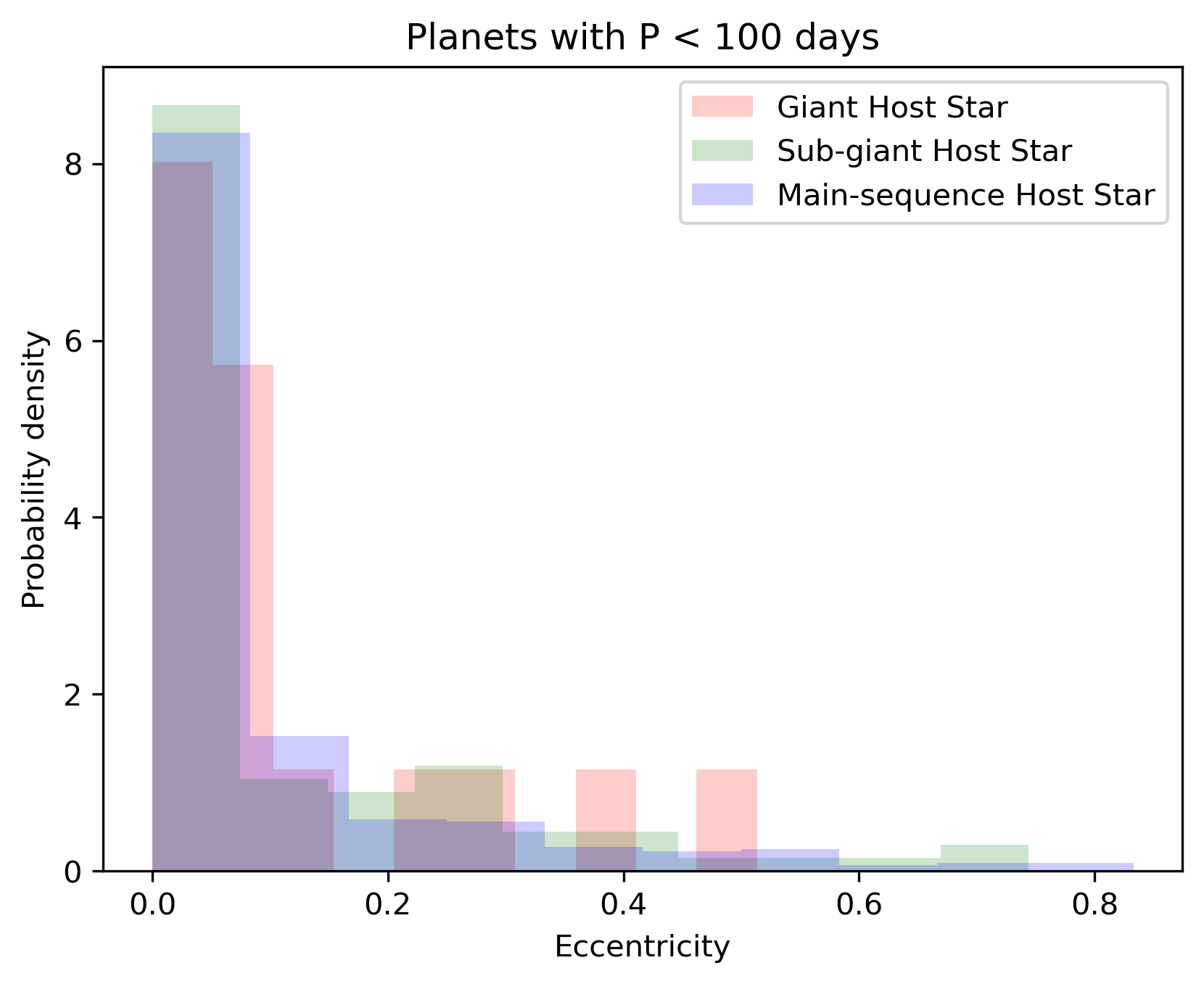} 
    \includegraphics[width=0.9\columnwidth]{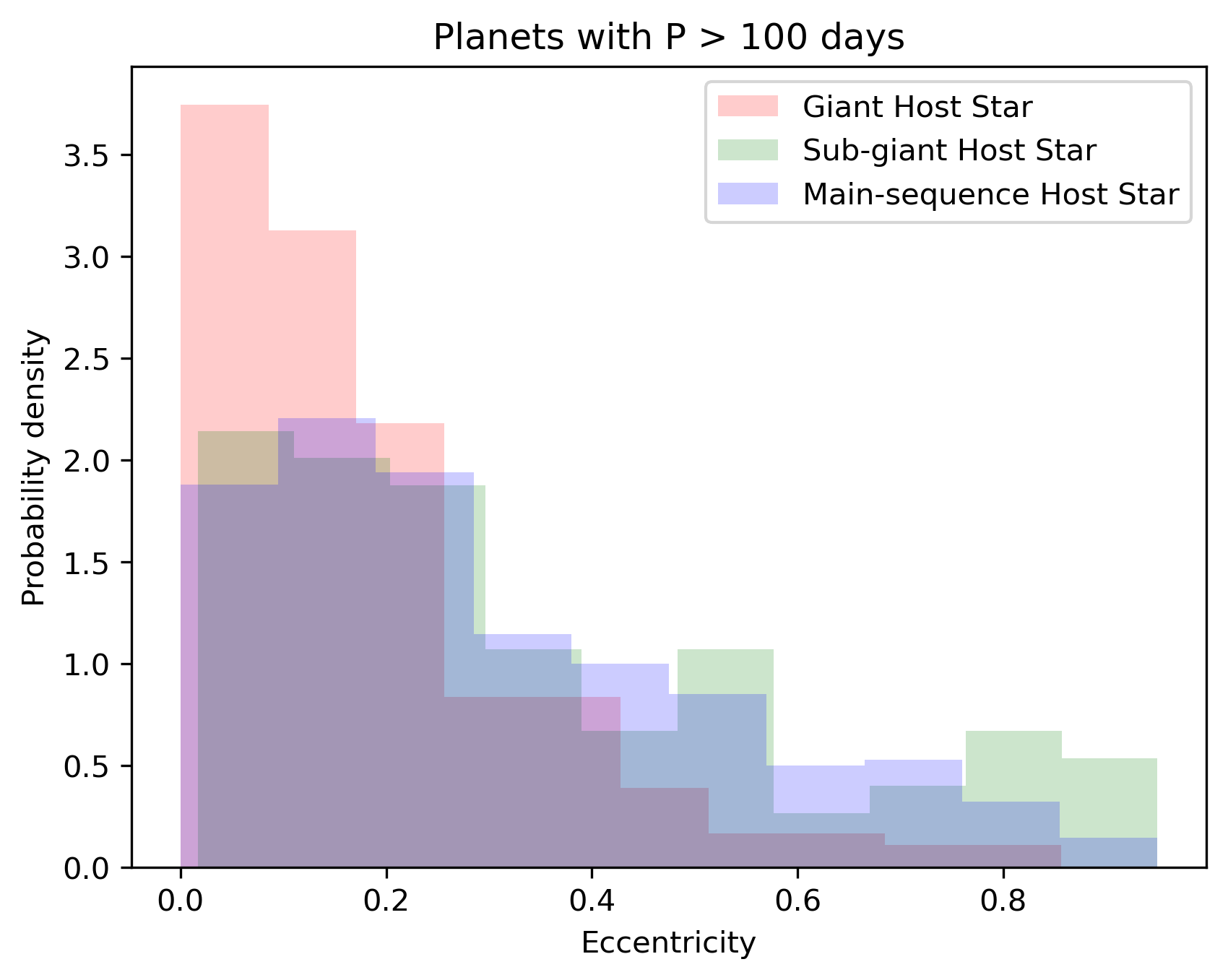} 
    \caption{Upper left: Host stars of giant exoplanets ($\text{M} \ \sin i > 0.3 \ \text{M}_J$) with either a true mass or M $sin i$ measurement divided into their evolutionary stage. Red points are giant stars, green points are sub-giants, and blue are main-sequence stars. Red points with black edges are giant stars at the beginning of their ascension up the red giant branch ($\log g > 3.0$). Upper right: The population of giant exoplanets in mass-period space sorted by the evolutionary stage of their host star. Lower: Eccentricity distribution of planets with periods shorter than 100 days (left) and larger than 100 days (right) orbiting main-sequence, sub-giant, and giant stars.}
    \label{fig:popstar}
\end{figure*}

The six planet hosts in M67 are distributed over different evolutionary stages with the three hot-Jupiters orbiting main-sequence stars and the longer period planets orbiting a turn-off point star and two red giants. The evolution of the host away from the main-sequence and into a red giant is expected to significantly influence the surrounding planets and their orbits through tidal interaction, evaporation, or engulfment \citep[see e.g.][]{Villaver2014,Veras2016}. However, it is still not well understood, when the effects from stellar evolution start to become dominant and alter the planet population around them. We expect to learn a lot from studying planet hosts in different evolutionary states located in the same cluster, as not only can it be difficult to determine the exact point when a star evolves off the main-sequence for field stars, but we also eliminate other factors such as system age and chemical composition which can also affect the planet properties.
For M67 we find three short-period planets around main-sequence stars but none around the more evolved stars. Unfortunately, our sample of planets in M67 is far too small to make general conclusions on the influence of stellar evolutionary state on planet properties.
\\
To look for trends in the larger exoplanet population, we divide the list of planets from the previous section into three groups corresponding to host-stars on the main-sequence, stars that are in the process of evolving off the main-sequence (sub-giants), and stars ascending the red-giant branch. We distinguish between the three evolutionary stages using the data-driven boundaries derived by \cite{Huber2016}. Figure \ref{fig:popstar} shows the distribution of planet hosts into the three groups. We again focus on giant planets (M $\sin i>0.3\text{M}_J$) and find a total of 971 planets around main-sequence stars, 177 around sub-giants, and 230 around giant stars with either a mass or M $\sin i$ measurement. 
\\
Looking at the distribution of exoplanets in mass-period space (Figure \ref{fig:popstar}) there seems to be no significant difference between the planet population around main-sequence and sub-giant stars. However, for the planets orbiting giant stars, there is a distinct lack of short-period planets. Additionally, except for one, all of the planets with periods inside 10 days orbit giant stars at the beginning of the giant branch. In the plot planets around these stars are marked with black edges around their markers. These results are consistent with recent occurrence rate analyses which found that the main-sequence and sub-giant occurrence rates are statistically indistinguishable and a paucity of short-period planets is only observed for stars that have significantly ascended on the giant branch \citep{grunblatt2019giant,temmink2023occurrence,Chontos2024}. 
\\
This also seems to hold true for changes in the orbits of planets. As shown in Figure \ref{fig:popstar}, the orbital eccentricity follows a similar distribution for both main-sequence and sub-giant host stars with mean eccentricity and standard deviation values of $0.1 \pm 0.16$ for main-sequence stars and $0.11 \pm 0.17$ for sub-giants for planets with shorter periods ($P<100$ days) and $0.34 \pm 0.27$ (sub-giant hosts) and $0.30\pm 0.23$ (main-sequence hosts) for planets with periods larger than 100 days. In the case of planet-hosting red giants, the mean eccentricities with their standard deviation are $0.12 \pm 0.14$ ($P<100$ days) and $0.18 \pm 0.17$ ($P>100$ days). While there is not much difference for the close-in planets, at larger separations, circularization effects of the orbits seem to be stronger for highly evolved stars possibly due to stronger stellar tides. 
\\
These trends suggest that planets around main-sequence and sub-giant stars are very similar and are only affected significantly once the star has evolved farther into the giant phase. However, as the exact evolutionary stage can be difficult to determine especially for the sub-giant phase, a sample of well-characterized planet hosts across different stages of stellar evolution is needed to find the mechanisms that are responsible for shaping the different planet populations.
\subsection{Hot-Jupiter occurrence in M67}
Lastly, we update the hot-Jupiter occurrence rate in M67 using the method of \cite{brucalassi2016search} to include our five new targets. As discussed in the previous section, the planet populations around main-sequence and turn-off/sub-giant stars are very similar which is why we consider both stellar types when calculating the hot-Jupiter occurrence rate. With no hot-Jupiters detected around the five new stars added to the sample, we derive an occurrence rate of $4.2_{-2.3}^{+4.1} \%$ (compared to $4.5 \% $ before) and $5.4_{-3.0}^{+5.1}  \%$ (compared to $5.6 \% $ before) only considering single stars. These values are still higher than the $0.5 - 1 \%$ derived for field stars \citep{wright2012frequency,petigura2018california,zhou2019two}. We calculate both occurrence rates to compare to different survey strategies. Blind radial velocity searches for planets often exclude binary star systems a priori while transit surveys mostly do not make this kind of preselection. Because of this occurrence rates from RV surveys have been higher than for transit surveys.
\\
Given that M67 was observed by TESS in sectors 34, 44, 45, and 46, it is possible to investigate whether the photometric data corroborates the higher hot-Jupiter occurrence rate of our RV survey. 
\\
We crosscheck a list of 1113 potential members of M67 with the TESS input catalog and find that 320 have a lightcurve produced either by the SPOC \citep{jenkins2016tess} or the QLP \citep{huang2020photometry} pipelines. Out of these 320 stars, none have a planet candidate in the TESS target of interest list. Assuming the occurrence rate of $4.2 \% $ for giant planets with periods between $1 - 10$ days derived from our RV survey and a transit probability $\sim \frac{R_*}{a}$ with $R_*$ being the stellar radius and $a$ the semi-major axis we would expect an average of $\sim 1.2$ transiting hot-Jupiters in the sample of 320 stars with TESS lightcurves. The calculation does not account for the ability to detect transit signals of hot-Jupiters in the given light curves as a detailed completeness analysis for the 320 stars is beyond the scope of this work. 
\\
Given that M67 is a cluster, blending of neighboring stars in the TESS apertures can complicate the detection of shallow transit signals so it might not be possible to recover planetary signals for all of the 320 stars, and the number of 1.2 expected transiting hot-Jupiters would be closer to an upper limit. There are however techniques to reduce the blending in crowded fields which could mitigate this issue \citep[see ][]{Libralato2016,Nardiello2019}
\\
Therefore, the non-detection of hot-Jupiter signals in the TESS data of M67 does not definitively disprove the higher occurrence rate that was derived during this survey.

\section{Summary} \label{five}
We report spectroscopic data from an extension of the "Search for giant planets in M67" \citep{pasquini2012search,brucalassi2014three,brucalassi2016search,brucalassi2017search} survey collected with the HPF spectrograph between 2019 and 2022. These data result in the detection of a new planet around the turn-off point star S1429. S1429 b is almost twice as massive as Jupiter at $\text{M} \sin i = 1.80 \pm 0.2$ M$_J$ and has a period of $77.48_{-0.19}^{+0.18}$ days assuming a circular orbit, but we can not rule out a small eccentricity of the orbit. Given the derived properties, S1429 b belongs to the class of warm-Jupiters with an equilibrium temperature of $683 \pm 9$ K
\\
We also report two new candidates for binary star systems in S995 and S2207. For S2207 we find an orbital solution with a period of $1882 \pm 24$ days and a minimum mass of the companion of $0.17 \pm 0.05$ M$_\odot$. Due to the incomplete sampling, we were not able to derive a definitive orbital solution for S995. Furthermore, the additional data confirms that a potential planetary signal in the RV of YBP2018 is caused by a stellar-mass companion with a minimum mass of $0.22 \pm 0.06$ M$_\odot$.
\\
M67 remains an interesting target for studying exoplanets with a relatively homogeneous sample of host stars. Especially the increased hot-Jupiter occurrence and the higher eccentricity distribution for the hot-Jupiters in M67 warrant further investigation either with detailed analysis of the available photometric data or additional radial velocity observations.

\begin{acknowledgements}
We thank the anonymous referee for corrections and suggestions that helped us improve the presentation of the paper. LT acknowledges support from the Excellence Cluster ORIGINS funded by the Deutsche Forschungsgemeinschaft (DFG, German Research Foundation) under Germany´s Excellence Strategy – EXC 2094 – 390783311. RPS, LP, AB thank ESO. RPS thanks the Hobby Eberly Telescope (HET) project and CNRS for allocating the observations and the technical support. These results are based on observations obtained with the Habitable-zone Planet Finder Spectrograph on the HET. The HPF team acknowledges support from NSF grants AST-1006676, AST-1126413, AST-1310885, AST-1517592, AST-1310875, ATI 2009889, ATI-2009982, AST-2108512, and the NASA Astrobiology Institute (NNA09DA76A) in the pursuit of precision radial velocities in the NIR. The HPF team also acknowledges support from the Heising-Simons Foundation via grant 2017-0494. The Hobby-Eberly Telescope is a joint project of the University of Texas at Austin, the Pennsylvania State University, Ludwig-Maximilians-Universität München, and Georg-August Universität Gottingen. The HET is named in honor of its principal benefactors, William P. Hobby and Robert E. Eberly. The HET Collaboration acknowledges the support and resources from the Texas Advanced Computing Center. We thank the Resident Astronomers and Telescope Operators at the HET for the skillful execution of our observations with HPF. We would like to acknowledge that the HET is built on Indigenous land. Moreover, we would like to acknowledge and pay our respects to the Carrizo \& Comecrudo, Coahuiltecan, Caddo, Tonkawa, Comanche, Lipan Apache, Alabama-Coushatta, Kickapoo, Tigua Pueblo, and all the American Indian and Indigenous Peoples and communities who have been or have become a part of these lands and territories in Texas, here on Turtle Island. AB thanks the GAPS-TNG community. HLR acknowledges the support of the DFG priority program SPP 1992 {\it Exploring the Diversity of Extrasolar Planets (RE 1664/20-1)}. We acknowledge support from the Programme National de Physique Stellaire and the Programme National de Plan{\'e}tologie of the Institut National des Sciences de l'Univers -- CNRS that allocated the observation time at the OHP 1.93\,m telescope. J.R.M., B.L.C.M., and I.C.L. acknowledge partial financial support from the Brazilian funding agencies CNPq, Print/CAPES/UFRN, and CAPES (Finance Code 001). The HARPS data used in this work was collected via the programs 106.215E.002 and 106.215E.004. This research has made use of the NASA Exoplanet Archive, which is operated by the California Institute of Technology, under contract with the National Aeronautics and Space Administration under the Exoplanet Exploration Program. We acknowledge the use of public TESS data from pipelines at the TESS Science Office and at the TESS Science Processing Operations Center. Funding for the TESS mission is provided by NASA's Science Mission Directorate. This research has made use of the Exoplanet Follow-up Observation Program website, which is operated by the California Institute of Technology, under contract with the National Aeronautics and Space Administration under the Exoplanet Exploration Program. Resources supporting this work were provided by the NASA High-End Computing (HEC) Program through the NASA Advanced Supercomputing (NAS) Division at Ames Research Center for the production of the SPOC data products. This paper includes data collected by the TESS mission that are publicly available from the Mikulski Archive for Space Telescopes (MAST).\\
\end{acknowledgements}

\bibliographystyle{aa}
\bibliography{bib} 

\begin{thebibliography}{121}
\expandafter\ifx\csname natexlab\endcsname\relax\def\natexlab#1{#1}\fi

\bibitem[{Allard {et~al.}(2012)Allard, Homeier, \& Freytag}]{Allard2012}
Allard, F., Homeier, D., \& Freytag, B. 2012, Philosophical Transactions of the Royal Society A: Mathematical, Physical and Engineering Sciences, 370, 2765

\bibitem[{Anglada-Escud{\'e} \& Butler(2012)}]{anglada2012harps}
Anglada-Escud{\'e}, G. \& Butler, R.~P. 2012, The Astrophysical Journal Supplement Series, 200, 15

\bibitem[{Barrag{\'a}n {et~al.}(2019)Barrag{\'a}n, Aigrain, Kubyshkina, Gandolfi, Livingston, Fridlund, Fossati, Korth, Parviainen, Malavolta, {et~al.}}]{barragan2019radial}
Barrag{\'a}n, O., Aigrain, S., Kubyshkina, D., {et~al.} 2019, Monthly Notices of the Royal Astronomical Society, 490, 698

\bibitem[{Bitsch {et~al.}(2020)Bitsch, Trifonov, \& Izidoro}]{bitsch2020eccentricity}
Bitsch, B., Trifonov, T., \& Izidoro, A. 2020, Astronomy \& Astrophysics, 643, A66

\bibitem[{Borucki {et~al.}(2010)Borucki, Koch, Basri, Batalha, Brown, Caldwell, Caldwell, Christensen-Dalsgaard, Cochran, DeVore, {et~al.}}]{borucki2010kepler}
Borucki, W.~J., Koch, D., Basri, G., {et~al.} 2010, Science, 327, 977

\bibitem[{{Bouchy} {et~al.}(2013){Bouchy}, {D{\'\i}az}, {H{\'e}brard}, {Arnold}, {Boisse}, {Delfosse}, {Perruchot}, \& {Santerne}}]{bouchy2013}
{Bouchy}, F., {D{\'\i}az}, R.~F., {H{\'e}brard}, G., {et~al.} 2013, \aap, 549, A49

\bibitem[{Brucalassi {et~al.}(2017)Brucalassi, Koppenhoefer, Saglia, Pasquini, Ruiz, Bonifacio, Bedin, Libralato, Biazzo, Melo, {et~al.}}]{brucalassi2017search}
Brucalassi, A., Koppenhoefer, J., Saglia, R., {et~al.} 2017, Astronomy \& Astrophysics, 603, A85

\bibitem[{Brucalassi {et~al.}(2014)Brucalassi, Pasquini, Saglia, Ruiz, Bonifacio, Bedin, Biazzo, Melo, Lovis, \& Randich}]{brucalassi2014three}
Brucalassi, A., Pasquini, L., Saglia, R., {et~al.} 2014, Astronomy \& Astrophysics, 561, L9

\bibitem[{Brucalassi {et~al.}(2016)Brucalassi, Pasquini, Saglia, Ruiz, Bonifacio, Le{\~a}o, Martins, De~Medeiros, Bedin, Biazzo, {et~al.}}]{brucalassi2016search}
Brucalassi, A., Pasquini, L., Saglia, R., {et~al.} 2016, Astronomy \& Astrophysics, 592, L1

\bibitem[{Bryan {et~al.}(2016)Bryan, Knutson, Howard, Ngo, Batygin, Crepp, Fulton, Hinkley, Isaacson, Johnson, {et~al.}}]{bryan2016statistics}
Bryan, M.~L., Knutson, H.~A., Howard, A.~W., {et~al.} 2016, The Astrophysical Journal, 821, 89

\bibitem[{Caldwell {et~al.}(2020)Caldwell, Tenenbaum, Twicken, Jenkins, Ting, Smith, Hedges, Fausnaugh, Rose, \& Burke}]{caldwell2020tess}
Caldwell, D.~A., Tenenbaum, P., Twicken, J.~D., {et~al.} 2020, Research Notes of the AAS, 4, 201

\bibitem[{{Castelli} \& {Kurucz}(2003)}]{Castelli2004}
{Castelli}, F. \& {Kurucz}, R.~L. 2003, in Modelling of Stellar Atmospheres, ed. N.~{Piskunov}, W.~W. {Weiss}, \& D.~F. {Gray}, Vol. 210, A20

\bibitem[{Choi {et~al.}(2016)Choi, Dotter, Conroy, Cantiello, Paxton, \& Johnson}]{choi2016mesa}
Choi, J., Dotter, A., Conroy, C., {et~al.} 2016, The Astrophysical Journal, 823, 102

\bibitem[{{Chontos} {et~al.}(2024){Chontos}, {Huber}, {Grunblatt}, {Saunders}, {Winn}, {McCormack}, {Knudstrup}, {Albrecht}, {Crossfield}, {Rodriguez}, {Ciardi}, {Collins}, {Jenkins}, {Bieryla}, {Batalha}, {Beard}, {Dai}, {Dalba}, {Fetherolf}, {Giacalone}, {Hill}, {Howard}, {Isaacson}, {Kane}, {Lubin}, {MacDougall}, {Mo{\v{c}}nik}, {Akana Murphy}, {Petigura}, {Pidhorodetska}, {Polanski}, {Robertson}, {Rubenzahl}, {Turtelboom}, {Weiss}, {Van Zandt}, {Rocker}, {Vanderspek}, {Latham}, {Seager}, {Quinn}, {Shporer}, {Eisner}, {Goeke}, {Levine}, {Ting}, {Howell}, {Schlieder}, {Benni}, {Boyle}, {Gan}, {Girardin}, {Gonzalez}, {Gregorio}, {Horne}, {Livingston}, {Lund}, {Mann}, {Massey}, {Matthews}, {McLeod}, {Palle}, {Popowicz}, {Relles}, {Schwarz}, {Sefako}, {Srdoc}, {Tan}, {Wang}, \& {Ziegler}}]{Chontos2024}
{Chontos}, A., {Huber}, D., {Grunblatt}, S.~K., {et~al.} 2024, arXiv e-prints, arXiv:2402.07893

\bibitem[{Currie(2009)}]{currie2009semimajor}
Currie, T. 2009, The Astrophysical Journal, 694, L171

\bibitem[{David {et~al.}(2019{\natexlab{a}})David, Cody, Hedges, Mamajek, Hillenbrand, Ciardi, Beichman, Petigura, Fulton, Isaacson, {et~al.}}]{david2019warm}
David, T.~J., Cody, A.~M., Hedges, C.~L., {et~al.} 2019{\natexlab{a}}, The Astronomical Journal, 158, 79

\bibitem[{David {et~al.}(2016)David, Hillenbrand, Petigura, Carpenter, Crossfield, Hinkley, Ciardi, Howard, Isaacson, Cody, {et~al.}}]{david2016neptune}
David, T.~J., Hillenbrand, L.~A., Petigura, E.~A., {et~al.} 2016, Nature, 534, 658

\bibitem[{David {et~al.}(2019{\natexlab{b}})David, Petigura, Luger, Foreman-Mackey, Livingston, Mamajek, \& Hillenbrand}]{david2019four}
David, T.~J., Petigura, E.~A., Luger, R., {et~al.} 2019{\natexlab{b}}, The Astrophysical Journal Letters, 885, L12

\bibitem[{Dawson \& Johnson(2018)}]{dawson2018origins}
Dawson, R.~I. \& Johnson, J.~A. 2018, Annual Review of Astronomy and Astrophysics, 56, 175

\bibitem[{De~Silva {et~al.}(2007)De~Silva, Freeman, Asplund, Bland-Hawthorn, Bessell, \& Collet}]{de2007chemical}
De~Silva, G., Freeman, K., Asplund, M., {et~al.} 2007, The Astronomical Journal, 133, 1161

\bibitem[{Dong {et~al.}(2021)Dong, Huang, Zhou, Dawson, Rodriguez, Eastman, Collins, Quinn, Shporer, Triaud, {et~al.}}]{dong2021toi}
Dong, J., Huang, C.~X., Zhou, G., {et~al.} 2021, The Astrophysical Journal Letters, 920, L16

\bibitem[{Dotter(2016)}]{dotter2016mesa}
Dotter, A. 2016, The Astrophysical Journal Supplement Series, 222, 8

\bibitem[{Espinoza {et~al.}(2019)Espinoza, Kossakowski, \& Brahm}]{espinoza2019juliet}
Espinoza, N., Kossakowski, D., \& Brahm, R. 2019, Monthly Notices of the Royal Astronomical Society, 490, 2262

\bibitem[{Fischer \& Valenti(2005)}]{fischer2005planet}
Fischer, D.~A. \& Valenti, J. 2005, The Astrophysical Journal, 622, 1102

\bibitem[{Foreman-Mackey {et~al.}(2013)Foreman-Mackey, Hogg, Lang, \& Goodman}]{foreman2013emcee}
Foreman-Mackey, D., Hogg, D.~W., Lang, D., \& Goodman, J. 2013, Publications of the Astronomical Society of the Pacific, 125, 306

\bibitem[{Fortney {et~al.}(2021)Fortney, Dawson, \& Komacek}]{fortney2021hot}
Fortney, J.~J., Dawson, R.~I., \& Komacek, T.~D. 2021, Journal of Geophysical Research: Planets, 126, e2020JE006629

\bibitem[{Fuhrmeister {et~al.}(2019)Fuhrmeister, Czesla, Schmitt, Johnson, Sch{\"o}fer, Jeffers, Caballero, Zechmeister, Reiners, Ribas, {et~al.}}]{fuhrmeister2019carmenes}
Fuhrmeister, B., Czesla, S., Schmitt, J.~H., {et~al.} 2019, Astronomy \& Astrophysics, 623, A24

\bibitem[{Fulton {et~al.}(2018)Fulton, Petigura, Blunt, \& Sinukoff}]{fulton2018radvel}
Fulton, B.~J., Petigura, E.~A., Blunt, S., \& Sinukoff, E. 2018, Publications of the Astronomical Society of the Pacific, 130, 044504

\bibitem[{Ghezzi {et~al.}(2018)Ghezzi, Montet, \& Johnson}]{ghezzi2018retired}
Ghezzi, L., Montet, B.~T., \& Johnson, J.~A. 2018, The Astrophysical Journal, 860, 109

\bibitem[{Goldreich \& Tremaine(1980)}]{goldreich1980disk}
Goldreich, P. \& Tremaine, S. 1980, Astrophysical Journal, 241, 425

\bibitem[{Gonzalez(1997)}]{gonzalez1997stellar}
Gonzalez, G. 1997, Monthly Notices of the Royal Astronomical Society, 285, 403

\bibitem[{Grunblatt {et~al.}(2019)Grunblatt, Huber, Gaidos, Hon, Zinn, \& Stello}]{grunblatt2019giant}
Grunblatt, S.~K., Huber, D., Gaidos, E., {et~al.} 2019, The Astronomical Journal, 158, 227

\bibitem[{{Hamers} \& {Tremaine}(2017)}]{hamers2017}
{Hamers}, A.~S. \& {Tremaine}, S. 2017, \aj, 154, 272

\bibitem[{Higson {et~al.}(2019)Higson, Handley, Hobson, \& Lasenby}]{higson2019dynamic}
Higson, E., Handley, W., Hobson, M., \& Lasenby, A. 2019, Statistics and Computing, 29, 891

\bibitem[{{Hill} {et~al.}(2021){Hill}, {Lee}, {MacQueen}, {Kelz}, {Drory}, {Vattiat}, {Good}, {Ramsey}, {Kriel}, {Peterson}, {DePoy}, {Gebhardt}, {Marshall}, {Tuttle}, {Bauer}, {Chonis}, {Fabricius}, {Froning}, {H{\"a}user}, {Indahl}, {Jahn}, {Landriau}, {Leck}, {Montesano}, {Prochaska}, {Snigula}, {Zeimann}, {Bryant}, {Damm}, {Fowler}, {Janowiecki}, {Martin}, {Mrozinski}, {Odewahn}, {Rostopchin}, {Shetrone}, {Spencer}, {Mentuch Cooper}, {Armandroff}, {Bender}, {Dalton}, {Hopp}, {Komatsu}, {Nicklas}, {Ramsey}, {Roth}, {Schneider}, {Sneden}, \& {Steinmetz}}]{het2021}
{Hill}, G.~J., {Lee}, H., {MacQueen}, P.~J., {et~al.} 2021, \aj, 162, 298

\bibitem[{Horne(1986)}]{horne1986optimal}
Horne, K. 1986, Publications of the Astronomical Society of the Pacific, 98, 609

\bibitem[{Howell {et~al.}(2014)Howell, Sobeck, Haas, Still, Barclay, Mullally, Troeltzsch, Aigrain, Bryson, Caldwell, {et~al.}}]{howell2014k2}
Howell, S.~B., Sobeck, C., Haas, M., {et~al.} 2014, Publications of the Astronomical Society of the Pacific, 126, 398

\bibitem[{Huang {et~al.}(2020)Huang, Vanderburg, P{\'a}l, Sha, Yu, Fong, Fausnaugh, Shporer, Guerrero, Vanderspek, {et~al.}}]{huang2020photometry}
Huang, C.~X., Vanderburg, A., P{\'a}l, A., {et~al.} 2020, Research Notes of the AAS, 4, 204

\bibitem[{{Huber} {et~al.}(2016){Huber}, {Bryson}, {Haas}, {Barclay}, {Barentsen}, {Howell}, {Sharma}, {Stello}, \& {Thompson}}]{Huber2016}
{Huber}, D., {Bryson}, S.~T., {Haas}, M.~R., {et~al.} 2016, \apjs, 224, 2

\bibitem[{Husser {et~al.}(2013)Husser, Wende-von Berg, Dreizler, Homeier, Reiners, Barman, \& Hauschildt}]{husser2013new}
Husser, T.-O., Wende-von Berg, S., Dreizler, S., {et~al.} 2013, Astronomy \& Astrophysics, 553, A6

\bibitem[{Ida \& Lin(2008)}]{ida2008toward}
Ida, S. \& Lin, D. 2008, The Astrophysical Journal, 673, 487

\bibitem[{Jenkins {et~al.}(2016)Jenkins, Twicken, McCauliff, Campbell, Sanderfer, Lung, Mansouri-Samani, Girouard, Tenenbaum, Klaus, {et~al.}}]{jenkins2016tess}
Jenkins, J.~M., Twicken, J.~D., McCauliff, S., {et~al.} 2016, in Software and Cyberinfrastructure for Astronomy IV, Vol. 9913, SPIE, 1232--1251

\bibitem[{Jenkins {et~al.}(2017)Jenkins, Jones, Tuomi, D{\'\i}az, Cordero, Aguayo, Pantoja, Arriagada, Mahu, Brahm, {et~al.}}]{jenkins2017new}
Jenkins, J.~S., Jones, H., Tuomi, M., {et~al.} 2017, Monthly Notices of the Royal Astronomical Society, 466, 443

\bibitem[{Johnson {et~al.}(2010)Johnson, Aller, Howard, \& Crepp}]{johnson2010giant}
Johnson, J.~A., Aller, K.~M., Howard, A.~W., \& Crepp, J.~R. 2010, Publications of the Astronomical Society of the Pacific, 122, 905

\bibitem[{J{\"o}nsson {et~al.}(2020)J{\"o}nsson, Holtzman, Prieto, Cunha, Garc{\'\i}a-Hern{\'a}ndez, Hasselquist, Masseron, Osorio, Shetrone, Smith, {et~al.}}]{jonsson2020apogee}
J{\"o}nsson, H., Holtzman, J.~A., Prieto, C.~A., {et~al.} 2020, The Astronomical Journal, 160, 120

\bibitem[{Kanodia \& Wright(2018)}]{Kanodia_2018}
Kanodia, S. \& Wright, J. 2018, Research Notes of the AAS, 2, 4

\bibitem[{Kennedy \& Kenyon(2009)}]{kennedy2009stellar}
Kennedy, G.~M. \& Kenyon, S.~J. 2009, The Astrophysical Journal, 695, 1210

\bibitem[{Knutson {et~al.}(2014)Knutson, Fulton, Montet, Kao, Ngo, Howard, Crepp, Hinkley, Bakos, Batygin, {et~al.}}]{knutson2014friends}
Knutson, H.~A., Fulton, B.~J., Montet, B.~T., {et~al.} 2014, The Astrophysical Journal, 785, 126

\bibitem[{{Kurucz}(1993)}]{1993KurCD..13.....K}
{Kurucz}, R. 1993, ATLAS9 Stellar Atmosphere Programs and 2 km/s grid. Kurucz CD-ROM No. 13. Cambridge, 13

\bibitem[{Laliotis {et~al.}(2023)Laliotis, Burt, Mamajek, Li, Perdelwitz, Zhao, Butler, Holden, Rosenthal, Fulton, {et~al.}}]{laliotis2023doppler}
Laliotis, K., Burt, J.~A., Mamajek, E.~E., {et~al.} 2023, The Astronomical Journal, 165, 176

\bibitem[{Li {et~al.}(2023{\natexlab{a}})Li, Mustill, Davies, \& Gong}]{li2023makingb}
Li, D., Mustill, A.~J., Davies, M.~B., \& Gong, Y.-X. 2023{\natexlab{a}}, Monthly Notices of the Royal Astronomical Society, stad3207

\bibitem[{Li {et~al.}(2023{\natexlab{b}})Li, Mustill, Davies, \& Gong}]{li2023making}
Li, D., Mustill, A.~J., Davies, M.~B., \& Gong, Y.-X. 2023{\natexlab{b}}, Monthly Notices of the Royal Astronomical Society, 518, 4265

\bibitem[{{Libralato} {et~al.}(2016){Libralato}, {Bedin}, {Nardiello}, \& {Piotto}}]{Libralato2016}
{Libralato}, M., {Bedin}, L.~R., {Nardiello}, D., \& {Piotto}, G. 2016, \mnras, 456, 1137

\bibitem[{{Lightkurve Collaboration} {et~al.}(2018){Lightkurve Collaboration}, {Cardoso}, {Hedges}, {Gully-Santiago}, {Saunders}, {Cody}, {Barclay}, {Hall}, {Sagear}, {Turtelboom}, {Zhang}, {Tzanidakis}, {Mighell}, {Coughlin}, {Bell}, {Berta-Thompson}, {Williams}, {Dotson}, \& {Barentsen}}]{2018ascl.soft12013L}
{Lightkurve Collaboration}, {Cardoso}, J.~V.~d.~M., {Hedges}, C., {et~al.} 2018, {Lightkurve: Kepler and TESS time series analysis in Python}, Astrophysics Source Code Library

\bibitem[{Lin \& Papaloizou(1986)}]{lin1986tidal}
Lin, D.~N. \& Papaloizou, J. 1986, Astrophysical Journal, Part 1 (ISSN 0004-637X), vol. 309, Oct. 15, 1986, p. 846-857., 309, 846

\bibitem[{Livingston {et~al.}(2018)Livingston, Dai, Hirano, Gandolfi, Nowak, Endl, Velasco, Fukui, Narita, Prieto-Arranz, {et~al.}}]{livingston2018three}
Livingston, J.~H., Dai, F., Hirano, T., {et~al.} 2018, The Astronomical Journal, 155, 115

\bibitem[{Livingston {et~al.}(2019)Livingston, Dai, Hirano, Gandolfi, Trani, Nowak, Cochran, Endl, Albrecht, Barragan, {et~al.}}]{livingston2019k2}
Livingston, J.~H., Dai, F., Hirano, T., {et~al.} 2019, Monthly Notices of the Royal Astronomical Society, 484, 8

\bibitem[{Lovis \& Mayor(2007)}]{lovis2007planets}
Lovis, C. \& Mayor, M. 2007, Astronomy \& Astrophysics, 472, 657

\bibitem[{Mahadevan {et~al.}(2018)Mahadevan, Anderson, Balderrama, Bender, Bevins, Blakeslee, Cole, Conran, Diddams, Dykhouse, {et~al.}}]{mahadevan2018habitable}
Mahadevan, S., Anderson, T., Balderrama, E., {et~al.} 2018, in Ground-based and airborne instrumentation for astronomy VII, Vol. 10702, SPIE, 1070214

\bibitem[{Mahadevan {et~al.}(2012)Mahadevan, Ramsey, Bender, Terrien, Wright, Halverson, Hearty, Nelson, Burton, Redman, {et~al.}}]{mahadevan2012habitable}
Mahadevan, S., Ramsey, L., Bender, C., {et~al.} 2012, in Ground-based and airborne instrumentation for astronomy IV, Vol. 8446, SPIE, 624--637

\bibitem[{Mahadevan {et~al.}(2014)Mahadevan, Ramsey, Terrien, Halverson, Roy, Hearty, Levi, Stefansson, Robertson, Bender, {et~al.}}]{mahadevan2014habitable}
Mahadevan, S., Ramsey, L.~W., Terrien, R., {et~al.} 2014, in Ground-based and airborne instrumentation for astronomy v, Vol. 9147, SPIE, 543--552

\bibitem[{Malavolta {et~al.}(2016)Malavolta, Nascimbeni, Piotto, Quinn, Borsato, Granata, Bonomo, Marzari, Bedin, Rainer, {et~al.}}]{malavolta2016gaps}
Malavolta, L., Nascimbeni, V., Piotto, G., {et~al.} 2016, Astronomy \& Astrophysics, 588, A118

\bibitem[{{Malmberg} {et~al.}(2011){Malmberg}, {Davies}, \& {Heggie}}]{malmberg2011}
{Malmberg}, D., {Davies}, M.~B., \& {Heggie}, D.~C. 2011, \mnras, 411, 859

\bibitem[{Mann {et~al.}(2016{\natexlab{a}})Mann, Gaidos, Mace, Johnson, Bowler, LaCourse, Jacobs, Vanderburg, Kraus, Kaplan, {et~al.}}]{mann2016zodiacal}
Mann, A.~W., Gaidos, E., Mace, G.~N., {et~al.} 2016{\natexlab{a}}, The Astrophysical Journal, 818, 46

\bibitem[{Mann {et~al.}(2017)Mann, Gaidos, Vanderburg, Rizzuto, Ansdell, Medina, Mace, Kraus, \& Sokal}]{mann2017zodiacal}
Mann, A.~W., Gaidos, E., Vanderburg, A., {et~al.} 2017, The Astronomical Journal, 153, 64

\bibitem[{Mann {et~al.}(2016{\natexlab{b}})Mann, Newton, Rizzuto, Irwin, Feiden, Gaidos, Mace, Kraus, James, Ansdell, {et~al.}}]{mann2016zodiacalb}
Mann, A.~W., Newton, E.~R., Rizzuto, A.~C., {et~al.} 2016{\natexlab{b}}, The Astronomical Journal, 152, 61

\bibitem[{Mathys {et~al.}(2023)Mathys, Holdsworth, \& Kurtz}]{mathys2023long}
Mathys, G., Holdsworth, D., \& Kurtz, D. 2023, arXiv preprint arXiv:2312.01140

\bibitem[{{Mayor} {et~al.}(2003){Mayor}, {Pepe}, {Queloz}, {Bouchy}, {Rupprecht}, {Lo Curto}, {Avila}, {Benz}, {Bertaux}, {Bonfils}, {Dall}, {Dekker}, {Delabre}, {Eckert}, {Fleury}, {Gilliotte}, {Gojak}, {Guzman}, {Kohler}, {Lizon}, {Longinotti}, {Lovis}, {Megevand}, {Pasquini}, {Reyes}, {Sivan}, {Sosnowska}, {Soto}, {Udry}, {van Kesteren}, {Weber}, \& {Weilenmann}}]{mayor2003}
{Mayor}, M., {Pepe}, F., {Queloz}, D., {et~al.} 2003, The Messenger, 114, 20

\bibitem[{Mayor \& Queloz(1995)}]{mayor1995jupiter}
Mayor, M. \& Queloz, D. 1995, nature, 378, 355

\bibitem[{Meibom {et~al.}(2013)Meibom, Torres, Fressin, Latham, Rowe, Ciardi, Bryson, Rogers, Henze, Janes, {et~al.}}]{meibom2013same}
Meibom, S., Torres, G., Fressin, F., {et~al.} 2013, Nature, 499, 55

\bibitem[{{Metcalf} {et~al.}(2019){Metcalf}, {Anderson}, {Bender}, {Blakeslee}, {Brand}, {Carlson}, {Cochran}, {Diddams}, {Endl}, {Fredrick}, {Halverson}, {Hickstein}, {Hearty}, {Jennings}, {Kanodia}, {Kaplan}, {Levi}, {Lubar}, {Mahadevan}, {Monson}, {Ninan}, {Nitroy}, {Osterman}, {Papp}, {Quinlan}, {Ramsey}, {Robertson}, {Roy}, {Schwab}, {Sigurdsson}, {Srinivasan}, {Stefansson}, {Sterner}, {Terrien}, {Wolszczan}, {Wright}, \& {Ycas}}]{metcalf2019}
{Metcalf}, A.~J., {Anderson}, T., {Bender}, C.~F., {et~al.} 2019, Optica, 6, 233

\bibitem[{Morton(2015)}]{morton2015isochrones}
Morton, T.~D. 2015, Astrophysics Source Code Library, ascl

\bibitem[{{Nardiello} {et~al.}(2019){Nardiello}, {Borsato}, {Piotto}, {Colombo}, {Manthopoulou}, {Bedin}, {Granata}, {Lacedelli}, {Libralato}, {Malavolta}, {Montalto}, \& {Nascimbeni}}]{Nardiello2019}
{Nardiello}, D., {Borsato}, L., {Piotto}, G., {et~al.} 2019, \mnras, 490, 3806

\bibitem[{Ninan {et~al.}(2018)Ninan, Bender, Mahadevan, Ford, Monson, Kaplan, Terrien, Roy, Robertson, Kanodia, {et~al.}}]{ninan2018habitable}
Ninan, J., Bender, C.~F., Mahadevan, S., {et~al.} 2018, in High Energy, Optical, and Infrared Detectors for Astronomy VIII, Vol. 10709, SPIE, 694--704

\bibitem[{Obermeier {et~al.}(2016)Obermeier, Henning, Schlieder, Crossfield, Petigura, Howard, Sinukoff, Isaacson, Ciardi, David, {et~al.}}]{obermeier2016k2}
Obermeier, C., Henning, T., Schlieder, J.~E., {et~al.} 2016, The Astronomical Journal, 152, 223

\bibitem[{{\"O}nehag {et~al.}(2014){\"O}nehag, Gustafsson, \& Korn}]{onehag2014abundances}
{\"O}nehag, A., Gustafsson, B., \& Korn, A. 2014, Astronomy \& Astrophysics, 562, A102

\bibitem[{{\"O}nehag {et~al.}(2011){\"O}nehag, Korn, Gustafsson, Stempels, \& VandenBerg}]{onehag2011m67}
{\"O}nehag, A., Korn, A., Gustafsson, B., Stempels, E., \& VandenBerg, D.~A. 2011, Astronomy \& Astrophysics, 528, A85

\bibitem[{Pace \& Pasquini(2004)}]{pace2004age}
Pace, G. \& Pasquini, L. 2004, Astronomy \& Astrophysics, 426, 1021

\bibitem[{Pasquini {et~al.}(2012)Pasquini, Brucalassi, Ruiz, Bonifacio, Lovis, Saglia, Melo, Biazzo, Randich, \& Bedin}]{pasquini2012search}
Pasquini, L., Brucalassi, A., Ruiz, M., {et~al.} 2012, Astronomy \& Astrophysics, 545, A139

\bibitem[{Pasquini {et~al.}(2004)Pasquini, Randich, Zoccali, Hill, Charbonnel, \& Nordstr{\"o}m}]{pasquini2004detailed}
Pasquini, L., Randich, S., Zoccali, M., {et~al.} 2004, Astronomy \& Astrophysics, 424, 951

\bibitem[{Paulson {et~al.}(2004)Paulson, Cochran, \& Hatzes}]{paulson2004searching}
Paulson, D.~B., Cochran, W.~D., \& Hatzes, A.~P. 2004, The Astronomical Journal, 127, 3579

\bibitem[{{Pepe} {et~al.}(2002){Pepe}, {Mayor}, {Galland}, {Naef}, {Queloz}, {Santos}, {Udry}, \& {Burnet}}]{pepe2002}
{Pepe}, F., {Mayor}, M., {Galland}, F., {et~al.} 2002, \aap, 388, 632

\bibitem[{Perruchot {et~al.}(2008)Perruchot, Kohler, Bouchy, Richaud, Richaud, Moreaux, Merzougui, Sottile, Hill, Knispel, {et~al.}}]{perruchot2008sophie}
Perruchot, S., Kohler, D., Bouchy, F., {et~al.} 2008, in Ground-based and Airborne Instrumentation for Astronomy II, Vol. 7014, SPIE, 235--246

\bibitem[{Petigura {et~al.}(2018)Petigura, Marcy, Winn, Weiss, Fulton, Howard, Sinukoff, Isaacson, Morton, \& Johnson}]{petigura2018california}
Petigura, E.~A., Marcy, G.~W., Winn, J.~N., {et~al.} 2018, The Astronomical Journal, 155, 89

\bibitem[{Petrovich(2015)}]{petrovich2015hot}
Petrovich, C. 2015, The Astrophysical Journal, 805, 75

\bibitem[{Quinn {et~al.}(2012)Quinn, White, Latham, Buchhave, Cantrell, Dahm, F{\H{u}}r{\'e}sz, Szentgyorgyi, Geary, Torres, {et~al.}}]{quinn2012two}
Quinn, S.~N., White, R.~J., Latham, D.~W., {et~al.} 2012, The Astrophysical Journal Letters, 756, L33

\bibitem[{Quinn {et~al.}(2014)Quinn, White, Latham, Buchhave, Torres, Stefanik, Berlind, Bieryla, Calkins, Esquerdo, {et~al.}}]{quinn2014hd}
Quinn, S.~N., White, R.~J., Latham, D.~W., {et~al.} 2014, The Astrophysical Journal, 787, 27

\bibitem[{{Ramsey} {et~al.}(1998){Ramsey}, {Adams}, {Barnes}, {Booth}, {Cornell}, {Fowler}, {Gaffney}, {Glaspey}, {Good}, {Hill}, {Kelton}, {Krabbendam}, {Long}, {MacQueen}, {Ray}, {Ricklefs}, {Sage}, {Sebring}, {Spiesman}, \& {Steiner}}]{het1998}
{Ramsey}, L.~W., {Adams}, M.~T., {Barnes}, T.~G., {et~al.} 1998, in Society of Photo-Optical Instrumentation Engineers (SPIE) Conference Series, Vol. 3352, Advanced Technology Optical/IR Telescopes VI, ed. L.~M. {Stepp}, 34--42

\bibitem[{Randich {et~al.}(2005)Randich, Bragaglia, Pastori, Prisinzano, Sestito, Span{\`o}, Villanova, Carraro, Carretta, Romano, {et~al.}}]{randich2005flames}
Randich, S., Bragaglia, A., Pastori, L., {et~al.} 2005, The Messenger, 121, 18

\bibitem[{Randich {et~al.}(2006)Randich, Sestito, Primas, Pallavicini, \& Pasquini}]{randich2006element}
Randich, S., Sestito, P., Primas, F., Pallavicini, R., \& Pasquini, L. 2006, Astronomy \& Astrophysics, 450, 557

\bibitem[{Rasio \& Ford(1996)}]{rasio1996dynamical}
Rasio, F.~A. \& Ford, E.~B. 1996, Science, 274, 954

\bibitem[{Reffert {et~al.}(2015)Reffert, Bergmann, Quirrenbach, Trifonov, \& K{\"u}nstler}]{reffert2015precise}
Reffert, S., Bergmann, C., Quirrenbach, A., Trifonov, T., \& K{\"u}nstler, A. 2015, Astronomy \& Astrophysics, 574, A116

\bibitem[{Rizzuto {et~al.}(2018)Rizzuto, Vanderburg, Mann, Kraus, Dressing, Ag{\"u}eros, Douglas, \& Krolikowski}]{rizzuto2018zodiacal}
Rizzuto, A.~C., Vanderburg, A., Mann, A.~W., {et~al.} 2018, The Astronomical Journal, 156, 195

\bibitem[{Rosenthal {et~al.}(2021)Rosenthal, Fulton, Hirsch, Isaacson, Howard, Dedrick, Sherstyuk, Blunt, Petigura, Knutson, {et~al.}}]{rosenthal2021california}
Rosenthal, L.~J., Fulton, B.~J., Hirsch, L.~A., {et~al.} 2021, The Astrophysical Journal Supplement Series, 255, 8

\bibitem[{Santos {et~al.}(2004)Santos, Israelian, \& Mayor}]{santos2004spectroscopic}
Santos, N.~C., Israelian, G., \& Mayor, M. 2004, Astronomy \& Astrophysics, 415, 1153

\bibitem[{Sato {et~al.}(2008)Sato, Izumiura, Toyota, Kambe, Ikoma, Omiya, Masuda, Takeda, Murata, Itoh, {et~al.}}]{sato2008planetary}
Sato, B., Izumiura, H., Toyota, E., {et~al.} 2008, Publications of the Astronomical Society of Japan, 60, 539

\bibitem[{Sato {et~al.}(2007)Sato, Izumiura, Toyota, Kambe, Takeda, Masuda, Omiya, Murata, Itoh, Ando, {et~al.}}]{sato2007planetary}
Sato, B., Izumiura, H., Toyota, E., {et~al.} 2007, The Astrophysical Journal, 661, 527

\bibitem[{Sato {et~al.}(2005)Sato, Kambe, Takeda, Izumiura, Masuda, \& Ando}]{sato2005radial}
Sato, B., Kambe, E., Takeda, Y., {et~al.} 2005, Publications of the Astronomical Society of Japan, 57, 97

\bibitem[{{Shara} {et~al.}(2016){Shara}, {Hurley}, \& {Mardling}}]{shara2016}
{Shara}, M.~M., {Hurley}, J.~R., \& {Mardling}, R.~A. 2016, \apj, 816, 59

\bibitem[{Smith {et~al.}(2012)Smith, Stumpe, Van~Cleve, Jenkins, Barclay, Fanelli, Girouard, Kolodziejczak, McCauliff, Morris, {et~al.}}]{smith2012kepler}
Smith, J.~C., Stumpe, M.~C., Van~Cleve, J.~E., {et~al.} 2012, Publications of the Astronomical Society of the Pacific, 124, 1000

\bibitem[{Speagle(2020)}]{speagle2020dynesty}
Speagle, J.~S. 2020, Monthly Notices of the Royal Astronomical Society, 493, 3132

\bibitem[{Stumpe {et~al.}(2014)Stumpe, Smith, Catanzarite, Van~Cleve, Jenkins, Twicken, \& Girouard}]{stumpe2014multiscale}
Stumpe, M.~C., Smith, J.~C., Catanzarite, J.~H., {et~al.} 2014, Publications of the Astronomical Society of the Pacific, 126, 100

\bibitem[{Stumpe {et~al.}(2012)Stumpe, Smith, Van~Cleve, Twicken, Barclay, Fanelli, Girouard, Jenkins, Kolodziejczak, McCauliff, {et~al.}}]{stumpe2012kepler}
Stumpe, M.~C., Smith, J.~C., Van~Cleve, J.~E., {et~al.} 2012, Publications of the Astronomical Society of the Pacific, 124, 985

\bibitem[{Takarada {et~al.}(2020)Takarada, Sato, Omiya, Hori, \& Fujii}]{takarada2020radial}
Takarada, T., Sato, B., Omiya, M., Hori, Y., \& Fujii, M.~S. 2020, Publications of the Astronomical Society of Japan, 72, 104

\bibitem[{Temmink \& Snellen(2023)}]{temmink2023occurrence}
Temmink, M. \& Snellen, I.~A. 2023, Astronomy \& Astrophysics, 670, A26

\bibitem[{Trotta(2008)}]{trotta2008bayes}
Trotta, R. 2008, Contemporary Physics, 49, 71

\bibitem[{Udry {et~al.}(2003)Udry, Mayor, \& Santos}]{udry2003statistical}
Udry, S., Mayor, M., \& Santos, N. 2003, Astronomy \& Astrophysics, 407, 369

\bibitem[{Vanderburg {et~al.}(2018)Vanderburg, Mann, Rizzuto, Bieryla, Kraus, Berlind, Calkins, Curtis, Douglas, Esquerdo, {et~al.}}]{vanderburg2018zodiacal}
Vanderburg, A., Mann, A.~W., Rizzuto, A., {et~al.} 2018, The Astronomical Journal, 156, 46

\bibitem[{{Veras}(2016)}]{Veras2016}
{Veras}, D. 2016, Royal Society Open Science, 3, 150571

\bibitem[{Villaver \& Livio(2009)}]{villaver2009orbital}
Villaver, E. \& Livio, M. 2009, The Astrophysical Journal, 705, L81

\bibitem[{{Villaver} {et~al.}(2014){Villaver}, {Livio}, {Mustill}, \& {Siess}}]{Villaver2014}
{Villaver}, E., {Livio}, M., {Mustill}, A.~J., \& {Siess}, L. 2014, \apj, 794, 3

\bibitem[{{Vines} \& {Jenkins}(2022)}]{vines2022}
{Vines}, J.~I. \& {Jenkins}, J.~S. 2022, \mnras, 513, 2719

\bibitem[{Wang {et~al.}(2020)Wang, Leigh, Perna, \& Shara}]{wang2020hot}
Wang, Y.-H., Leigh, N.~W., Perna, R., \& Shara, M.~M. 2020, The Astrophysical Journal, 905, 136

\bibitem[{Wang {et~al.}(2022)Wang, Perna, Leigh, \& Shara}]{wang2022hot}
Wang, Y.-H., Perna, R., Leigh, N.~W., \& Shara, M.~M. 2022, Monthly Notices of the Royal Astronomical Society, 509, 5253

\bibitem[{Wolthoff {et~al.}(2022)Wolthoff, Reffert, Quirrenbach, Jones, Wittenmyer, \& Jenkins}]{wolthoff2022precise}
Wolthoff, V., Reffert, S., Quirrenbach, A., {et~al.} 2022, Astronomy \& Astrophysics, 661, A63

\bibitem[{Wright {et~al.}(2012)Wright, Marcy, Howard, Johnson, Morton, \& Fischer}]{wright2012frequency}
Wright, J., Marcy, G., Howard, A., {et~al.} 2012, The Astrophysical Journal, 753, 160

\bibitem[{Wu \& Lithwick(2011)}]{wu2011secular}
Wu, Y. \& Lithwick, Y. 2011, The Astrophysical Journal, 735, 109

\bibitem[{Yadav {et~al.}(2008)Yadav, Bedin, Piotto, Anderson, Cassisi, Villanova, Platais, Pasquini, Momany, \& Sagar}]{yadav2008ground}
Yadav, R., Bedin, L., Piotto, G., {et~al.} 2008, Astronomy \& Astrophysics, 484, 609

\bibitem[{{Zechmeister} {et~al.}(2018){Zechmeister}, {Reiners}, {Amado}, {Azzaro}, {Bauer}, {B{\'e}jar}, {Caballero}, {Guenther}, {Hagen}, {Jeffers}, {Kaminski}, {K{\"u}rster}, {Launhardt}, {Montes}, {Morales}, {Quirrenbach}, {Reffert}, {Ribas}, {Seifert}, {Tal-Or}, \& {Wolthoff}}]{zechmeister2018spectrum}
{Zechmeister}, M., {Reiners}, A., {Amado}, P.~J., {et~al.} 2018, \aap, 609, A12

\bibitem[{Zhou {et~al.}(2019)Zhou, Huang, Bakos, Hartman, Latham, Quinn, Collins, Winn, Wong, Kov{\'a}cs, {et~al.}}]{zhou2019two}
Zhou, G., Huang, C., Bakos, G., {et~al.} 2019, The Astronomical Journal, 158, 141

\bibitem[{Zink \& Howard(2023)}]{zink2023hot}
Zink, J.~K. \& Howard, A.~W. 2023, The Astrophysical Journal Letters, 956, L29

\end{thebibliography}
\onecolumn
\begin{appendix}
\section{Injection Recovery Plots}

\begin{minipage}{\textwidth}
    \centering
    \includegraphics[width=0.41\textwidth]{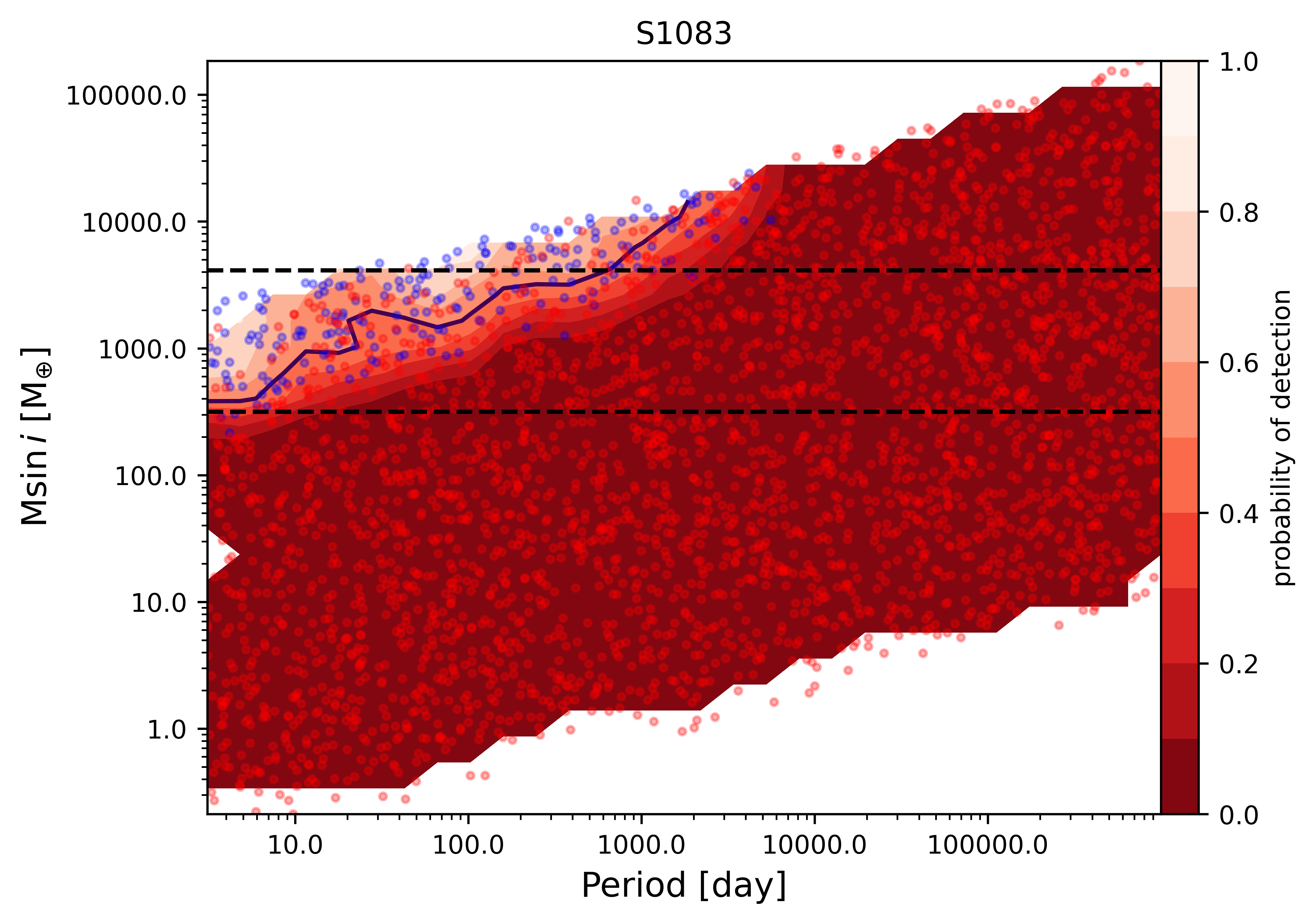}
    \includegraphics[width=0.41\textwidth]{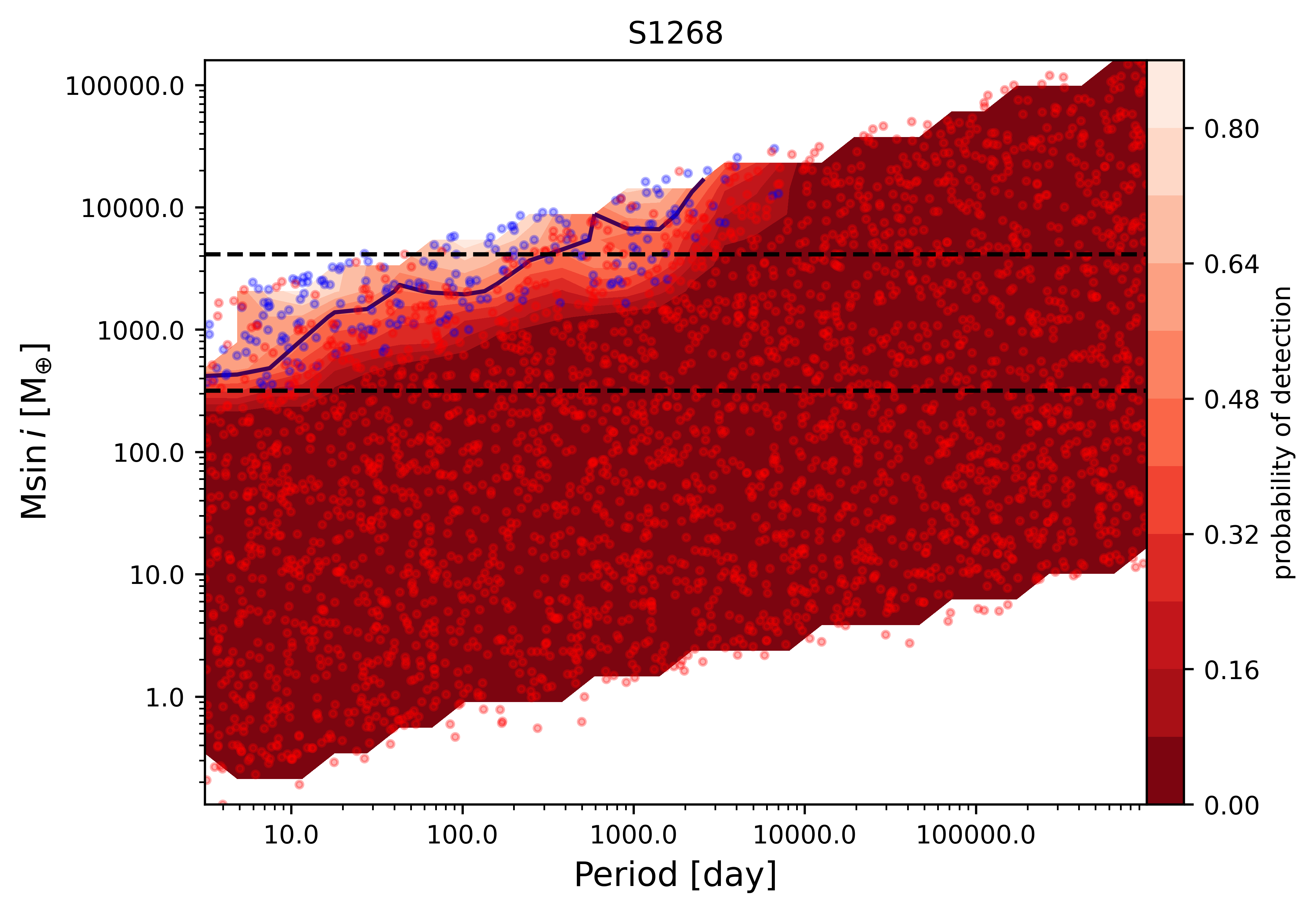}
    \includegraphics[width=0.41\textwidth]{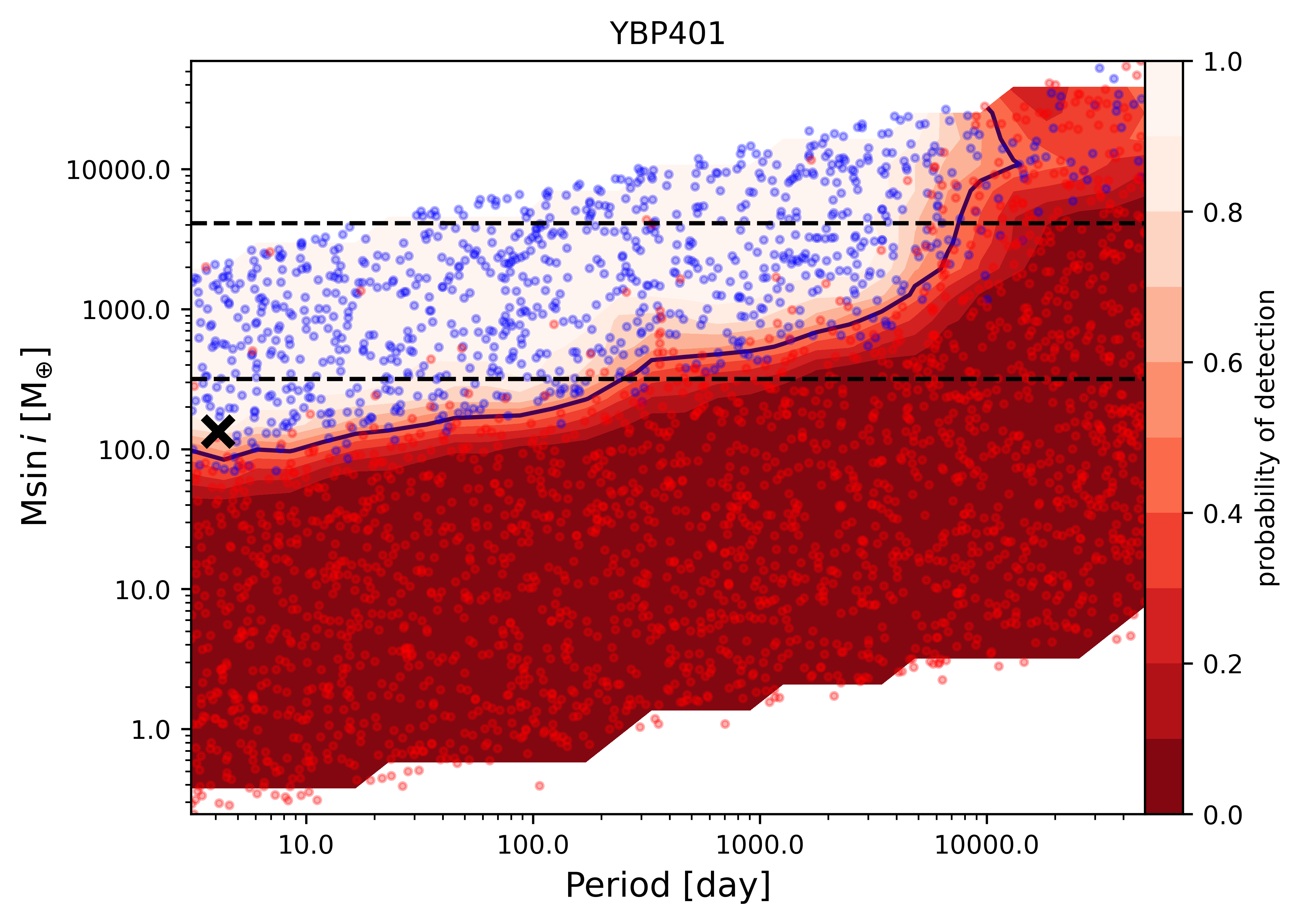}
    \includegraphics[width=0.41\textwidth]{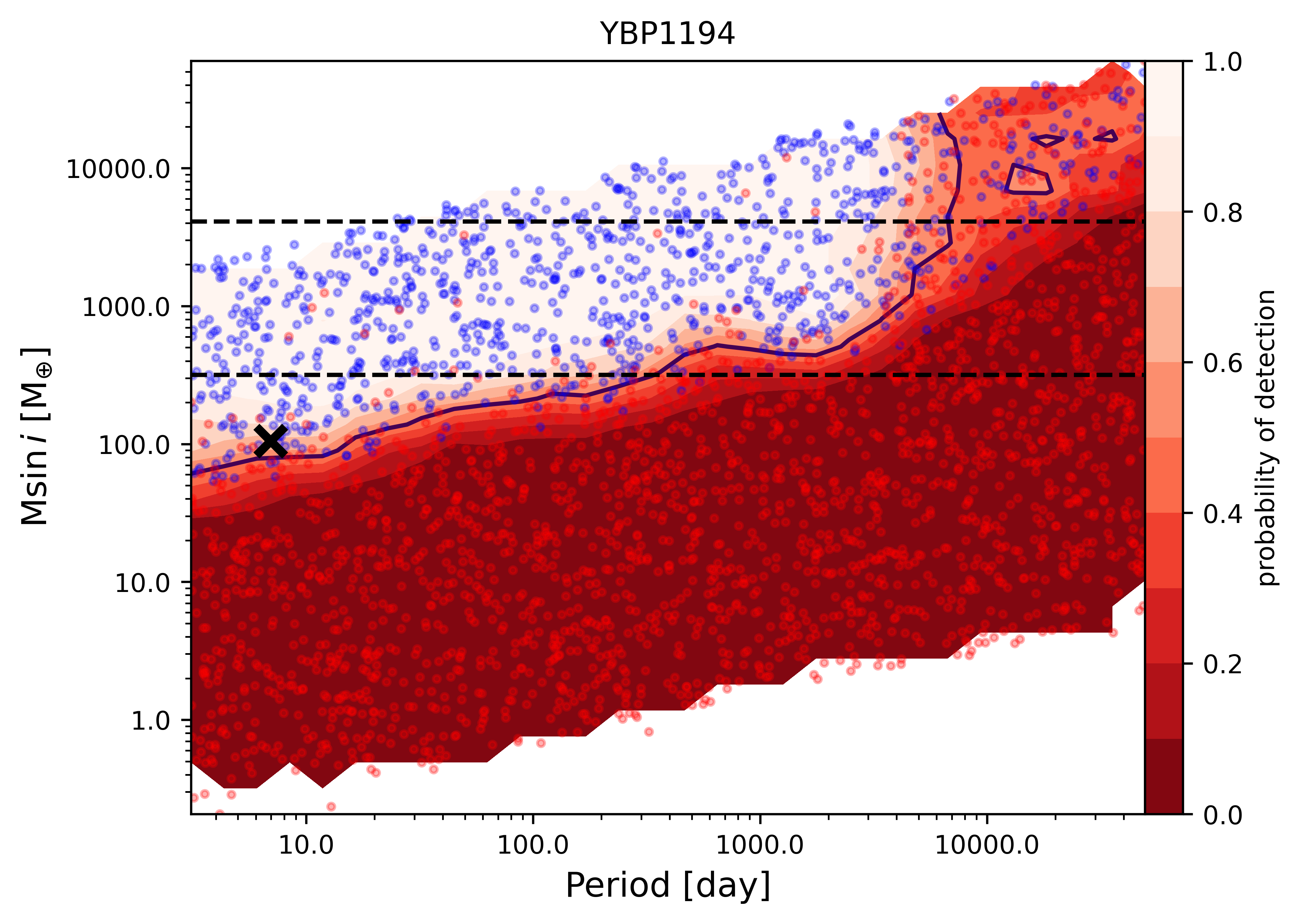}
    \includegraphics[width=0.41\textwidth]{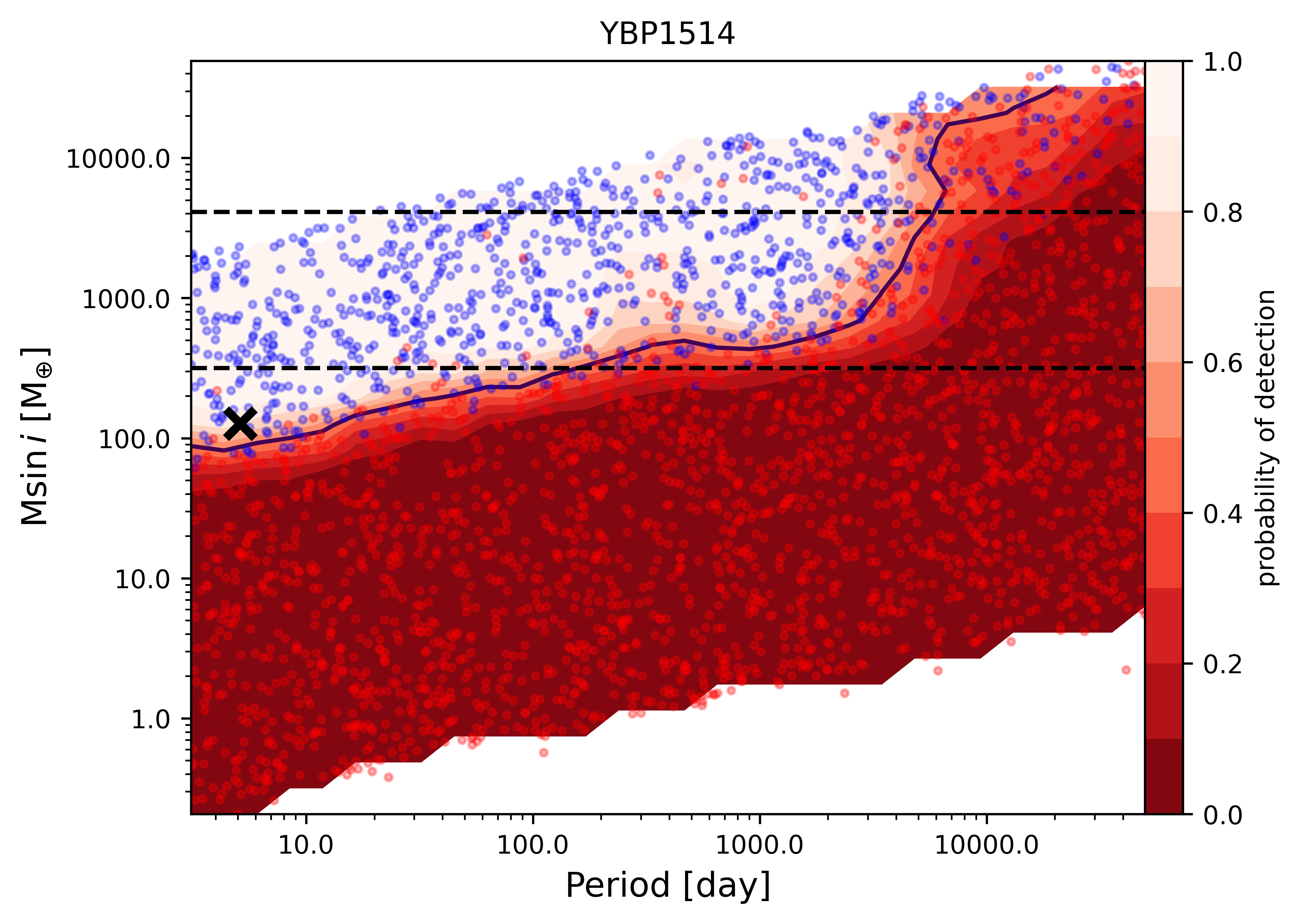}
    \includegraphics[width=0.41\textwidth]{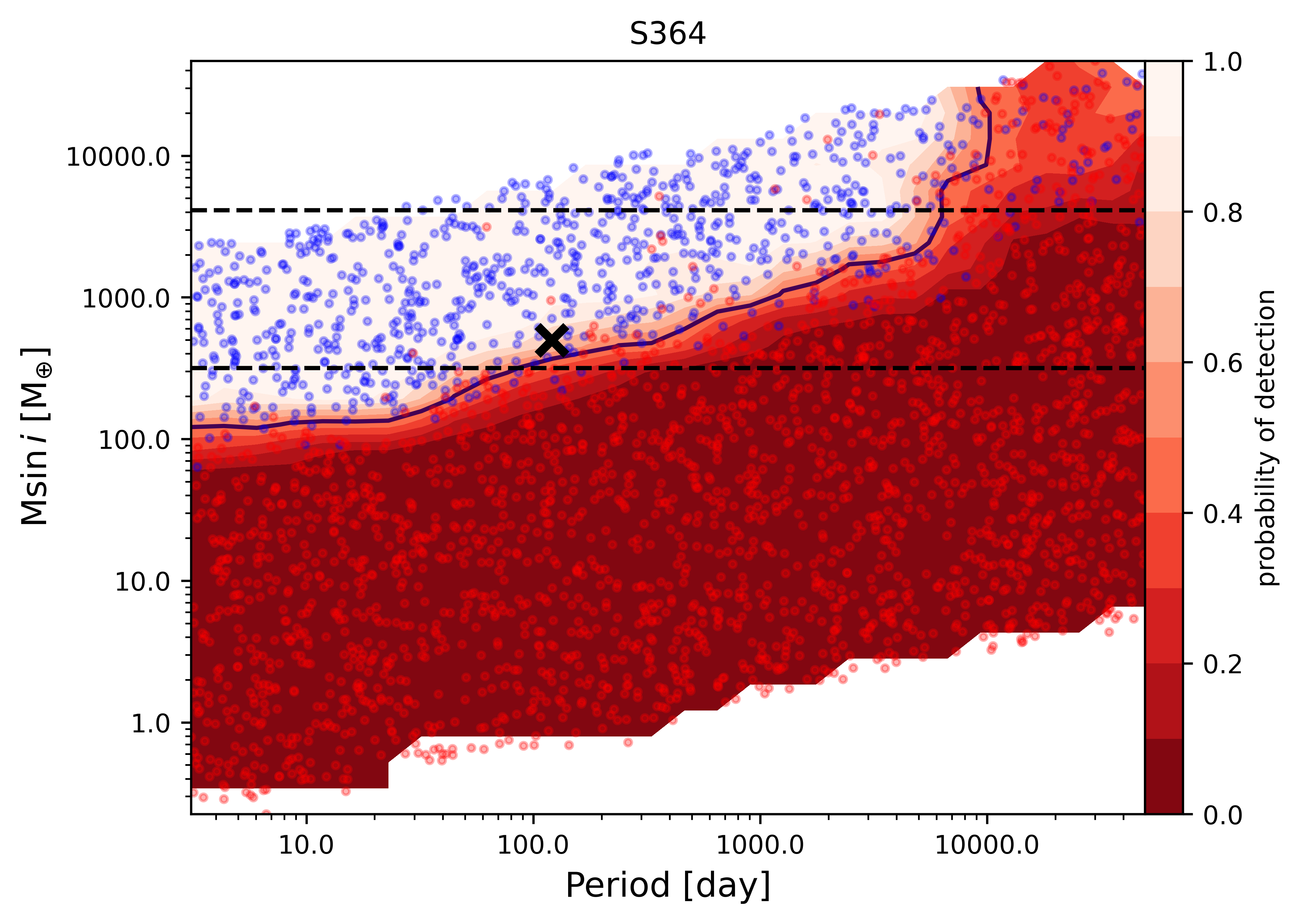}
    \includegraphics[width=0.41\textwidth]{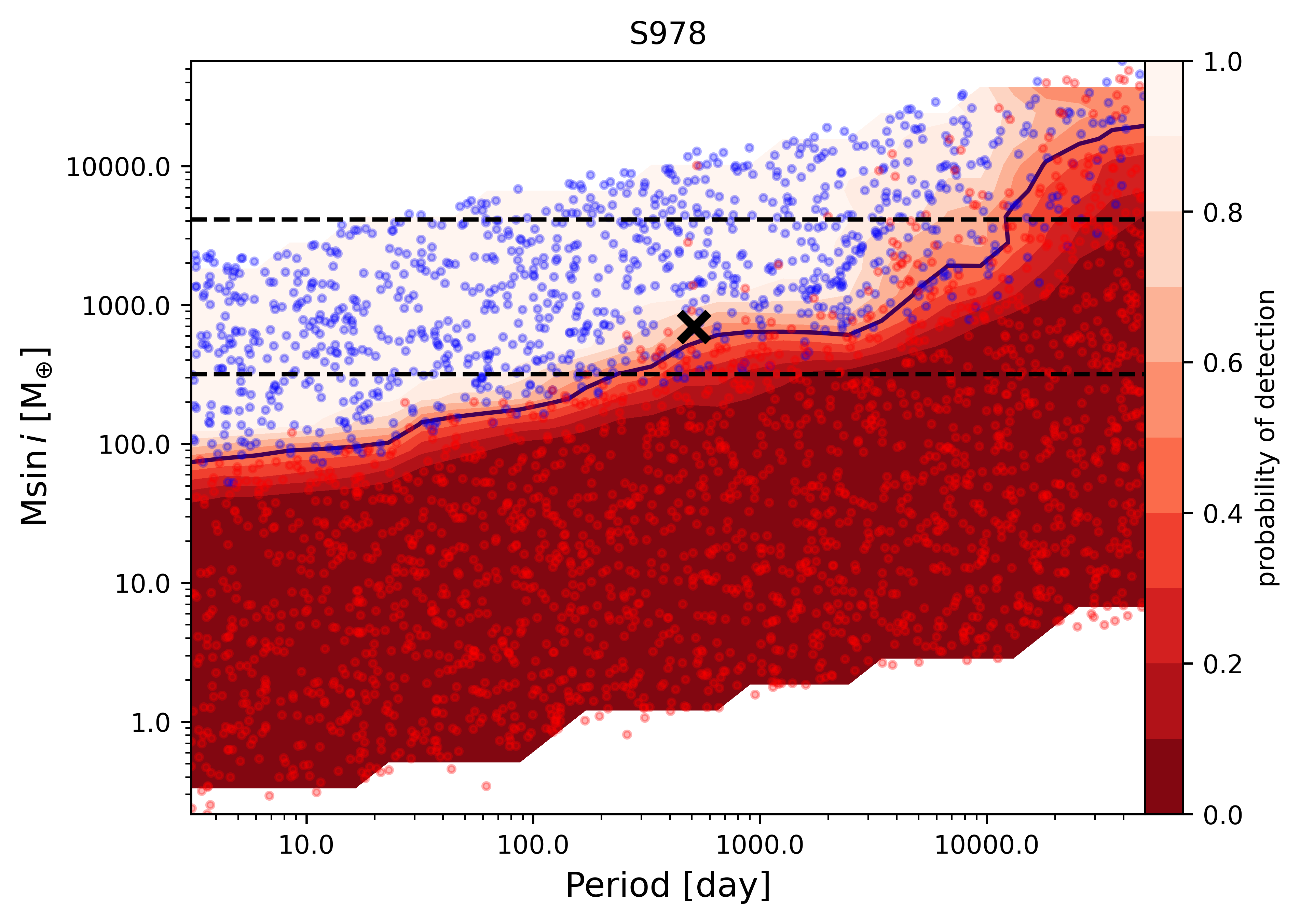}
    \includegraphics[width=0.41\textwidth]{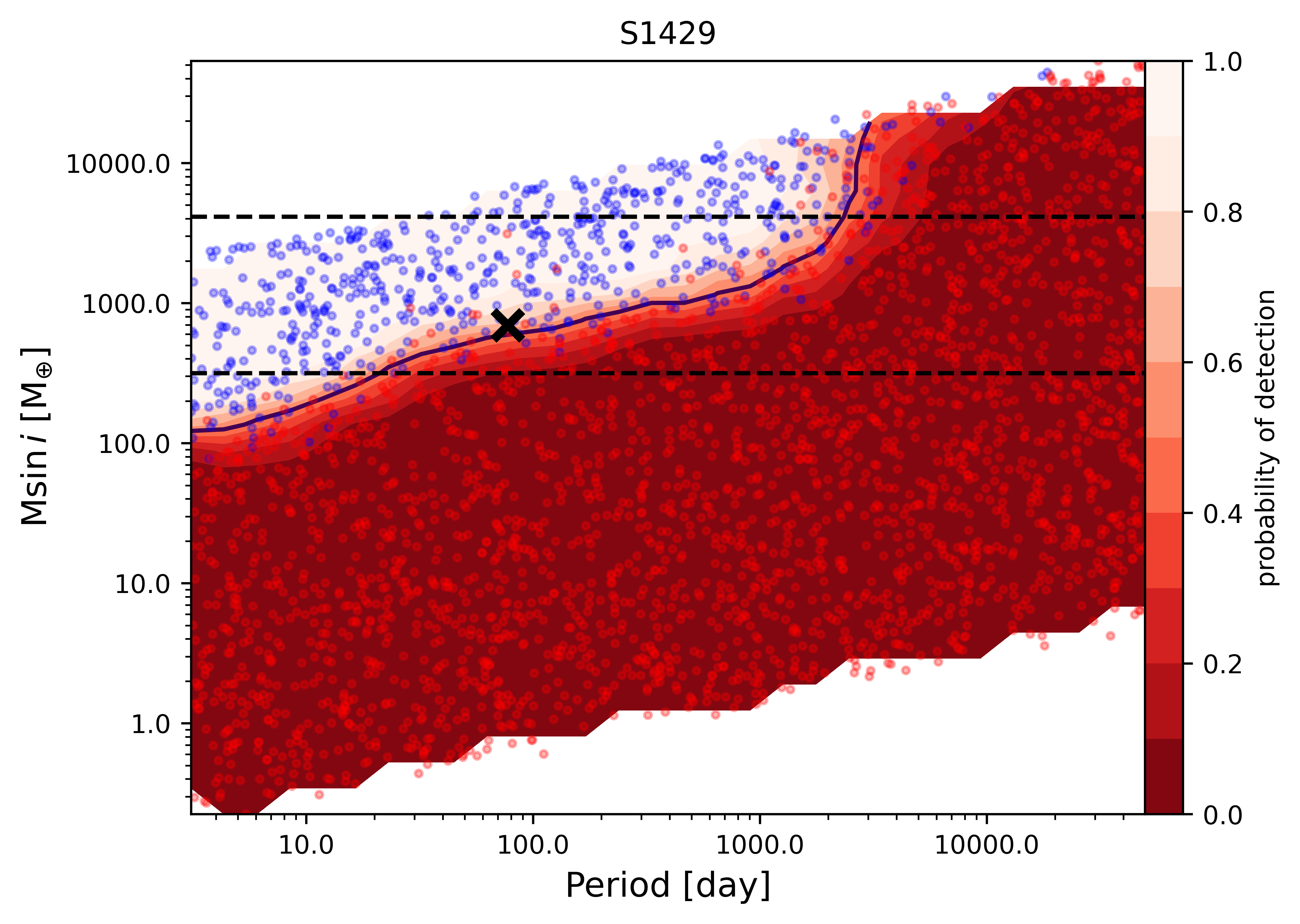}
    \captionof{figure}{Injection recovery tests for S1083 and S1268 (Top row) and all planet hosts in M67 (other rows). The blue datapoints are injected signals that could be recovered by RVSearch while the red datapoints were not recovered when injected into the RV data. The horizontal black dotted lines are at 1 Jupiter mass and at the brown dwarf limit ($13 \ \text{M}_J$). For detected planets, the location is indicated by the black cross.}
    \label{fig:recov}
\end{minipage}
\FloatBarrier
\section{Periodogram and RV plots} \label{appb}
\begin{minipage}{\textwidth}
    \centering
    \includegraphics[width=0.48\textwidth]{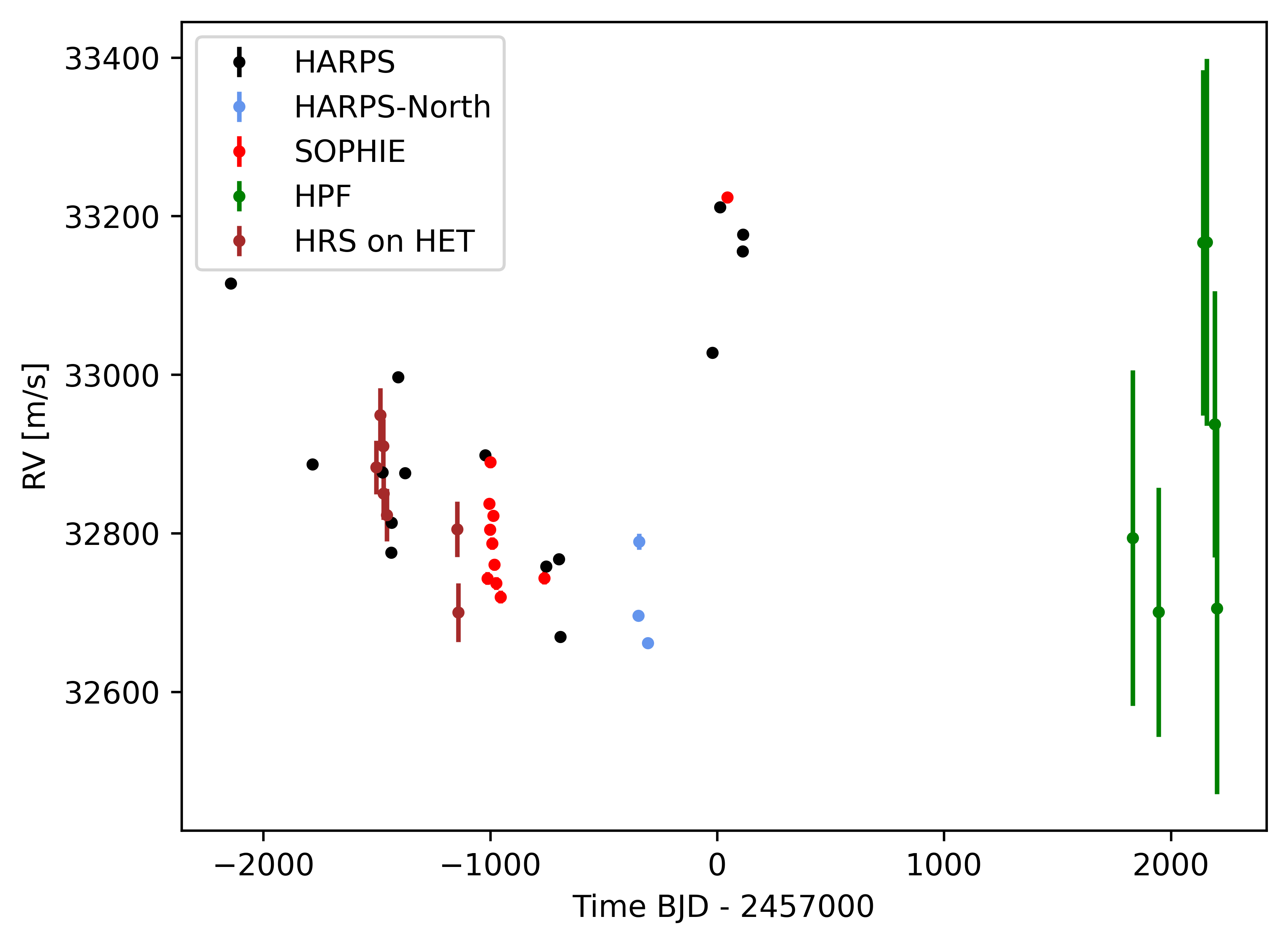} 
    \includegraphics[width=0.48\textwidth]{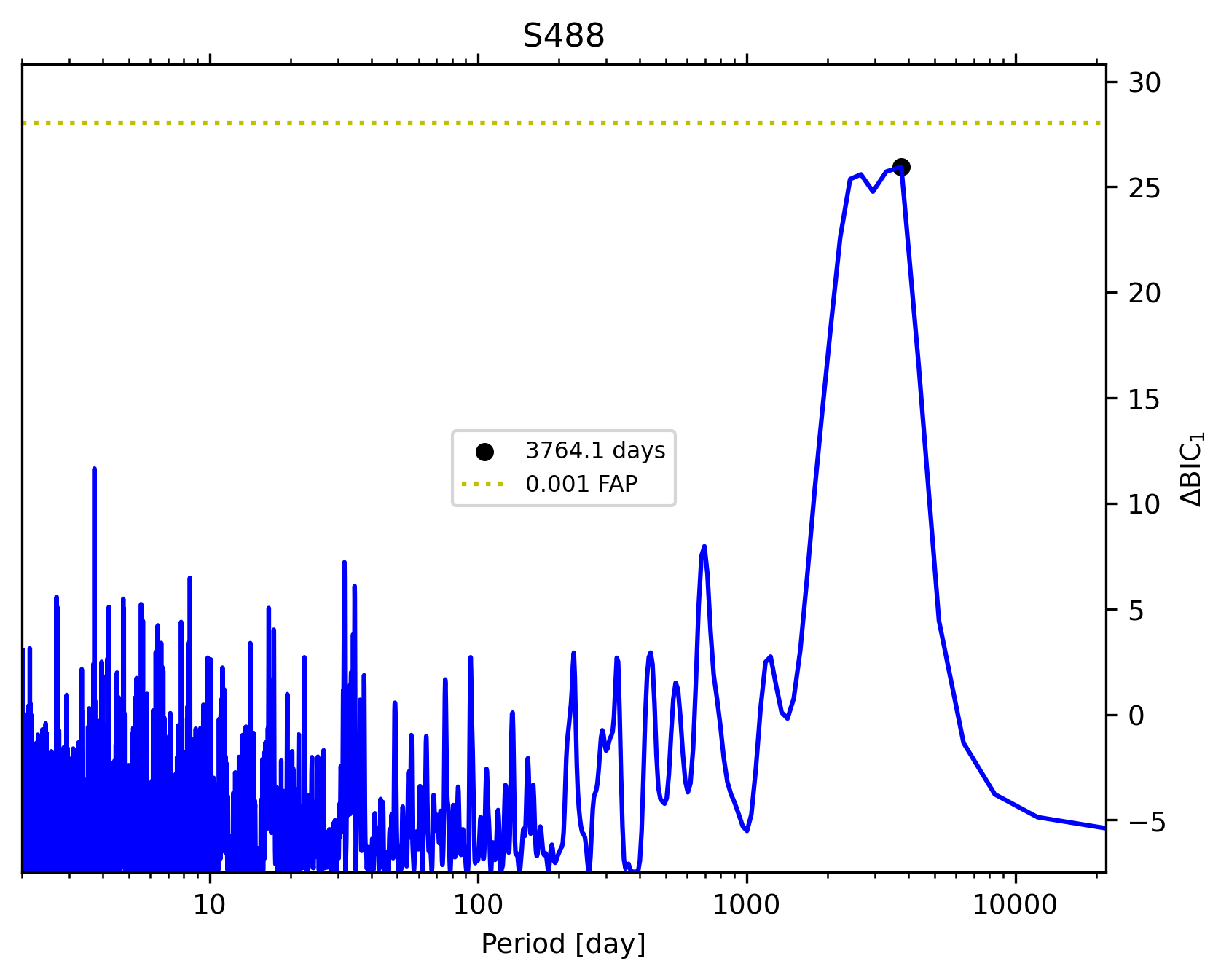} 
\captionof{figure}{Left: Time-series of the radial velocities for S488. All RV values are corrected to the zero-point of HARPS as described in the paper. Right: Result of the periodogram analysis with RVSearch.}
    \label{fig:488}
    \includegraphics[width=0.48\textwidth]{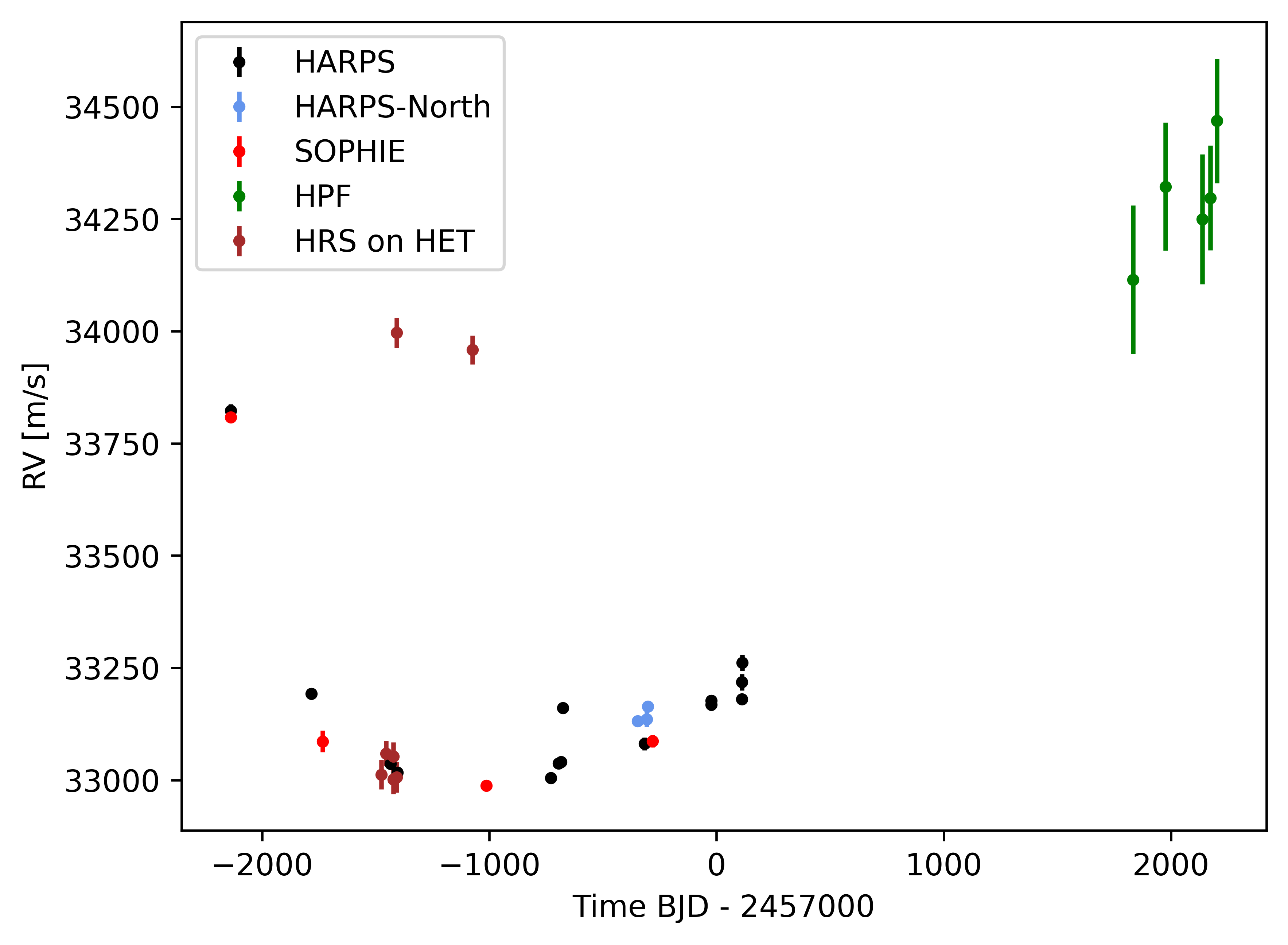}
    \includegraphics[width=0.48\textwidth]{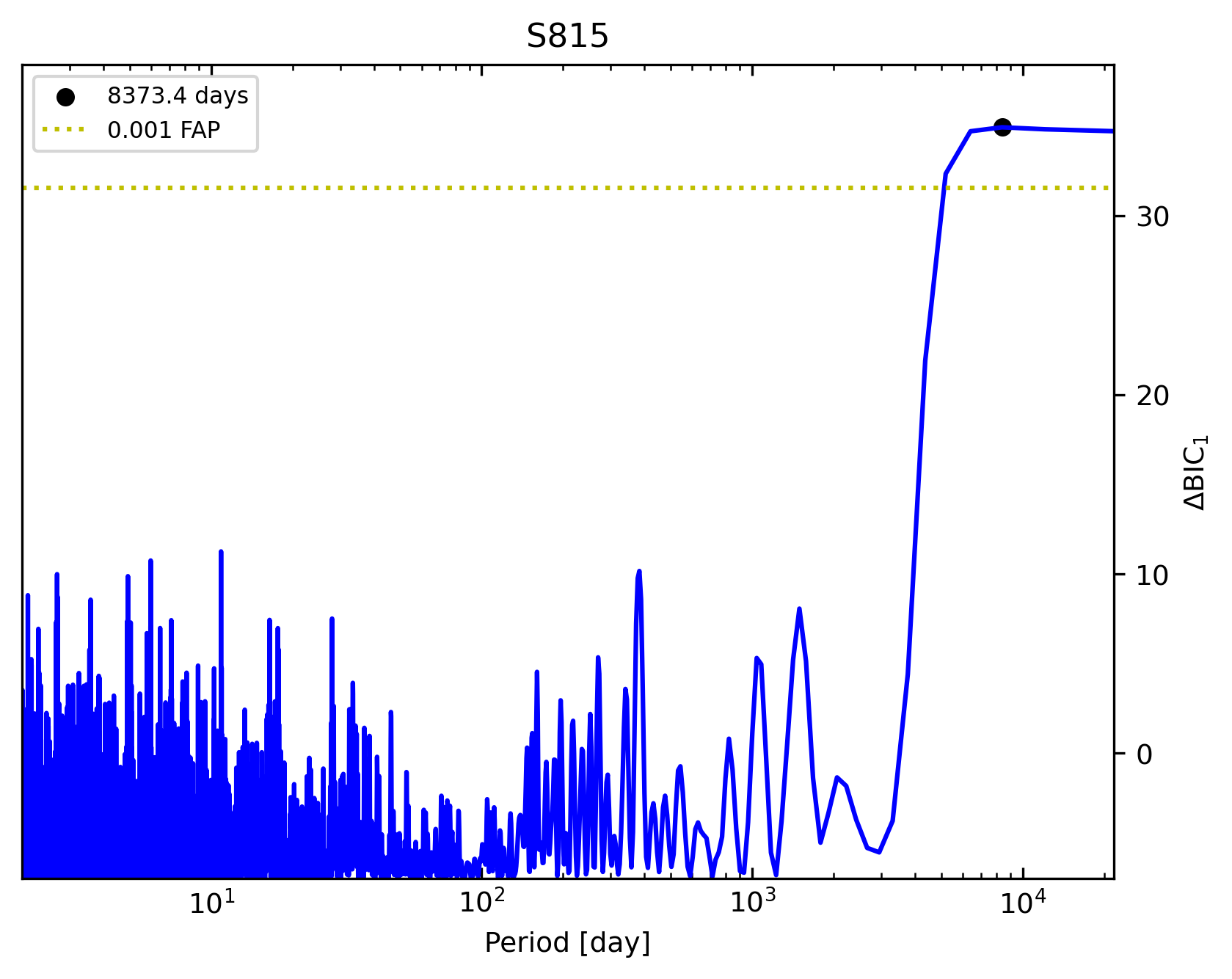} 
    \captionof{figure}{Same as Fig. \ref{fig:488} but for S815.}
    \label{fig:815}
    \includegraphics[width=0.48\textwidth]{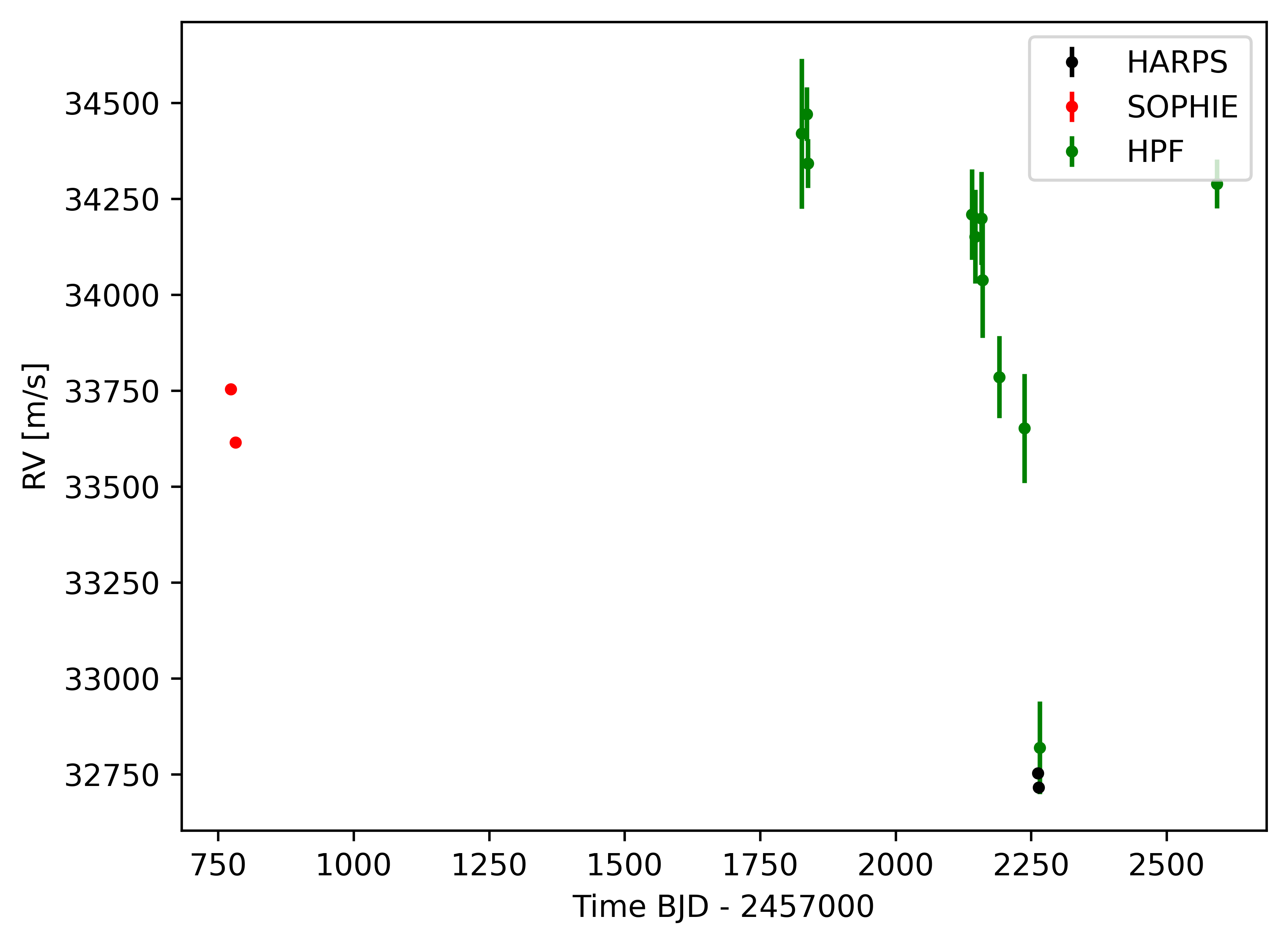} 
    \includegraphics[width=0.48\textwidth]{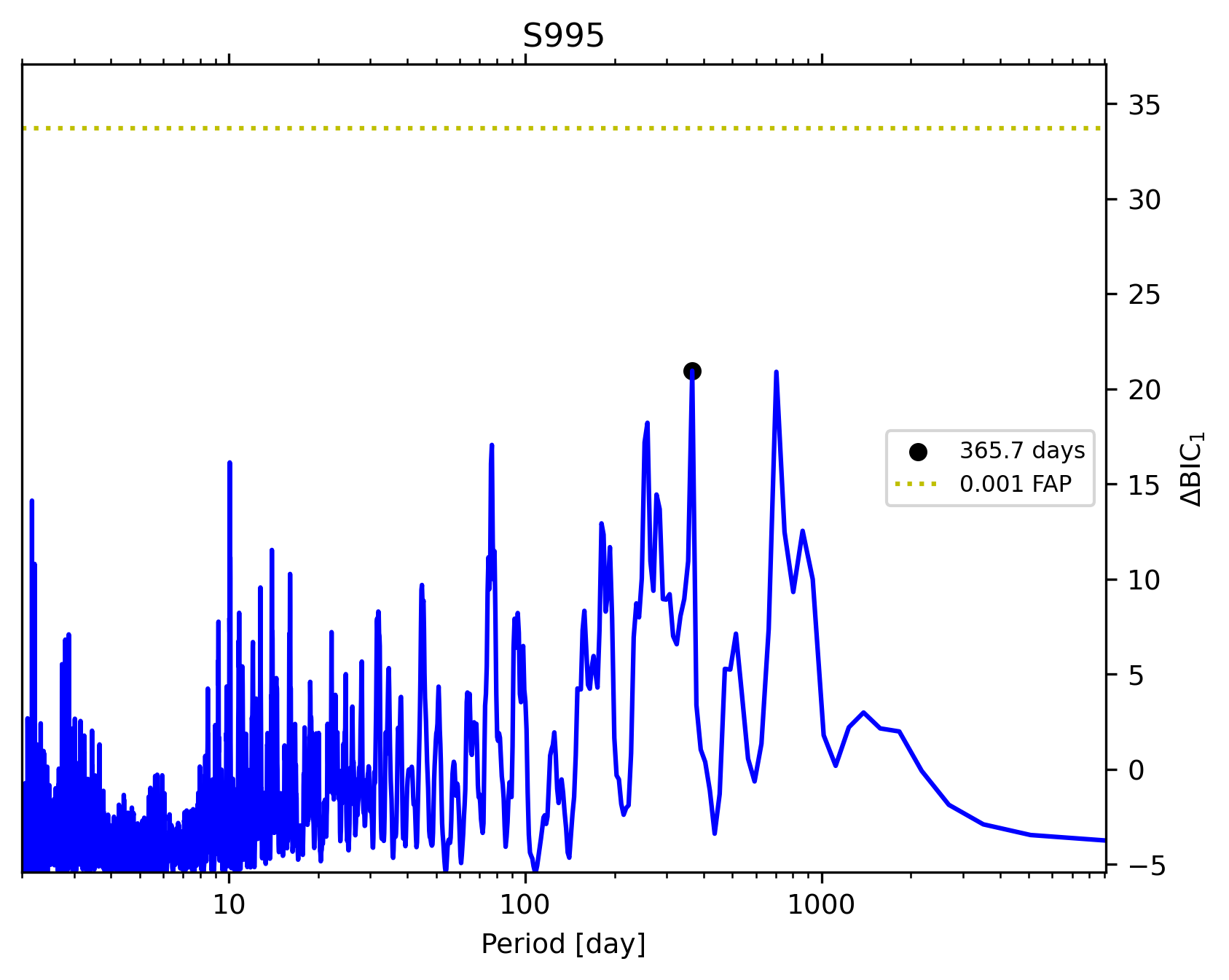} 
    \captionof{figure}{Same as Fig. \ref{fig:488} but for S995.}
    \label{fig:995}
\end{minipage}

\begin{figure}
    \centering
    \includegraphics[width=0.48\textwidth]{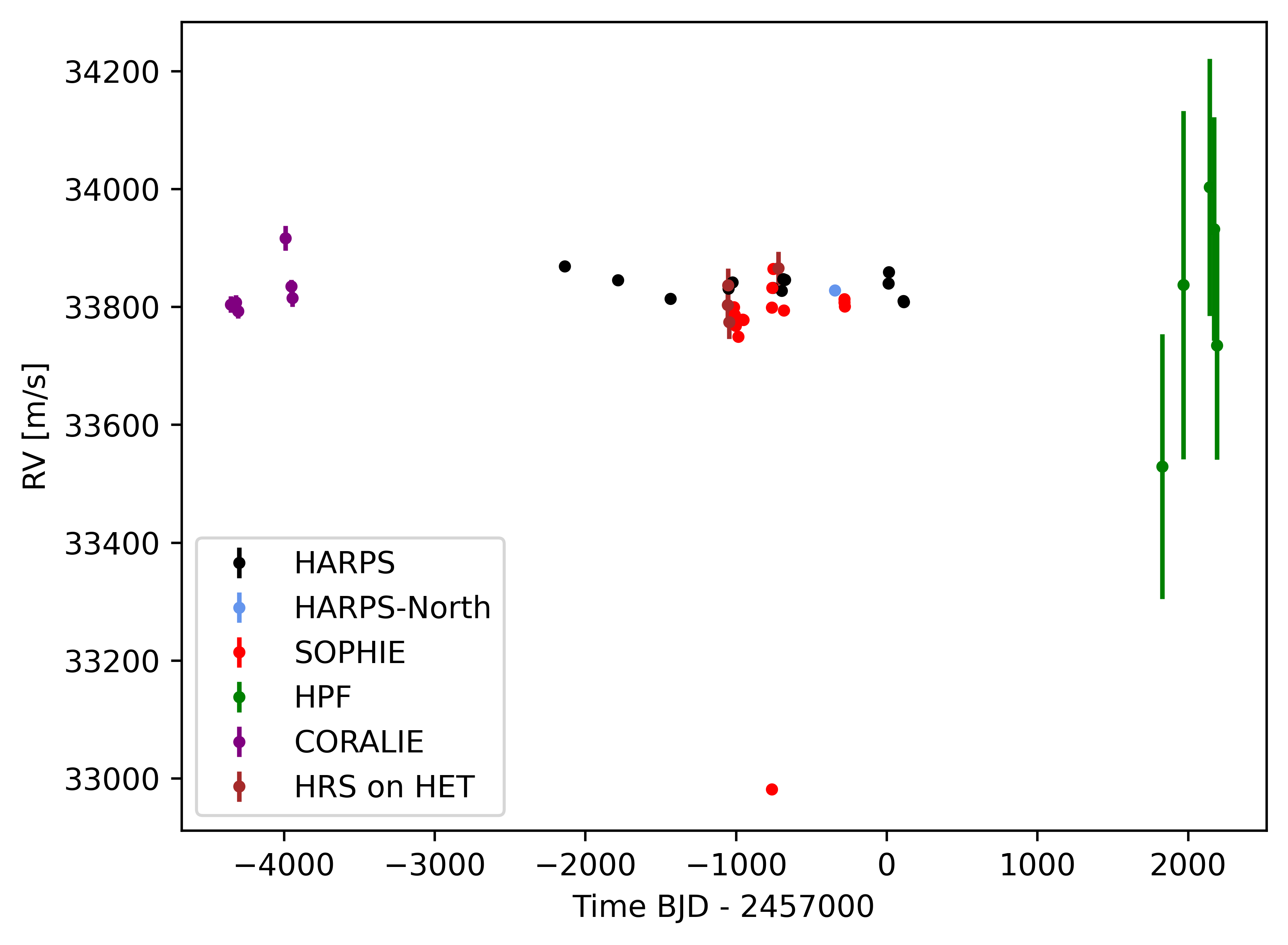} 
    \includegraphics[width=0.48\textwidth]{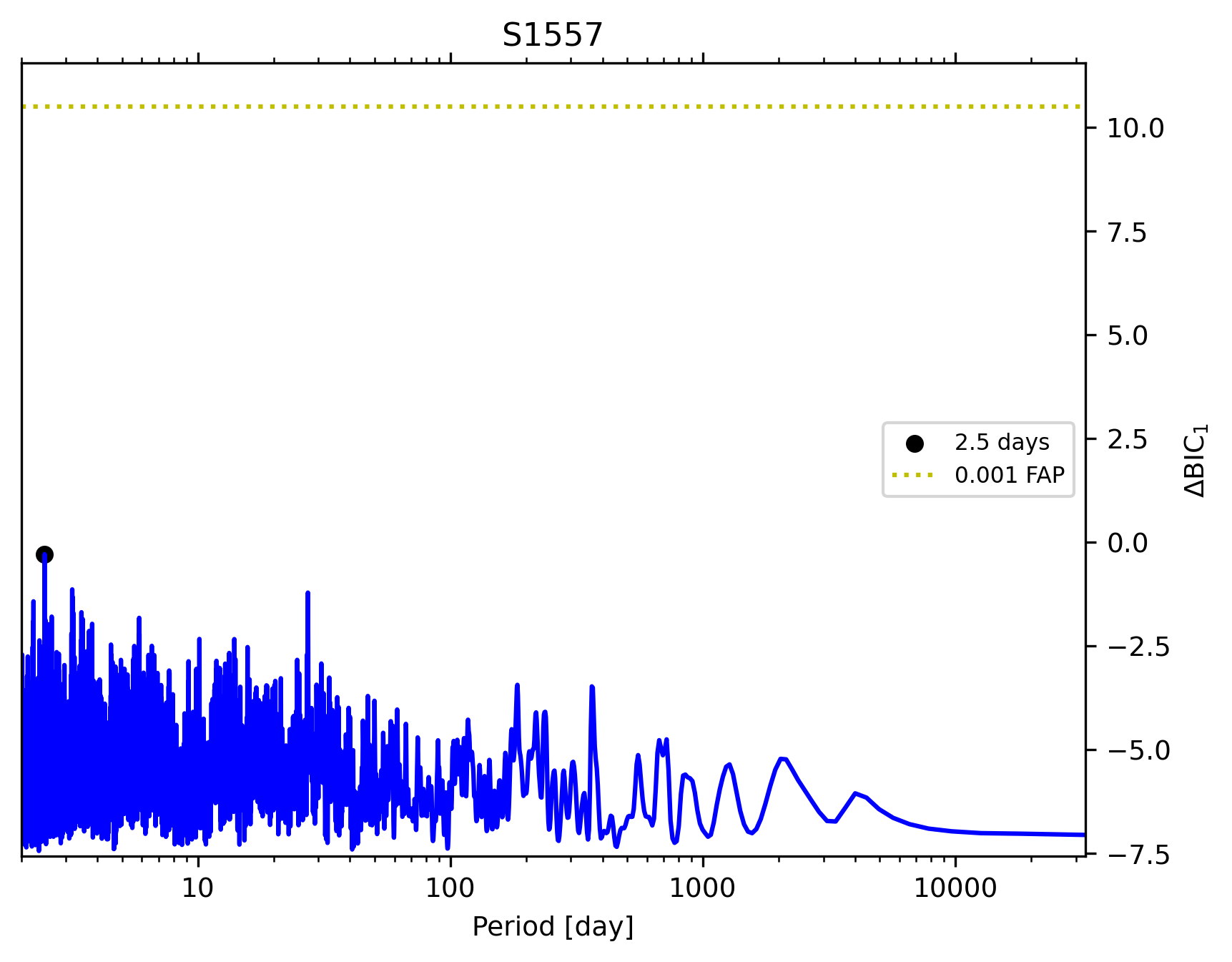} 
    \caption{Same as Fig. \ref{fig:488} but for S1557.}
    \label{fig:1557}
\end{figure}

\begin{figure}
    \centering
    \includegraphics[width=0.48\textwidth]{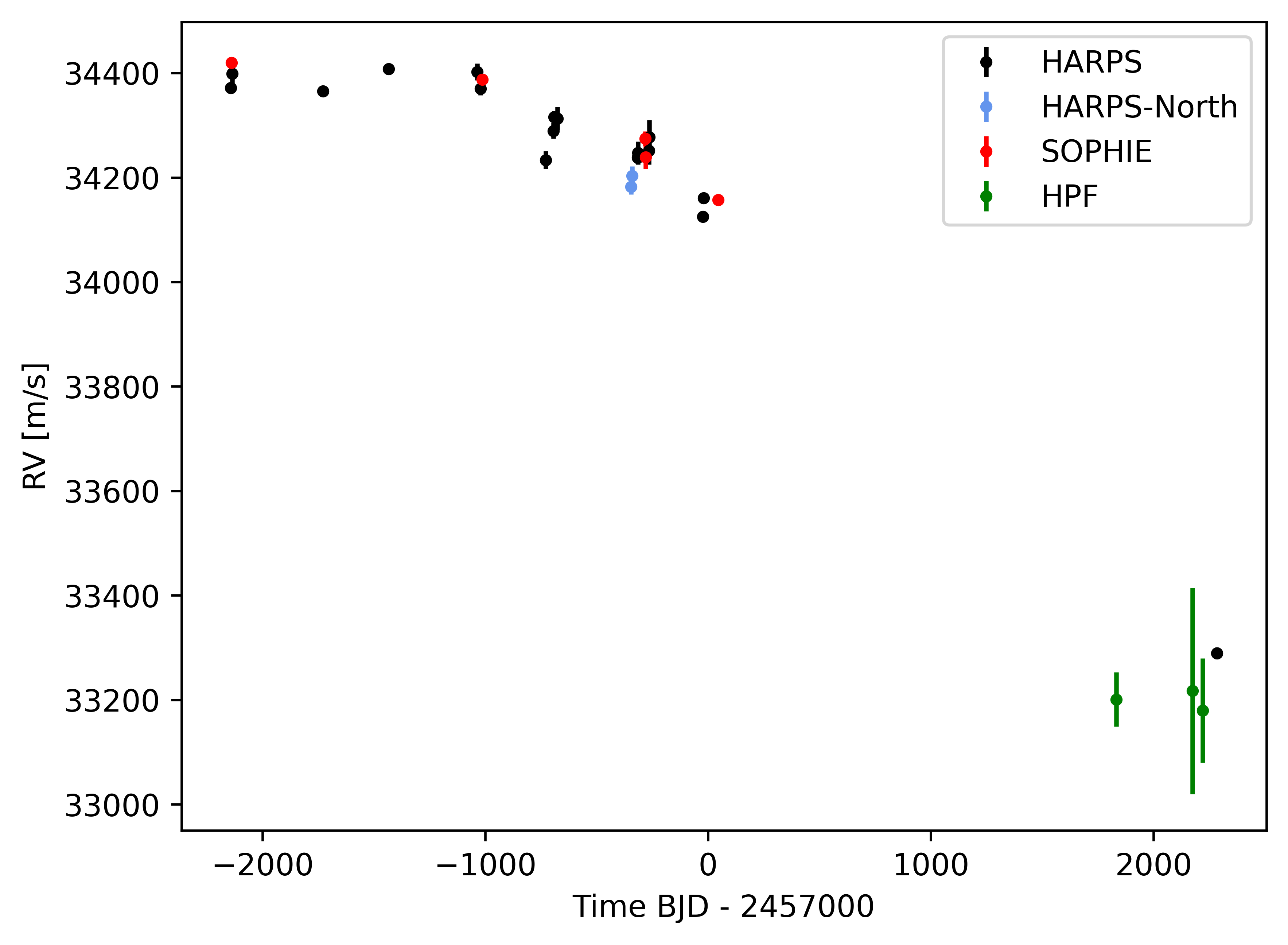} 
    \includegraphics[width=0.48\textwidth]{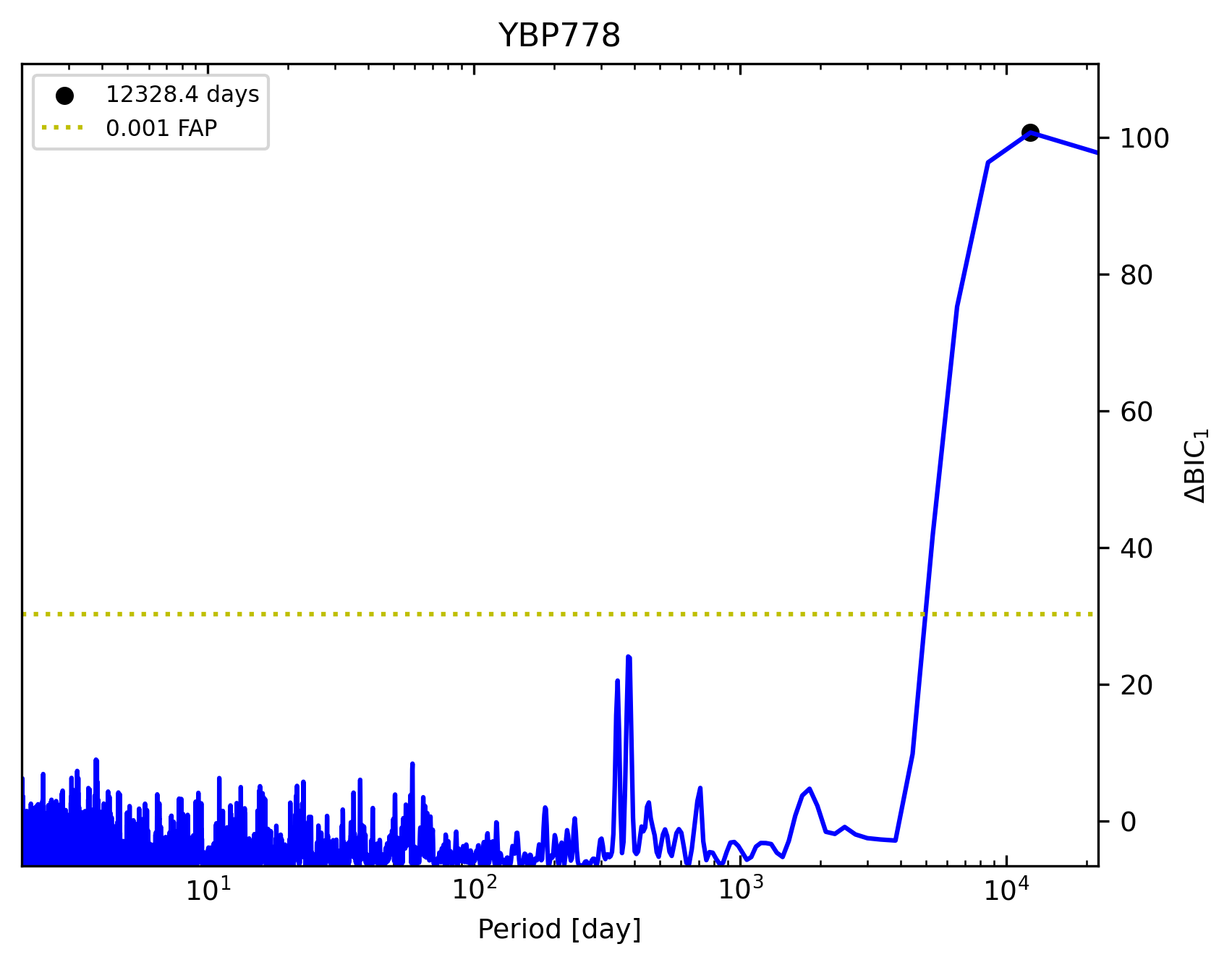}
    \caption{Same as Fig. \ref{fig:488} but for YBP778.}
    \label{fig:778}
\end{figure}

\begin{figure}
    \centering
    \includegraphics[width=0.48\textwidth]{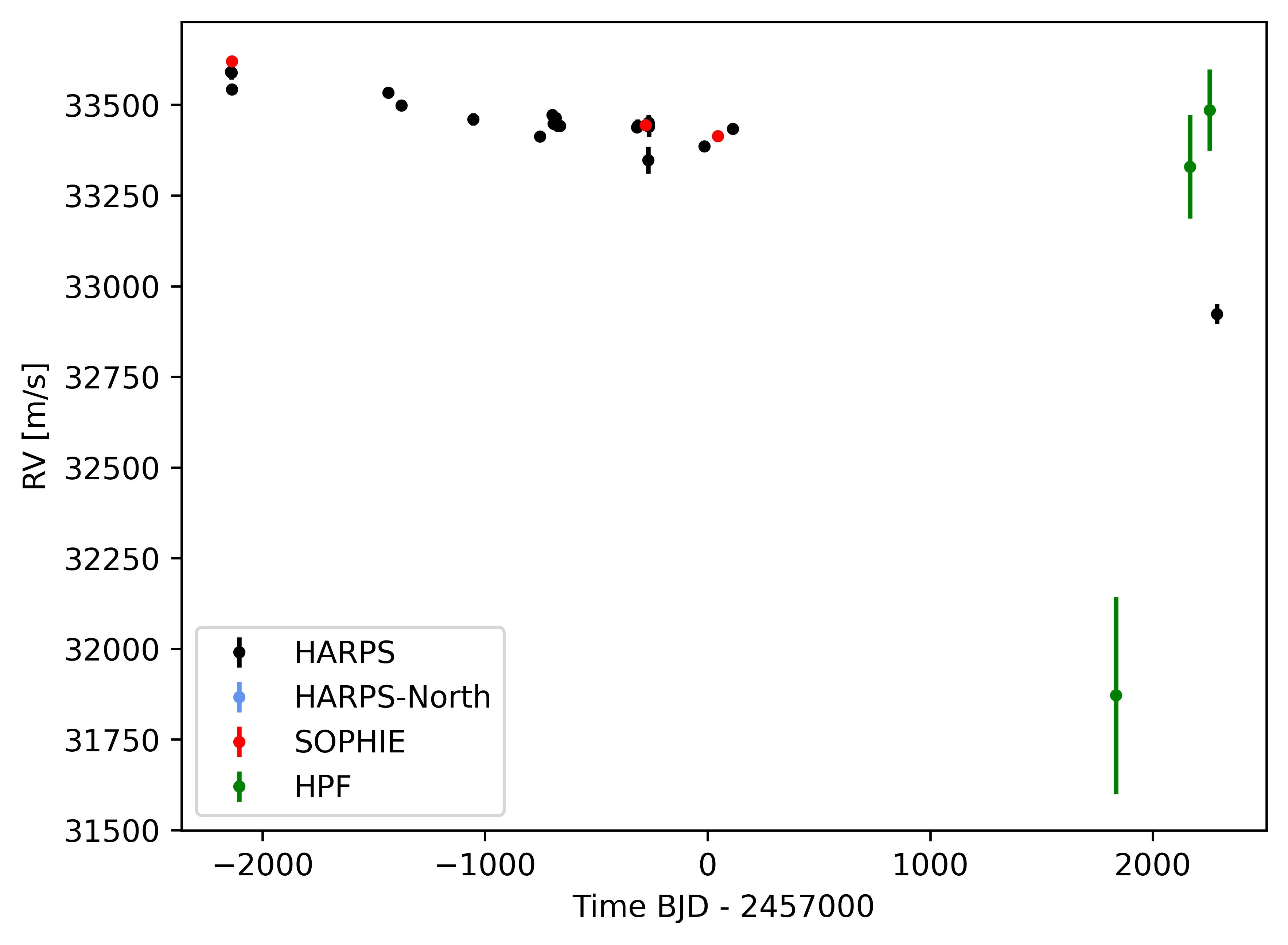} 
    \includegraphics[width=0.48\textwidth]{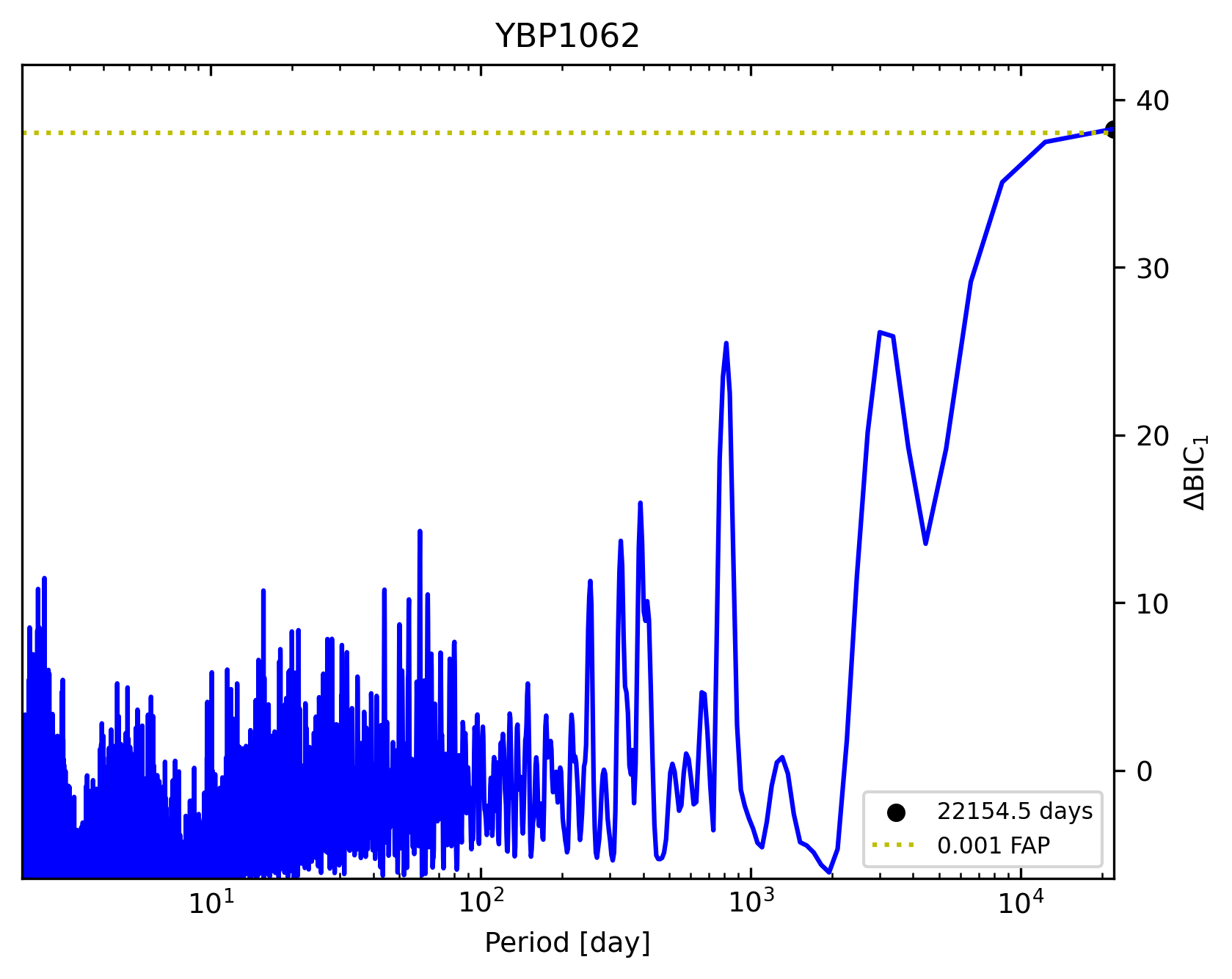}
    \caption{Same as Fig. \ref{fig:488} but for YBP1062.}
    \label{fig:1062}
\end{figure}

\begin{figure}
    \centering
    \includegraphics[width=0.48\textwidth]{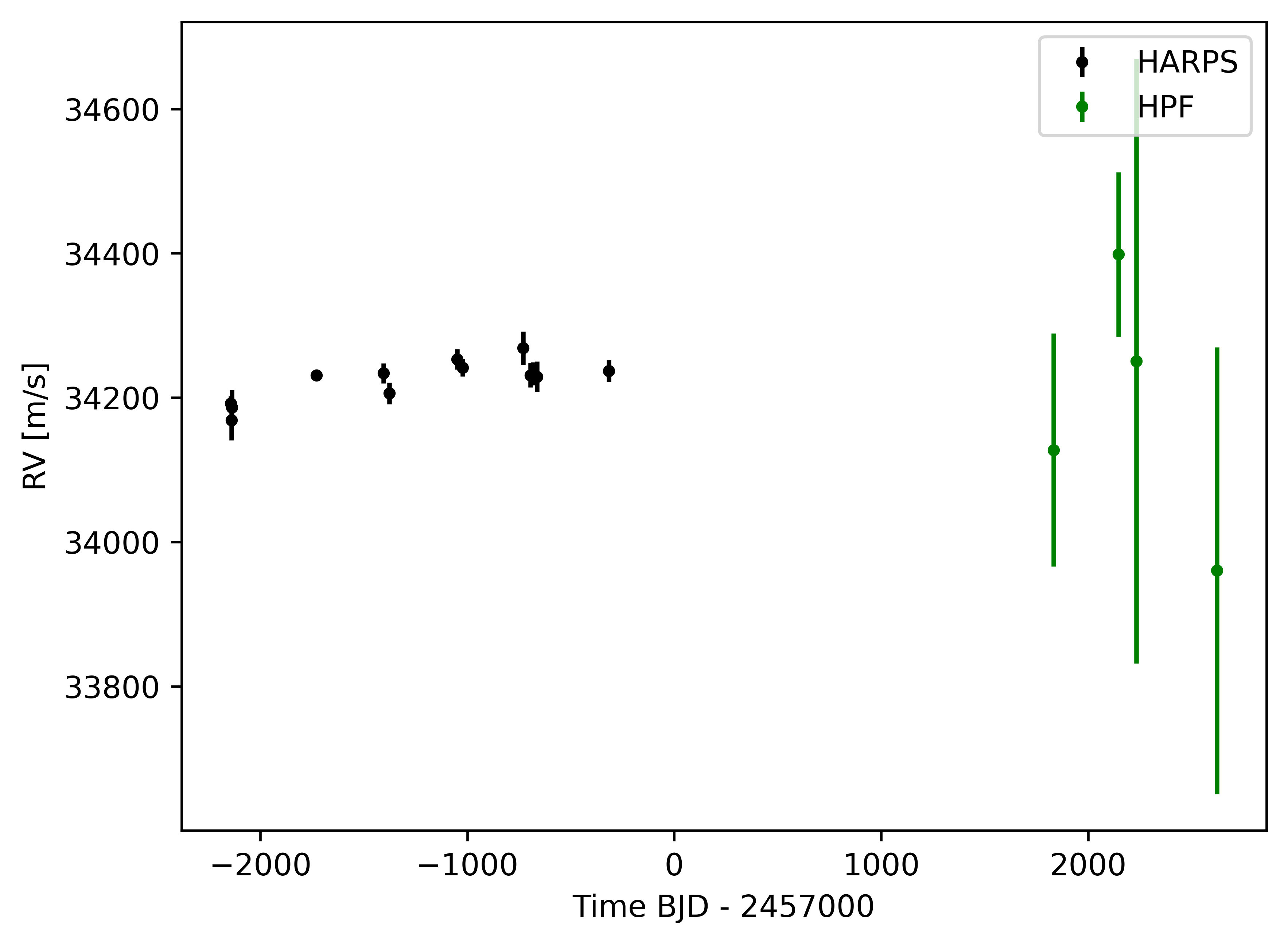}
    \includegraphics[width=0.48\textwidth]{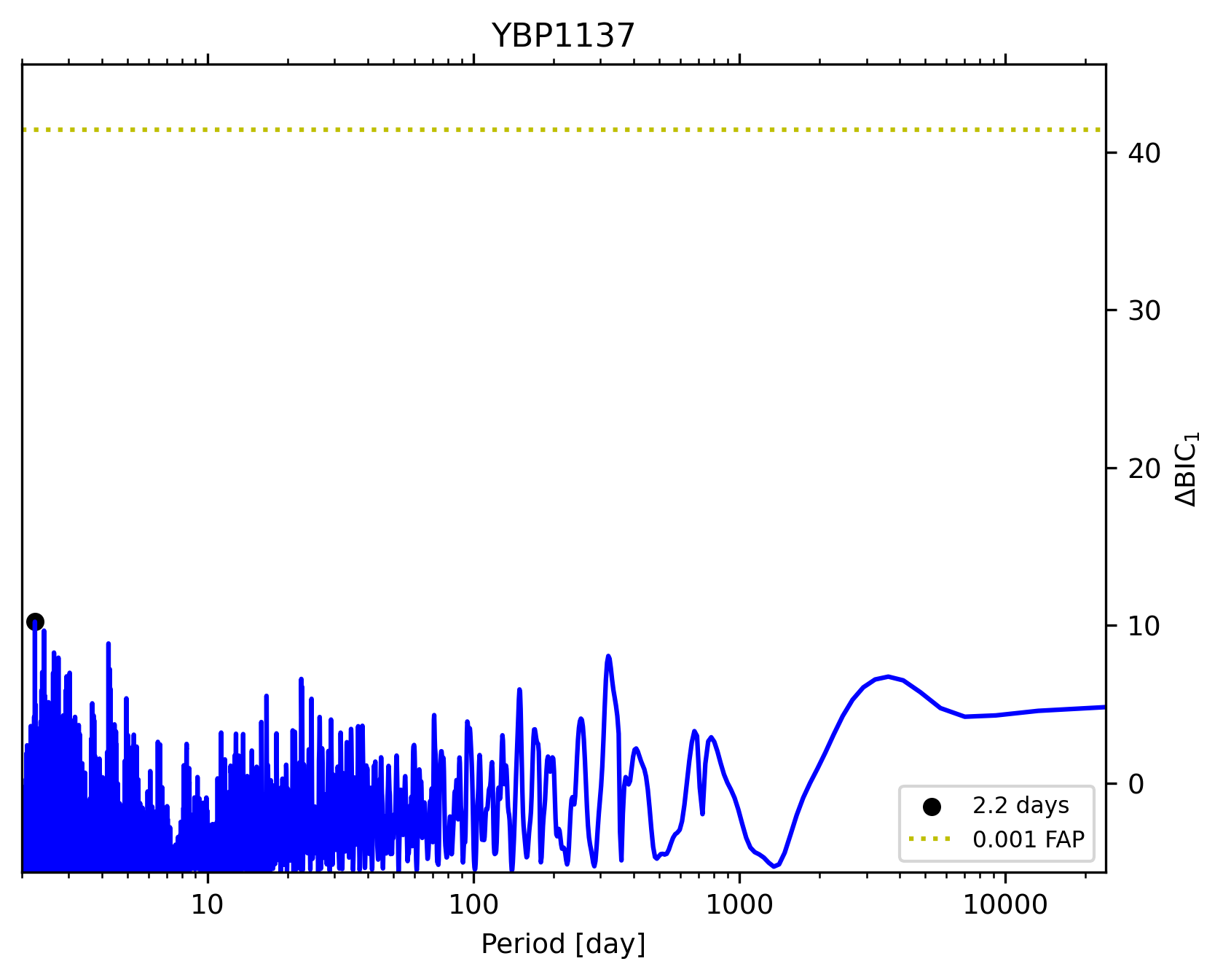}
    \caption{Same as Fig. \ref{fig:488} but for YBP1137.}
    \label{fig:1137}
\end{figure}

\FloatBarrier

\section{Juliet Corner Plots} \label{corn}
\begin{minipage}{\textwidth}
    \centering
    \includegraphics[width=0.8\textwidth]{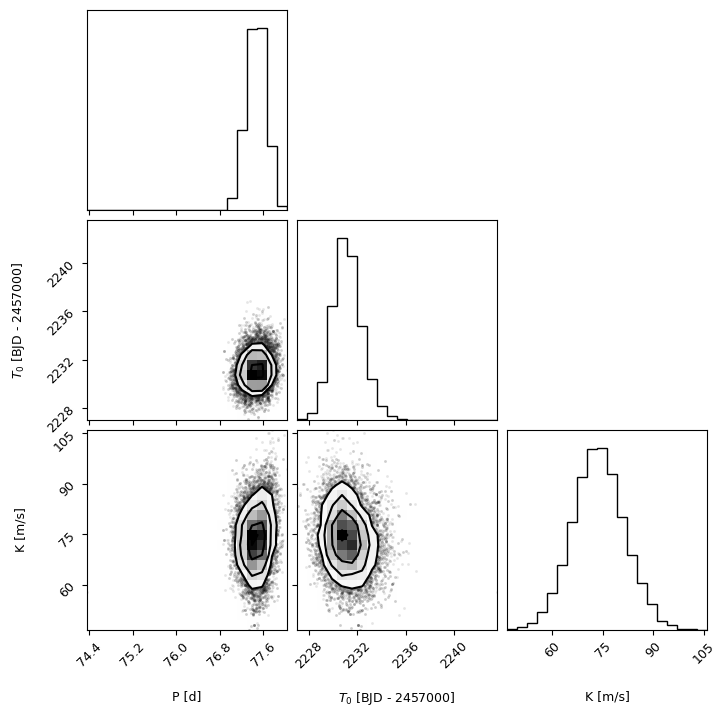}
    \captionof{figure}{Corner plot of the Juliet fit with fixed eccentricity and argument of periastron.}
    \label{fig:corn1}
\end{minipage}
\begin{figure}
    \centering
    \includegraphics[width=1.0\textwidth]{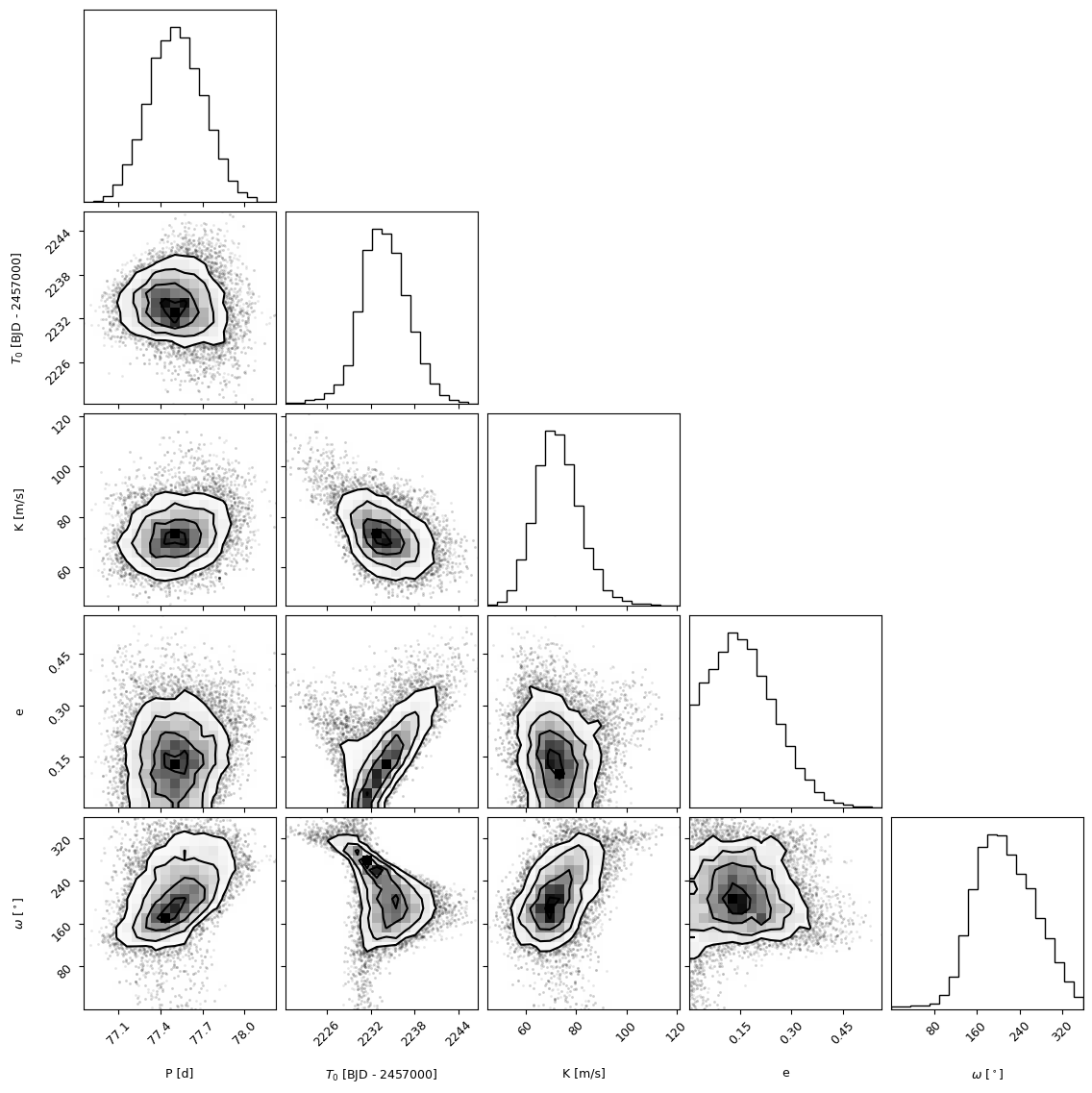}
    \caption{Corner plot of the Juliet fit where eccentricity and $\omega$ are allowed to vary freely.}
    \label{fig:corn2}
\end{figure}

\FloatBarrier

\section{Radial Velocity Tables} \label{tables}
\FloatBarrier
\begin{table*}
    \centering
    \caption{Radial Velocities for S1429. All radial velocities are corrected to the zero-point of HARPS using the offset derived with the Juliet fitting.}
    \begin{tabular}{l|l|l|c}
    \hline
      Time (BJD)   & RV (m/s) & RVerr (m/s) & Instrument \\
      \hline
      \hline
       2458828.867117 & 34005.8 & 36.4 & HPF\\
2458830.856013 & 33994.3 & 32.3 & HPF\\
2458859.776506 & 33942.9 & 46.1 & HPF\\
2458945.699714 & 33861.9 & 37.8 & HPF\\
2459147.985632 & 34010.8 & 29.4 & HPF\\
2459159.955035 & 33971.8 & 39.3 & HPF\\
2459161.941308 & 33958.0 & 35.9 & HPF\\
2459162.938455 & 33909.7 & 89.8 & HPF\\
2459172.92174 & 33990.5 & 43.3 & HPF\\
2459195.844298 & 33995.2 & 45.9 & HPF\\
2459213.816785 & 34026.6 & 37.4 & HPF\\
2459245.714999 & 33821.7 & 76.3 & HPF\\
2459264.657625 & 33936.6 & 39.5 & HPF\\
2459274.794775 & 33859.8 & 79.8 & HPF\\
2459299.740938 & 34030.7 & 33.8 & HPF\\
2459300.73939 & 34064.7 & 32.0 & HPF\\
2459306.717727 & 34017.7 & 33.8 & HPF\\
2459320.682886 & 33957.8 & 26.2 & HPF\\
2459329.651372 & 33938.3 & 33.5 & HPF\\
2459329.662351 & 33893.8 & 37.8 & HPF\\
2459503.006508 & 34030.0 & 19.1 & HPF\\
2459507.000525 & 33979.4 & 29.7 & HPF\\
2459518.952611 & 34175.0 & 32.6 & HPF\\
2459540.906436 & 33884.3 & 60.2 & HPF\\
2459548.882695 & 33989.6 & 30.5 & HPF\\
2459560.012561 & 33794.9 & 40.9 & HPF\\
2459561.843822 & 33924.7 & 36.5 & HPF\\
2459568.837382 & 33910.9 & 45.4 & HPF\\
2459571.821832 & 33974.5 & 45.6 & HPF\\
2459578.980509 & 33960.0 & 18.4 & HPF\\
2459616.708679 & 34029.7 & 35.6 & HPF\\
2459620.685981 & 34001.9 & 34.0 & HPF\\
2459620.863773 & 34024.5 & 31.8 & HPF\\
2459621.695099 & 33991.3 & 18.5 & HPF\\
2459622.692502 & 33950.3 & 24.5 & HPF\\
2459623.684597 & 33881.0 & 20.3 & HPF\\
2459198.758447 & 34065.5 & 9.7 & HARPS\\
2459219.795652 & 34062.2 & 8.3 & HARPS\\
2459222.729577 & 34079.9 & 8.2 & HARPS\\
2457775.765668 & 33972.3 & 25.2 & HARPSN\\
2457776.49722 & 33954.1 & 12.6 & HARPSN\\
2457778.459009 & 33943.2 & 11.4 & HARPSN\\
2457779.48129 & 33960.7 & 12.7 & HARPSN\\
2457780.48637 & 33938.5 & 9.5 & HARPSN\\
2457780.708583 & 33952.8 & 6.3 & HARPSN\\
2457780.770841 & 33952.4 & 9.1 & HARPSN\\
2457781.431425 & 33952.8 & 8.0 & HARPSN\\
2457773.387611 & 33907.5 & 8.6 & SOPHIE\\
2457781.465175 & 34006.1 & 17.7 & SOPHIE\\

       \hline
    \end{tabular}
    \label{tab:rv1429}
\end{table*}

\begin{table*}
    \centering
    \caption{Radial Velocities for S1083.  All radial velocities are corrected to the zero-point of HARPS using the offset derived from the periodogram analysis with RVSearch.}
    \begin{tabular}{l|l|l|c}
    \hline
      Time (BJD)   & RV (m/s) & RVerr (m/s) & Instrument \\
      \hline
      \hline
       2458824.864623 & 33337.7 & 36.5 & HPF\\
2458862.755668 & 33506.6 & 72.6 & HPF\\
2458894.675127 & 33484.4 & 61.6 & HPF\\
2459135.006272 & 33495.7 & 44.0 & HPF\\
2459164.927214 & 33455.1 & 40.9 & HPF\\
2459169.912316 & 33380.7 & 49.5 & HPF\\
2459171.912769 & 33382.0 & 57.3 & HPF\\
2459178.884257 & 33396.2 & 29.2 & HPF\\
2459189.862508 & 33422.3 & 50.7 & HPF\\
2459219.778943 & 33434.7 & 39.3 & HPF\\
2459230.752283 & 33476.0 & 75.4 & HPF\\
2459249.701833 & 33412.8 & 66.0 & HPF\\
2459299.755111 & 33343.9 & 59.5 & HPF\\
2459300.754056 & 33337.1 & 112.5 & HPF\\
2459303.731384 & 33410.9 & 58.0 & HPF\\
2459305.709891 & 33336.1 & 84.8 & HPF\\
2459306.731775 & 33407.6 & 33.8 & HPF\\
2459324.668689 & 33433.9 & 36.0 & HPF\\
2459332.644789 & 33305.9 & 57.9 & HPF\\
2459332.655769 & 33413.0 & 54.1 & HPF\\
2459509.981625 & 33439.4 & 23.4 & HPF\\
2459513.968129 & 33562.1 & 77.3 & HPF\\
2459533.927667 & 33366.6 & 87.5 & HPF\\
2459535.913327 & 33369.4 & 54.5 & HPF\\
2459550.870572 & 33342.3 & 52.0 & HPF\\
2459552.87052 & 33452.8 & 59.2 & HPF\\
2459562.849355 & 33325.9 & 42.4 & HPF\\
2459563.019511 & 33413.9 & 49.8 & HPF\\
2459563.840682 & 33352.1 & 50.4 & HPF\\
2459564.83596 & 33385.5 & 62.0 & HPF\\
2459621.680908 & 33417.6 & 69.6 & HPF\\
2457776.756743 & 33384.0 & 20.4 & HARPSN\\
2457777.74405 & 33391.0 & 15.0 & HARPSN\\
2457778.746702 & 33409.9 & 8.5 & HARPSN\\
2457780.7281 & 33414.0 & 6.8 & HARPSN\\
2457774.582653 & 33426.8 & 8.2 & SOPHIE\\
2457783.681696 & 33380.6 & 10.2 & SOPHIE\\
       \hline
    \end{tabular}
    \label{tab:rv1083}
\end{table*}

\begin{table*}
    \centering
    \caption{Radial Velocities for S1268. All radial velocities are corrected to the zero-point of HARPS using the offset derived from the periodogram analysis with RVSearch.}
    \begin{tabular}{l|l|l|c}
    \hline
      Time (BJD)   & RV (m/s) & RVerr (m/s) & Instrument \\
      \hline
      \hline
       2458832.842448 & 32785.0 & 44.3 & HPF\\
2458918.79 & 32742.2 & 83.8 & HPF\\
2458965.66473 & 32776.9 & 43.1 & HPF\\
2459145.979875 & 32769.8 & 50.6 & HPF\\
2459165.925582 & 32761.2 & 51.7 & HPF\\
2459170.920553 & 32821.6 & 54.0 & HPF\\
2459175.912693 & 32776.1 & 23.8 & HPF\\
2459177.912031 & 32880.8 & 75.5 & HPF\\
2459256.691575 & 32783.5 & 30.8 & HPF\\
2459268.82989 & 32737.7 & 26.6 & HPF\\
2459305.724335 & 32749.1 & 21.6 & HPF\\
2459308.715345 & 32754.4 & 27.2 & HPF\\
2459316.69445 & 32729.6 & 38.1 & HPF\\
2459326.665316 & 32677.1 & 53.5 & HPF\\
2459326.676296 & 32652.4 & 99.3 & HPF\\
2459512.991327 & 32822.7 & 41.6 & HPF\\
2459513.981815 & 32833.8 & 50.5 & HPF\\
2459532.938283 & 32786.2 & 79.4 & HPF\\
2459539.901405 & 32713.7 & 105.0 & HPF\\
2459554.867492 & 32797.6 & 54.4 & HPF\\
2459556.857563 & 32702.8 & 57.5 & HPF\\
2459560.026759 & 32842.8 & 89.0 & HPF\\
2459565.022153 & 32752.2 & 50.0 & HPF\\
2459565.835669 & 32746.5 & 77.5 & HPF\\
2459572.993549 & 32741.3 & 62.7 & HPF\\
2459575.988949 & 32803.5 & 59.4 & HPF\\
2459577.981434 & 32822.6 & 52.9 & HPF\\
2459621.847966 & 32812.5 & 23.7 & HPF\\
2459622.861286 & 32723.7 & 59.7 & HPF\\
2459625.684702 & 32765.8 & 33.3 & HPF\\
2459629.666377 & 32725.9 & 64.2 & HPF\\
2459634.652938 & 32802.7 & 102.4 & HPF\\
2459643.629668 & 32788.9 & 47.2 & HPF\\
2459647.619611 & 32772.7 & 57.4 & HPF\\
2459648.6208 & 32795.5 & 27.5 & HPF\\
2459651.60757 & 32686.4 & 35.4 & HPF\\
2459652.606782 & 32734.3 & 74.9 & HPF\\
2459262.732283 & 32778.0 & 7.2 & HARPS\\
2459263.624862 & 32761.1 & 6.2 & HARPS\\
2457777.565723 & 32745.7 & 17.6 & HARPSN\\
2457778.551779 & 32779.3 & 5.9 & HARPSN\\
2457779.597335 & 32747.8 & 11.8 & HARPSN\\
2457780.561026 & 32766.4 & 5.1 & HARPSN\\
2457774.674516 & 32745.7 & 21.5 & SOPHIE\\
2457782.673345 & 32775.9 & 12.3 & SOPHIE\\

       \hline
    \end{tabular}
    \label{tab:rv1268}
\end{table*}

\begin{table*}
    \centering
    \caption{Radial Velocities for YBP778. HPF radial velocities are corrected to the zero-point of HARPS using the offset derived in Section \ref{obsrv}.}
    \begin{tabular}{l|l|l|c}
    \hline
      Time (BJD)   & RV (m/s) & RVerr (m/s) & Instrument \\
      \hline
      \hline
    2459284.657134 & 33289.4 & 11.1 & HARPS\\
    2458832.017021 & 33200.9 & 51.9 & HPF\\
    2459173.897188 & 33217.0 & 196.9 & HPF\\
    2459219.94823 & 33179.5 & 99.8 & HPF\\

       \hline
    \end{tabular}
    \label{tab:rv778}
\end{table*}

\begin{table*}
    \centering
    \caption{Radial Velocities for YBP1062. HPF radial velocities are corrected to the zero-point of HARPS using the offset derived in Section \ref{obsrv}.}
    \begin{tabular}{l|l|l|c}
    \hline
      Time (BJD)   & RV (m/s) & RVerr (m/s) & Instrument \\
      \hline
      \hline
       2458834.007889 & 31871.9 & 271.9 & HPF\\
2459288.637082 & 32923.5 & 27.3 & HARPS\\
2459166.921989 & 33329.9 & 142.8 & HPF\\
2459255.683439 & 33485.8 & 112.4 & HPF\\
       \hline
    \end{tabular}
    \label{tab:rv1062}
\end{table*}

\begin{table*}
    \centering
    \caption{Radial Velocities for YBP1137. HPF radial velocities are corrected to the zero-point of HARPS using the offset derived in Section \ref{obsrv}.}
    \begin{tabular}{l|l|l|c}
    \hline
      Time (BJD)   & RV (m/s) & RVerr (m/s) & Instrument \\
      \hline
      \hline
       2458833.014716 & 34127.6 & 161.2 & HPF\\
2459145.994203 & 34398.5 & 113.9 & HPF\\
2459231.74911 & 34250.8 & 418.9 & HPF\\
2459621.862769 & 33960.5 & 309.3 & HPF\\
       \hline
    \end{tabular}
    \label{tab:rv1137}
\end{table*}

\begin{table*}
    \centering
    \caption{Radial Velocities for YBP2018. HPF radial velocities are corrected to the zero-point of HARPS using the offset derived in Section \ref{obsrv}.}
    \begin{tabular}{l|l|l|c}
    \hline
      Time (BJD)   & RV (m/s) & RVerr (m/s) & Instrument \\
      \hline
      \hline
       2459286.670183 & 34226.5 & 22.3 & HARPS\\
2458835.826672 & 33597.2 & 121.4 & HPF\\
2459141.983661 & 34091.7 & 123.4 & HPF\\
2459172.907692 & 34138.2 & 122.1 & HPF\\
2459622.67856 & 34350.6 & 121.4 & HPF\\
    \hline
    \end{tabular}
    \label{tab:rv2018}
\end{table*}

\begin{table*}
    \centering
    \caption{Radial Velocities for S488. HPF radial velocities are corrected to the zero-point of HARPS using the offset derived in Section \ref{obsrv}.}
    \begin{tabular}{l|l|l|c}
    \hline
      Time (BJD)   & RV (m/s) & RVerr (m/s) & Instrument \\
      \hline
      \hline
       2458832.026981 & 32794.1 & 211.7 & HPF\\
2458945.715585 & 32700.6 & 157.2 & HPF\\
2459141.994246 & 33166.5 & 217.8 & HPF\\
2459158.947758 & 33167.2 & 231.5 & HPF\\
2459194.02845 & 32937.7 & 167.9 & HPF\\
2459203.82112 & 32705.3 & 233.9 & HPF\\
       \hline
    \end{tabular}
    \label{tab:rv488}
\end{table*}

\begin{table*}
    \centering
    \caption{Radial Velocities for S815. HPF radial velocities are corrected to the zero-point of HARPS using the offset derived in Section \ref{obsrv}.}
    \begin{tabular}{l|l|l|c}
    \hline
      Time (BJD)   & RV (m/s) & RVerr (m/s) & Instrument \\
      \hline
      \hline
       2458832.828524 & 34114.8 & 165.2 & HPF\\
2458975.628515 & 34322.1 & 142.8 & HPF\\
2459137.99363 & 34249.8 & 144.7 & HPF\\
2459172.893022 & 34297.0 & 116.6 & HPF\\
2459202.824545 & 34468.5 & 138.9 & HPF\\
       \hline
    \end{tabular}
    \label{tab:rv815}
\end{table*}

\begin{table*}
    \centering
    \caption{Radial Velocities for S995. HPF radial velocities are corrected to the zero-point of HARPS using the offset derived in Section \ref{obsrv}.}
    \begin{tabular}{l|l|l|c}
    \hline
      Time (BJD)   & RV (m/s) & RVerr (m/s) & Instrument \\
      \hline
      \hline
       2457773.457212 & 33753.7 & 6.9 & SOPHIE\\
2457782.433587 & 33614.6 & 10.2 & SOPHIE\\
2458826.858757 & 34419.7 & 195.3 & HPF\\
2458836.002384 & 34470.4 & 69.6 & HPF\\
2458837.826586 & 34341.5 & 63.6 & HPF\\
2459140.988252 & 34208.6 & 117.7 & HPF\\
2459146.981718 & 34151.0 & 122.1 & HPF\\
2459157.950757 & 34198.7 & 121.1 & HPF\\
2459159.941358 & 34037.6 & 150.2 & HPF\\
2459190.856064 & 33785.0 & 106.9 & HPF\\
2459237.903511 & 33651.6 & 142.0 & HPF\\
2459265.842299 & 32819.6 & 120.7 & HPF\\
2459592.933403 & 34288.6 & 63.8 & HPF\\
2459262.715833 & 32752.5 & 11.3 & HARPS\\
2459263.608192 & 32715.7 & 8.4 & HARPS\\
       \hline
    \end{tabular}
    \label{tab:rv995}
\end{table*}

\begin{table*}
    \centering
    \caption{Radial Velocities for S1557. HPF radial velocities are corrected to the zero-point of HARPS using the offset derived in Section \ref{obsrv}.}
    \begin{tabular}{l|l|l|c}
    \hline
      Time (BJD)   & RV (m/s) & RVerr (m/s) & Instrument \\
      \hline
      \hline
       2458829.853192 & 33529.0 & 224.6 & HPF\\
2458969.648162 & 33837.1 & 295.5 & HPF\\
2459142.998912 & 34003.0 & 218.1 & HPF\\
2459171.925322 & 33932.1 & 189.8 & HPF\\
2459192.853267 & 33734.7 & 193.9 & HPF\\
       \hline
    \end{tabular}
    \label{tab:rv1557}
\end{table*}

\begin{table*}
    \centering
    \caption{Radial Velocities for S2207. HPF radial velocities are corrected to the zero-point of HARPS using the offset derived in Section \ref{obsrv}.}
    \begin{tabular}{l|l|l|c}
    \hline
      Time (BJD)   & RV (m/s) & RVerr (m/s) & Instrument \\
      \hline
      \hline
       2457776.547733 & 34554.9 & 9.6 & HARPSN\\
2457777.533594 & 34556.9 & 14.1 & HARPSN\\
2457778.516941 & 34562.0 & 7.2 & HARPSN\\
2457779.539985 & 34572.2 & 7.0 & HARPSN\\
2457780.525817 & 34570.1 & 6.1 & HARPSN\\
2457781.46711 & 34589.5 & 11.4 & HARPSN\\
2457775.557763 & 34569.5 & 8.0 & SOPHIE\\
2457783.65828 & 34565.7 & 8.1 & SOPHIE\\
2458829.0281 & 31256.4 & 85.0 & HPF\\
2458858.768689 & 31274.2 & 153.0 & HPF\\
2458861.760188 & 31220.9 & 55.5 & HPF\\
2459142.984378 & 32173.5 & 138.7 & HPF\\
2459173.911128 & 32204.7 & 99.1 & HPF\\
2459175.897398 & 32254.9 & 101.8 & HPF\\
2459177.895872 & 32184.3 & 131.2 & HPF\\
2459178.898196 & 32084.4 & 65.4 & HPF\\
2459252.689992 & 32455.6 & 149.3 & HPF\\
2459299.726985 & 32657.8 & 93.0 & HPF\\
2459300.724449 & 32593.5 & 91.6 & HPF\\
2459303.715102 & 32673.6 & 95.7 & HPF\\
2459305.738517 & 32619.7 & 71.4 & HPF\\
2459596.761071 & 34197.5 & 105.0 & HPF\\
2459195.798301 & 32257.2 & 8.1 & HARPS\\
2459198.77405 & 32274.3 & 7.7 & HARPS\\
2459219.778586 & 32349.4 & 6.4 & HARPS\\
2459222.749411 & 32349.2 & 6.6 & HARPS\\
       \hline
    \end{tabular}
    \label{tab:rv2207}
\end{table*}

\end{appendix}
\end{document}